\documentclass[aps,prx,twocolumn,longbibliography,superscriptaddress]{revtex4-1}

\usepackage{amssymb}
\usepackage{multirow}
\usepackage{bm}
\usepackage{amsmath}
\usepackage{graphicx}
\usepackage{epstopdf}
\usepackage{subfigure}
\usepackage{natbib}
\usepackage{epsfig}
\usepackage{amsfonts}
\usepackage{mathrsfs}
\usepackage[toc,page,title,titletoc,header]{appendix}
\usepackage[colorlinks,linkcolor=blue,citecolor=blue,anchorcolor=blue]{hyperref}
\newcommand{\bea}{\begin{equation} \begin{aligned}}
\newcommand{\eea}{\end{aligned} \end{equation} }

\newcommand{\be}{\beta}
\newcommand{\al}{\alpha}
\usepackage{braket}
\DeclareRobustCommand{\Eq}[1]{Eq.~(\ref{#1})}

\usepackage{dsfont,amsthm,amsbsy}

\def\tr{{\rm tr}}

\newcommand{\mbf}[1]{\mathbf{#1}}

\usepackage{fancyhdr}

\usepackage{bbm}

\begin{document}
\title{``Quantum Geometric Nesting'' and Solvable Model  Flat-Band Systems}

\author{Zhaoyu~Han}
\affiliation{Department of Physics, Stanford University, Stanford, CA 94305, USA}

\author{Jonah~Herzog-Arbeitman}
\affiliation{Department of Physics, Princeton University, Princeton, NJ 08544, USA}

\author{B. Andrei Bernevig}
\affiliation{Department of Physics, Princeton University, Princeton, NJ 08544, USA}
\affiliation{Donostia International Physics Center, P. Manuel de Lardizabal 4, 20018 Donostia-San Sebastian, Spain}
\affiliation{IKERBASQUE, Basque Foundation for Science, Bilbao, Spain}

\author{Steven~A.~Kivelson}
\affiliation{Department of Physics, Stanford University, Stanford, CA 94305, USA}

\begin{abstract} 
We introduce the concept of ``quantum geometric nesting'' (QGN) to characterize the idealized ordering tendencies of certain flat-band systems implicit in the geometric structure of the flat-band subspace.  Perfect QGN implies the existence of an infinite class of local interactions that can be explicitly constructed and give rise to solvable ground states with various forms of possible fermion bi-linear order, including flavor ferromagnetism, density waves, and superconductivity. For the ideal Hamiltonians constructed in this way, we show that certain aspects of the low-energy spectrum can also be exactly computed including, in the superconducting case, the phase stiffness. Examples of perfect QGN include flat bands with certain symmetries (e.g. chiral or time-reversal), and non-symmetry-related cases exemplified with an engineered model for pair-density-wave. Extending this approach, we obtain exact superconducting ground states with nontrivial pairing symmetry. 
\end{abstract}
\maketitle

The properties of a correlated electronic system are largely determined by its electronic structure. One such familiar manifestation is the role of Fermi surface nesting conditions~\cite{lomer1962electronic,PhysRev.143.245}: if one portion of a Fermi surface overlaps with another under a shift by a wavevector $\bm{Q}$ (i.e. $\bm{k} \to \bm{k}+\bm{Q}$), the Fermi surface is said to be ``perfectly nested'' and consequently the system is prone to density wave (DW) order at $\bm{Q}$ with particle-hole (p-h) fermion bilinear order. If the overlap occurs under a momentum-space inversion about a certain wavevector $\bm{Q}/2$ (i.e. $\bm{k} \to \bm{Q}-\bm{k}$), the favored order is superconductivity (SC) with particle-particle (p-p) fermion bilinear order and a Cooper pair center of mass momentum $\bm{Q}$. (Such perfect nesting always exists at $\bm{Q}=0$ in time-reversal symmetric systems, where it leads to the familiar Cooper instability.) The thinking behind Fermi surface nesting is based on weak coupling mean-field analysis, in which ordering tendencies are diagnosed by divergent bare susceptibilities in corresponding channels. These concepts are indeed convenient and powerful (see e.g. Refs.~\cite{doi:10.1126/science.252.5002.96,  doi:10.1126/science.1059255, RevModPhys.75.1201,    doi:10.1126/science.1103627,  RevModPhys.78.17,  PhysRevB.77.165135, terashima2009Fermi, PhysRevB.79.014505,   nandkishore2012chiral, feng2013incommensurate,  PhysRevX.4.031017, PhysRevLett.129.167001} for typical applications). However, their validity is limited to weakly coupled systems in which the interaction strength is small compared to the bandwidth. In general, all forms of non-interacting susceptibilities grow as decreasing $T$ with the same asymptotic behavior $\chi \sim 1/T$ for $T$ greater than the bandwidth, which thus indicates comparable ordering tendencies for all possible order parameters. Significantly, this concept, defined solely by the electronic dispersion, overlooks another important piece of information in the electronic structure - the quantum geometry~\cite{provost1980riemannian,2011EPJB...79..121R} and topology~\cite{2017Natur.547..298B,2017NatCo...8...50P,PhysRevX.7.041069} encoded in the single-particle wavefunctions. An entirely different concept of ``nesting'' is thus needed for flat-band systems, whose bandwidths are small compared to the interaction strengths but which often have non-trivial quantum geometry and rich ordering phenomena~\cite{2015NatCo...6.8944P, PhysRevB.94.245149, PhysRevLett.122.246401,PhysRevB.99.045107,  PhysRevB.102.201112,  PhysRevX.10.031034, PhysRevX.11.041063, doi:10.1126/sciadv.abf5299, PhysRevB.103.205414,   PhysRevB.103.205415, ledwith2021strong, PhysRevLett.126.137601, PhysRevB.104.075143, PhysRevLett.126.027002, PhysRevResearch.3.013033, rossi2021quantum, 2021PNAS..11821826A,2022arXiv220906524W, PhysRevB.105.L140506,  2022PhRvB.106a4518H, 
2022NatRP...4..528T,PhysRevB.106.104514,
2022PhRvL.129g6401H, PhysRevB.106.165133, 
2022PhRvB.105l1110P, 2022PhRvB.106r4517Z, 2022PhRvB.106h5140H, 2022PhRvL.128o7201L, herzog2022many,2022arXiv221200030H,2022NJPh...24k3019P, PhysRevB.105.224508, PhysRevB.105.024502, PhysRevLett.122.246401, PhysRevLett.130.226001, PhysRevLett.130.216401, 2023arXiv230206250S, 2023arXiv230805686H, PhysRevB.96.064505,2023arXiv230315504C,2023PhRvB.107v4505I,2023arXiv230808248P,  2023PhRvA.107e3323I,  2023arXiv230810780T,
2023PhRvB.107x1105Z,2023PhRvL.130a6401P,crepel2023chiral,PhysRevLett.131.106801,kwan2023strong,sahay2023superconductivity,2023PhRvL.131a6002J,2023arXiv230519927S,2023arXiv230303352S,2023SciA....9I6063D,2023arXiv231112920H,2023arXiv231104958J,2023arXiv231116655W,2023SCPMA..6687212C,2023arXiv230710012S,PhysRevB.105.045112,PhysRevB.107.L201106,2023PhRvB.107l5116K,2023PhRvB.107x5145S,2023PhRvB.107s5403J,2023arXiv230502340Y,2023arXiv230902483P,2023arXiv230113870S,2023arXiv230701253V,setty2023electronic}.

In this paper, we propose such a concept, which we call ``Quantum Geometric Nesting'' (QGN), that focuses on the geometric aspects of the electronic structure in flat-band systems. In contrast to Fermi surface nesting, here we start from a strong-coupling perspective in which the band dispersion is assumed negligible in comparison with the interaction strength, and then ask: for a given quantum geometry of the flat bands, are there any natural forms of order? If so, what sort of interactions favor them? We will show that these questions can be addressed by examining whether a set of QGN conditions is satisfied. Intuitively, these conditions refer to certain forms of `overlap' of the {\it wavefunctions} in the flat band subspace at wavevectors related by a shift (p-h case) or inversion (p-p case) in momentum space, akin to the familiar case of Fermi surface nesting. For ``perfectly nested quantum geometry,'' we construct ideal Hamiltonians containing arbitrary combinations of an {\it infinite} class of engineered short-range two-body interactions that have {\it solvable} ground states with fermion bilinear order of a character determined by the quantum geometry. For these ideal Hamiltonians, the few-body excitations are solvable as well, including single-particle (quasi-particle), particle-hole (exciton), and two-particle (Cooper pair)  excitations. For the p-p case, we further show that these ideal Hamiltonians host a (fine-tuned) pseudospin $SU(2)$ symmetry, whose stiffness can be inferred from the excitation spectra. We prove that certain symmetries (e.g. chiral or time-reversal) can ensure particular forms of QGN in flat bands. Non-symmetry-related cases also exist, such as an engineered model for pair-density-wave order that we construct. Extending the construction scheme, we further obtain an SSH-type model with solvable $p$-wave pairing groundstates. 

Indeed, except for symmetry-enforced cases, it is unlikely for a physical system to have exact QGN - in common with the case of Fermi surface nesting. However, the use of QGN can be much more versatile: for systems with almost perfect QGN (quantification of the closeness to the ideal nesting will be defined below as `nestability'), one may engineer an idealized electronic structure with perfect QGN and then construct ideal interacting Hamiltonians that are close to the physical one. The solvability of these ideal models thus makes them excellent starting points for including perturbative corrections to reach the physical Hamiltonian. This type of intrinsically strong coupling viewpoint (in the sense that the dominant part of the interactions has been accounted for in the solvable models) has proven fruitful in recent studies of twisted bilayer graphene and other flat-band materials~\cite{PhysRevB.94.245149, PhysRevLett.126.137601, PhysRevB.104.075143, PhysRevLett.122.246401, doi:10.1126/sciadv.abf5299, PhysRevX.10.031034, PhysRevB.103.205414, PhysRevB.103.205415, ledwith2021strong, 2022PhRvB.105l1110P,2022PhRvL.128o7201L,herzog2022many, 2022PhRvB.106r4517Z,2022PhRvB.106h5140H,crepel2023chiral,2023PhRvB.107x1105Z,2023PhRvL.130a6401P,PhysRevLett.131.106801,kwan2023strong}. It is noteworthy that perfect QGN is an underlying feature of most of the known examples of solvable models for flat-band systems featuring fermion bi-linear order. We hence expect that QGN can greatly facilitate studies within this paradigm, in a way similar to the roles played by `ideal band conditions'~\cite{PhysRevB.90.165139,PhysRevLett.114.236802, PhysRevB.104.115160, PhysRevLett.128.176404, PhysRevLett.127.246403,PhysRevLett.128.176403,PhysRevResearch.5.023167,PhysRevResearch.5.023166} and ``vortexability''~\cite{PhysRevB.108.205144} in studies of fractional Chern insulator states.

We will present the conclusions in the main text and defer the detailed proofs and technical discussions to the corresponding sections in the Supplemental Materials (SM). In Fig.~\ref{fig: layout}, we provide a schematic flowchart for the central results as a guide for readers.

\begin{figure}[t!]
    \centering
    \includegraphics[width = 0.8 \linewidth]{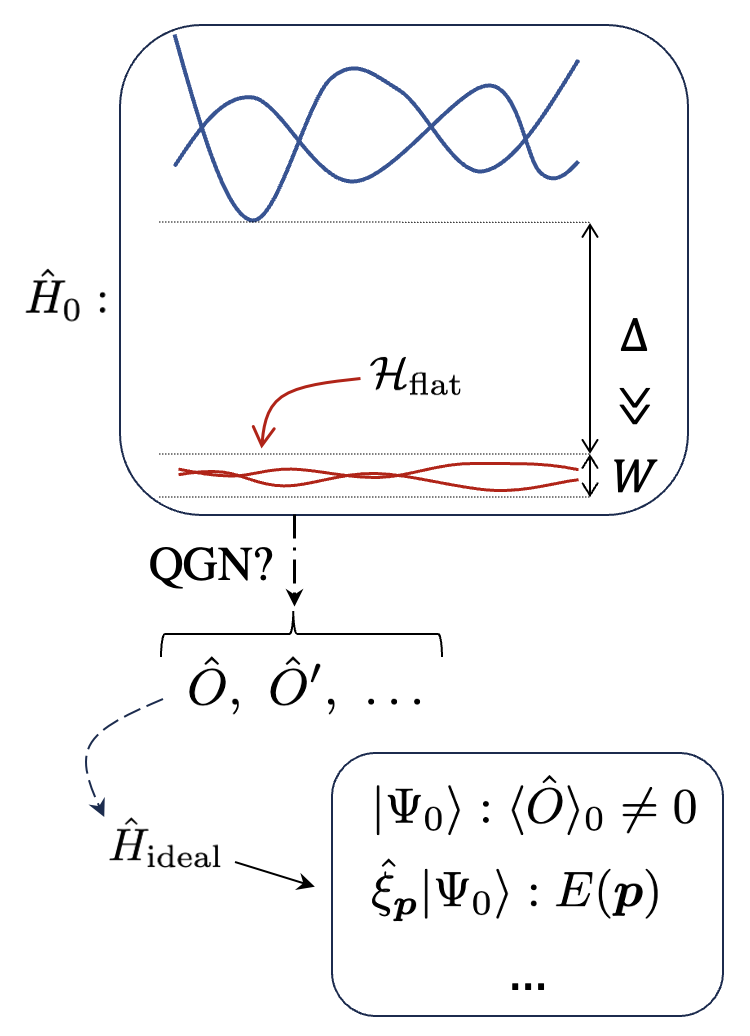}
    \caption{Schematic flowchart illustrating the structure of the theory: Given any electronic structure containing isolated flat bands, one can test whether a QGN condition is satisfied; if it is, one or more fermion bilinear order parameters (OP) will be suggested (Sec.~\ref{sec: formalism}). For each OP, infinitely many ``ideal'' interacting Hamiltonians can be systematically constructed (Sec.~\ref{sec: interactions}), which, after projecting onto the flat-band Hilbert space $\mathcal{H}_\text{flat}$, feature solvable ground states with the corresponding order (Sec.~\ref{sec: ground state}) and solvable aspects of the collective excitation spectrum (Sec.~\ref{sec: excitations}). A practical recipe for using QGN in perturbative analyses of physical systems is proposed in Sec.~\ref{sec: interactions}. Some examples of electronic structures with perfect QGN can be found in Sec.~\ref{sec: examples}. Possible extensions of this approach are commented on in Sec.~\ref{sec: extensions}.}
    \label{fig: layout}
\end{figure}

\section{Formalism}
\label{sec: formalism}

\subsection{Preliminaries}

We consider an electronic structure encoded in a tight-binding model~\footnote{One could also formulate the model in continuum and work in momentum space throughout the discussion; no conclusions will be affected so long as lattice is not considered. Here we start with an explicit tight-binding model to avoid the subtleties when discussing the locality of the interactions and to connect to some existing literature. } with charge conservation and translation symmetry among $\mathsf{V}$ unit cells in arbitrary dimension:
\begin{align}
\hat{H}_0 = \sum_{\bm{R}\bm{R}', \mu\nu} t_{\mu\nu}(\bm{R}-\bm{R}') \ \hat{c}^\dagger_{\bm{R},\mu} \hat{c}_{\bm{R}',\nu}
\end{align}
where $\bm{R}$ indicates the unit cell position, Greek letters in lowercase $\mu,\nu,\dots$ denote a ``flavor'' index which, for simplicity, includes both orbital and spin indices unless otherwise specified. The annihilation operators for Bloch states with a given flavor index are expressed as  $\hat{c}_{\bm{k},\mu}\equiv\sum_{\bm{R}}\mathrm{e}^{\mathrm{i} \bm{k}\cdot \bm{R}} \hat{c}_{\bm{R},\mu}/\sqrt{\mathsf{V}}$ in the periodic embedding.~\footnote{QGN can be defined for any embedding of the intra-unit-cell positions $\{\bm{r}_\mu\}$, but the periodic embedding $\{\bm{r}_\mu=\bm{0}\}$ is a special one since if a certain QGN condition is satisfied for {\it any} embedding, it is also satisfied in the periodic embedding. It thus suffices to always work with periodic embedding. In the sense that the embedding can be fixed, QGN is an {\it embedding-independent} concept, which is necessary for QGN to be a physical concept since embedding is purely a redundancy in the definition of Bloch basis \cite{2022PhRvB.106a4518H}. } In momentum space, the single particle Hamiltonian is diagonalized with a band basis
\begin{align}
\hat{H}_0 &= \sum_{\bm{k}, n} \epsilon_n(\bm{k}) \hat{\gamma}^\dagger_{\bm{k},n} \hat{\gamma}_{\bm{k},n}, \quad 
\hat{\gamma}_{\bm{k},n} = U_{n\mu}^\dagger(\bm{k}) \hat{c}_{\bm{k},\mu},
\label{gamma}
\end{align}
where $U_{n\mu}^\dagger(\bm{k})$ are unitary matrices and we use Latin letters in lowercase ($n,m,\dots$) to label the bands. We assume there are $\mathsf{N}$ flavors and thus $\mathsf{N}$ bands in total.

We will quantify the degree of QGN for flat-band systems by utilizing the geometric information in $U_{\mu n}(\bm{k})$. To do so, we consider an idealized setup (illustrated in Fig.~\ref{fig: layout}) where there are $\mathsf{N}_\text{flat}$ nearly degenerate and flat bands isolated from all the other bands by a large gap $\Delta$, such that the relevant interaction strength $V$ is small compared to $\Delta$ but large compared to the flat bands' bare bandwidth $W$,
\begin{align}
    W\ll V\ll \Delta.
\end{align}
Without loss of generality, we can relabel all the bands to make the $n=1\dots \mathsf{N}_\text{flat}$ bands the flat bands. We then define the $\bm{k}$-dependent projection matrix for the flat bands
\begin{align}\label{eq: projector}
    P_{\mu \nu} (\bm{k})\equiv  \sum_{n\le \mathsf{N}_\text{flat}} U_{\mu n }(\bm{k}) U_{n\nu}^\dagger(\bm{k}) 
\end{align}
and, for later convenience, the projection matrix onto the remaining bands, $Q_{\mu \nu} (\bm{k})\equiv \delta_{\mu\nu} - P_{\mu \nu} (\bm{k})$.

For the ideal scenario we are assuming, we define the ``flat-band subspace'', $\mathcal{H}_\text{flat}$, as the tensor product of the Fock space of the flat bands and a ``vacuum'' state of the other modes $\gamma_{\bm{k},n=\mathsf{N}_\text{flat}+1\dots \mathsf{N}}$ (i.e. all the modes with energy higher or lower than the flat bands are respectively empty or occupied). We formally define the projector onto $\mathcal{H}_\text{flat}$ as $\hat{P}$. To leading order of the perturbation series treating both $V/\Delta$ and $W/V$ as small parameters, we can set $\hat{H}_0 =0$ and project the interactions, $\hat{H}_\text{int}$, onto $\mathcal{H}_\text{flat}$. Formally, this means studying the effective Hamiltonian $\hat{P}\hat{H}_\text{int} \hat{P}$, but operationally this can be more conveniently done by retaining terms that only involve $\hat{\gamma}_{\bm{k}, n\le \mathsf{N}_\text{flat}}$, or equivalently, by replacing all $\hat{c}_{\bm{k}\mu}\rightarrow P_{\mu \nu}(\bm{k}) \hat{c}_{\bm{k} \nu}$ in the interacting Hamiltonian~\footnote{We note that this scheme in general will result in some additional quadratic terms when compared to $\hat{P} \hat{H}_\text{int}\hat{P}$, which can be operationally viewed as part of $\hat{H}_0$. See Sec.~\ref{sec: interactions} below and Sec.~II~D of SM for detailed discussions.}.

\subsection{General Definitions} 

In terms of the gauge-invariant projection matrices $P$ and $Q$, the key information of QGN at a given wavevector $\bm{Q}$ for p-p or p-h channel is encoded in the linear operator
\begin{align}
    \Pi^{\text{p-p},\bm{Q}}_{\mu'\nu';\mu\nu} =\frac{1}{\mathsf{V}}\sum_{\bm{k}}\left[P^\star_{\mu'\mu}(\frac{\bm{Q}}{2}+\bm{k}) Q_{\nu \nu' } (\frac{\bm{Q}}{2}-\bm{k})+ (P\leftrightarrow Q) \right], \label{eq: p-p Pi}\\
    \Pi^{\text{p-h},\bm{Q}}_{\mu'\nu';\mu\nu} =\frac{1}{\mathsf{V}}\sum_{\bm{k}}\left[P_{\mu'\mu}(\bm{k}+\frac{\bm{Q}}{2}) Q_{\nu \nu' } (\bm{k}-\frac{\bm{Q}}{2}) +  (P\leftrightarrow Q) \right]. \label{eq: p-h Pi}
\end{align}
where $^\star$ means complex conjugation. When viewed as a matrix (with $\mu\nu$ a grouped index and $\mu'\nu'$ the other), $\Pi$ is hermitian and positive semi-definite and can be diagonalized with non-negative eigenvalues. 

{\it By definition}, an electronic structure has ``perfect QGN'' at a wavevector $\bm{Q}$ in the p-p or p-h channel when the corresponding $\Pi$ has eigenvalue(s) equal to $0$. In other words, a perfect QGN is satisfied if there exists a $\mathsf{N}\times\mathsf{N}$ matrix $N^{\bm{Q}}_{\mu\nu}$, which we call the ``nesting matrix'', satisfying (for any $\mu'\nu'$)
\begin{align}\label{eq: QGN definition}
\sum_{\mu\nu}\Pi_{\mu'\nu';\mu\nu}^{\bm{Q}}N^{\bm{Q}}_{\mu\nu} =0.
\end{align}

We will show that the favored OP operator within the flat-band subspace can be expressed in terms of $N^{\bm{Q}}$ as: 
\begin{align}
    \hat{O}^{\text{p-p},\bm{Q}} &\equiv \frac{1}{\mathsf{V}} \sum_{\bm{k}; n,m\le \mathsf{N}_\text{flat}} F^{\text{p-p},\bm{Q}}_{nm} (\bm{k})\hat{\gamma}_{\bm{Q}/2+\bm{k}, n} \hat{\gamma}_{\bm{Q}/2-\bm{k},m}  \label{eq: pp order parameter} \\
  \hat{O}^{\text{p-h},\bm{Q}} &\equiv \frac{1}{\mathsf{V}} \sum_{\bm{k};n,m\le \mathsf{N}_\text{flat}} F^{\text{p-h},\bm{Q}}_{nm} (\bm{k})\hat{\gamma}^\dagger_{\bm{k}+\bm{Q}/2, n} \hat{\gamma}_{\bm{k}-\bm{Q}/2,m} \label{eq: ph order parameter}
\end{align}
where $F(\bm{k})$ is the form factor in the band basis corresponding to $N^{\bm{Q}}$ being the unprojected form factor in the original orbital basis, 
\begin{align}
 \ &F^{\text{p-p},\bm{Q}}_{nm}(\bm{k}) \equiv \sum_{\mu \nu} U_{\mu n } (\frac{\bm{Q}}{2}+\bm{k}) N^{\text{p-p},\bm{Q}}_{\mu\nu} U_{\nu m} (\frac{\bm{Q}}{2}-\bm{k})\label{eq: p-p form factor} \\
  \ &F^{\text{p-h},\bm{Q}}_{nm}(\bm{k}) \equiv \sum_{\mu \nu} U^\dagger_{ n\mu } (\bm{k}+\frac{\bm{Q}}{2}) N^{\text{p-h},\bm{Q}}_{\mu\nu} U_{\nu m} (\bm{k}-\frac{\bm{Q}}{2}). \label{eq: p-h form factor}
\end{align}

One important property (which can actually serve as an equivalent definition) of QGN is that for arbitrary $\bm{k}$, the form factor $F_{nm}(\bm{k})$ defined by a nesting matrix is a block-diagonal matrix that disconnects flat band indices from the others, i.e. $F_{nm}(\bm{k})=0$ as long as $n,m$ are not simultaneously $\le \mathsf{N}_\text{flat}$ or $> \mathsf{N}_\text{flat}$. In other words, QGN implies that the wavefunctions of the flat bands at two momenta related by a shift (p-h case) or inversion (p-p case) are fully compatible, in the sense $N^{\bm{Q}}$ maps the single-particle flat band null spaces to each other. As a direct result, the static susceptibility of the OP defined in Eqs.~\ref{eq: pp order parameter}\&\ref{eq: ph order parameter} can be shown (see Sec.~I~C of SM) to strictly saturate an upper bound when the corresponding QGN is satisfied, suggesting the harmony between the quantum geometry and the OP in the ideal circumstances. We will further show in Sec.~\ref{sec: interactions} that an infinite class of interacting solvable models can be constructed, which host ground states with the corresponding order. 

For QGN to be non-trivial in the p-p channel, it is additionally required that $N^{\bm{Q}}$ cannot be symmetric since 
$\hat{O}^{\text{p-p},\bm{Q}}$ vanishes in this case. (It is proven by explicit construction that valid, anti-symmetric solutions can indeed exist for $\Pi$.) On the other hand, the identity matrix is always an allowed solution for Eq.~\ref{eq: QGN definition} in the p-h case for $\bm{Q}=\bm{0}$, but this is a trivial one in the sense that the resulting $\hat{O}^{\text{p-h},\bm{Q}=\bm{0}}$ is simply the total density operator not associated with any broken symmetry. Excluding all the unwanted solutions, for a generic flat-band system, the value of the minimal eigenvalue of $\Pi^{\bm{Q}}$, $\varpi_0^{\bm{Q}}$, is a natural dimensionless measure of ``nestability'' that quantifies the `distance' of a system to perfect QGN in the corresponding channel at $\bm{Q}$~\footnote{We note that these concepts are spiritually similar to the ``ideal band condition''~\cite{PhysRevB.90.165139,PhysRevLett.114.236802, PhysRevB.104.115160, PhysRevLett.128.176404, PhysRevLett.127.246403,PhysRevLett.128.176403,PhysRevResearch.5.023167,PhysRevResearch.5.023166} and ``vortexability''~\cite{PhysRevB.108.205144} which are concerned with whether a Chern band is suitable for hosting fractional quantum Hall states.}. Later in Sec.~\ref{sec: examples} we will give several examples of electronic structures with perfect QGN, i.e. $\varpi_0^{\bm{Q}} =0$.

Depending on the channel, momentum $\bm{Q}$, and nesting matrix $N^{\bm{Q}}$, Eq.~\ref{eq: pp order parameter} or \ref{eq: ph order parameter} represent fermion bilinear orders with different natures. In the p-p channel, if $\bm{Q}=\bm{0}$, the order is a uniform SC order; while for $\bm{Q}\neq \bm{0}$ it is pair-density-wave (PDW, not to be confused with the other DWs in the p-h channel) order of the single-$\bm{Q}$, Fulde–Ferrell type~\cite{agterberg2020physics}. In the p-h channel, if $\bm{Q} = \bm{0}$, non-trivial QGN ($N^{\text{p-h}, \bm{0}}\neq \mathbbm{1}$) leads to flavor-symmetry-breaking orders, including flavor ferromagnetism and inter-flavor coherent states; the $\bm{Q}\neq \bm{0}$ cases include various forms of DWs (e.g., spin, charge, or current/flux DWs).

A system can satisfy multiple QGN conditions in different channels and/or at different $\bm{Q}$s. It is also possible that there can be more than one linearly independent nesting matrix for each channel and $\bm{Q}$, which typically signals a further internal symmetry breaking, e.g. triplet SC in spin invariant systems.

We mention that each satisfied QGN condition allows for constructing a {\it generalized} quantum geometry in the sense that it defines a notion of distance in momentum space and thus a generalized metric tensor, which can be related to the second spatial moment of the OP density {\it bare} correlation functions originating from the flat bands. The familiar definition of quantum distance $d(\bm{k},\bm{k}') \equiv \sqrt{\mathsf{N}_{\text{flat}}- \tr{\left[P(\bm{k})P(\bm{k}')\right]}}$~\cite{provost1980riemannian,2011EPJB...79..121R} using the Hilbert-Schmidt norm turns out to be one special case in this class of constructions, which is always valid since it corresponds to the `trivial QGN' we mentioned above with $N^{\text{p-h},\bm{Q}=\bm{0}}=\mathbbm{1}$. We refer interested readers to Sec.~I~D of SM for formal definitions and discussions.

\section{Ideal interacting Hamiltonians} 
\label{sec: interactions} 

We now present a general strategy to construct a solvable ``ideal'' interacting Hamiltonian, $\hat{H}_\text{int}$, for a flat-band system with perfect QGN. It is an engineered model in the sense that it targets the OP $\hat O$ defined as in Eq.~\ref{eq: pp order parameter} or \ref{eq: ph order parameter};  specifically, it is constructed to have ground states that exhibit a pattern of spontaneous symmetry breaking associated with an anomalous expectation value of $\hat O$. For reasons that will be made clear in the next section, we write $\hat{H}_\text{int}$ in the form:
\begin{align}\label{eq: interaction}
    \hat{H}_\text{int} = \sum_{\bm{R}\bm{R}'; IJ} V_{IJ} (\bm{R}-\bm{R}')   \left(\hat{S}^{(I)}_{\bm{R}}-\langle \hat{S}_{\bm{R}}^{(I)}\rangle\right) \left(\hat{S}^{(J)}_{\bm{R}'}-\langle \hat{S}_{\bm{R}'}^{(J)}\rangle\right)
\end{align}
where $V_{IJ}(\bm{R}-\bm{R}')$ is any function whose Fourier transform
\begin{align}
    V_{IJ}(\bm{q}) \equiv \frac{1}{\mathcal{V}} \sum_{\bm{R}} \mathrm{e}^{\mathrm{i} \bm{q}\cdot \bm{R}} \ V_{IJ} (\bm{R})
\end{align}
is positive semi-definite for all $\bm{q}$. Thus, by construction, the eigen-energies of $\hat{H}_\text{int}$ are non-negative. $\{\hat{S}^{(I)}_{\bm{R}}\}$ is an infinite set of linearly independent Hermitian operators centered in any unit cell $\bm{R}$, which, after projection onto $\mathcal{H}_\text{flat}$, commute with $\hat{O}$ (in Eq.~\ref{eq: pp order parameter}/\ref{eq: ph order parameter}):
\begin{align}\label{eq: S O commutator}
\hat{P}\  [ \hat{S}^{(I)}_{\bm{R}} ,\hat{O}] \ \hat{P} = 0\ . 
\end{align}
These hermitian operators, $\hat{S}^{(I)}_{\bm{R}}$, are labeled with Latin letters in uppercase, and $\langle \hat{S}_{\bm{R}}^{(I)}\rangle$ are the corresponding ground-state expectation values. It should be noted that these operators are defined in the entire Hilbert space, whereas the above commutation relation only holds after projection.

For generic flat-band systems, no set of local hermitian operators that satisfy Eq.~\ref{eq: S O commutator} exists. However, when the OP derives from a band geometry with perfect QGN, there is (at least) an infinite class of solutions that take a special flavor-space separable form
\begin{align}\label{eq: separable form}
    \hat{S}^{(I)}_{\bm{R}} = & \sum_{\bm{R}_1\bm{R}_2;\mu\nu} A^{(I)}(\bm{R}_1, \bm{R}_2)  B^{(I)}_{\mu\nu} \hat{c}^\dagger_{\bm{R}+\bm{R}_1,\mu} \hat{c}_{\bm{R}+\bm{R}_2,\nu},
\end{align}
where $A^{(I)}(\bm{R}_1,\bm{R}_2) = [A^{(I)}(\bm{R}_2,\bm{R}_1)]^\star$ is a spatial coefficient and $B^{(I)}$ is a hermitian matrix acting on orbital indices. In order for Eq.~\ref{eq: S O commutator} to hold, the constraints on the coefficient $A^{(I)}$ and matrix $B^{(I)}$ are not overly restrictive, in that they only need to obey
\begin{align}
    & A^{(I)}_{\bm{R}_1,\bm{R}_2}  = \left|A^{(I)}_{\bm{R}_1,\bm{R}_2}\right| \mathrm{e}^{-\mathrm{i}(\bm{R}_1-\bm{R}_2)\cdot \bm{Q}/2}, \\
    &N^{\text{p-p}, \bm{Q}}\cdot B^{(I)} + [B^{(I)}]^T\cdot N^{\text{p-p}, \bm{Q}}   = 0,
\end{align}
for the p-p case, and 
\begin{align}
    & A^{(I)}(\bm{R}_1,\bm{R}_2) = 0 \ \ \text{if} \ \ (\bm{R}_1-\bm{R}_2)\cdot \bm{Q} \neq 0 \text{ mod } 2\pi, \\
    &N^{\text{p-h}, \bm{Q}}\cdot B^{(I)} - B^{(I)} \cdot N^{\text{p-h}, \bm{Q}} = 0.
\end{align}
for the p-h case. We derive in Sec.~II~B\&C of SM a systematic way of constructing all possible $B^{(I)}$ and prove that there are at least $3\mathsf{N}_\text{flat}$ of them in the p-p case, and $\mathsf{N}_\text{flat}$ of them in the p-h case when $2\bm{Q}=\bm{0}$. (Some explicit worked examples are presented in Sec.~\ref{sec: examples}.) In particular, the identity matrix is always a solution for $B$ in the p-h case, which makes $\hat{S}_{\bm{R}}$ a local density operator for the unit cell at $\bm{R}$. Clearly, even given all the constraints, we are still left with an infinite class of possible interaction terms with considerable leeway, at least in choosing $V_{IJ}(\bm{R}-\bm{R}')$ and $A^{(I)}(\bm{R}_1, \bm{R}_2)$. 

Although the above construction scheme could encompass many realistic forms of interactions, it is admittedly unlikely that the electronic structure and interactions of any physical system are perfectly ideal. Therefore, the merit of QGN should be viewed in a more practical way: it can help identify solvable limits of physical flat band systems of interest. Specifically, we are proposing the following recipe for the use of QGN in the study of a physical Hamiltonian $\hat{H}_{\text{phys}}$: First, one computes its electronic structure and examine whether there is a channel and a wavevector on which QGN is nearly satisfied (i.e. the corresponding $\varpi_0$ is small compared to $1$). If there is indeed such a nearly nested possibility, one can consider its perfect nesting limit by slightly modifying the electronic structure either numerically or analytically (through recognizing certain approximate symmetries, e.g. those discussed below in Sec.~\ref{sec: examples}). Then, one can further search in the space of all possible ideal solvable interacting Hamiltonians validated by the QGN, to find one that is closest to the physical Hamiltonian, $\hat{H}_{\text{ideal}}$. This can be systematically achieved by minimizing the norm of the difference between them, $||\delta \hat{H}|| = ||\hat{H}_{\text{ideal}} - \hat{H}_\text{phys}||$. A small value of this operator distance justifies perturbative analyses starting from the ideal solvable limit. The flexibility of our construction allows the dominant part of many interactions to be captured. Furthermore, we remark that these nearby solvable limits can also facilitate numerical studies of $\hat{H}_\text{phys}$, in that one can benchmark the validity of the methods being used at these limits, and confirm the phase being observed by checking whether there is an intervening phase transition as interpolating between $\hat{H}_\text{phys}$ and $\hat{H}_\text{ideal}$.

One caveat on the terminology should be noted: After projecting onto $\mathcal{H}_\text{flat}$, there are terms quadratic in fermion operators, besides the quartic ones that are conventionally called ``interactions''. For the special case of Hubbard interactions, a property called the ``uniform pairing condition''~\cite{PhysRevB.94.245149} renders these terms trivial, and it was found that this condition can be guaranteed simply by certain point-group symmetries~\cite{herzog2022many}. It is possible to generalize this condition, which we will exemplify in Sec.~\ref{sec: engineered model}. However, these quadratic terms are not of central focus since the current work is concerned with the {\it existence} of ideal interacting Hamiltonians for a given set of isolated flat bands, whose quantum geometry is encoded in the projection matrix and is unaffected by these terms to the leading order in $V/\Delta$. An operational workaround is to simply regard these additional quadratic terms as parts of $\hat{H}_0$. For more detailed discussions about this nomenclature subtlety, see Sec.~II~D of SM.

We remark that the building blocks of the ideal Hamiltonians, $\{\hat{S}^{I}_{\bm{R}}\}$, do not necessarily mutually commute. As we will see in the next section, the solvability of these interactions results from the fact that they all commute with the order parameter and share certain eigenstates. The ideal interacting Hamiltonians are in this sense `frustration-free'.

\section{Solvable ground state} 
\label{sec: ground state}

In this section, we present a general recipe for obtaining the many-body ground states for an ideal Hamiltonian $\hat{H}_\text{int}$ constructed in the previous section. We first define a ``pseudo-Hamiltonian'' within $\mathcal{H}_\text{flat}$ with $\theta$ a free $U(1)$ phase
\begin{align}
\hat{\mathcal{E}}  =\mathsf{V} (\mathrm{e}^{\mathrm{i}\theta} \hat{O} + \mathrm{e}^{-\mathrm{i}\theta} \hat{O}^\dagger),
\end{align}
which can be viewed as a trial Hamiltonian of the original system. Since $\hat{\mathcal{E}}$ commutes with the ideal $\{\hat{S}^{(I)}_{\bm{R}}\}$ and thus the ideal $\hat{H}_\text{int}$ within $\mathcal{H}_\text{flat}$, any {\it non-degenerate} eigenstate of $\hat{\mathcal{E}}$, $|\Psi\rangle$, must also be an eigenstate of $\{\hat{S}^{(I)}_{\bm{R}}\}$ and $\hat{H}_\text{int}$ projected onto $\mathcal{H}_\text{flat}$. Taking $\langle \hat{S}_{\bm{R}}^{(I)}\rangle \equiv \langle \Psi | \hat{S}^{(I)}_{\bm{R}} | \Psi \rangle$ in Eq.~\ref{eq: interaction} implies $|\Psi\rangle$ is a zero-energy eigenstate of $\hat{H}_\text{int}$ within $\mathcal{H}_\text{flat}$ and thus a ground state.

Therefore, the task is to find the unique eigenstates of $\hat{\mathcal{E}}$ and show that they are ordered. Since $\hat{\mathcal{E}}$ is quadratic in fermion operators, its spectrum and eigenmodes can be easily solved. In general, the bands of $\hat{\mathcal{E}}$ are flat, and the values of the ``pseudo-energies'' are determined by $N^{\bm{Q}}$. 

If the QGN is in the p-p channel, the spectrum of $\hat{\mathcal{E}}$ is given by  $\mathsf{N}_\text{flat}$ Bogoliubov-de-Gennes quasiparticle pseudo-bands with perfect particle-hole symmetry (since no $\hat{\gamma}^\dagger \hat{\gamma}$ term is present). Fully occupying all the negative pseudo-energy bands gives rise to a state $|\Psi_\text{SC}(\theta)\rangle$ with eigenvalue $\Phi_0$, which is a sum of all the pseudo-energies of the occupied bands and thus is non-zero and extensive. $\Phi_0$ is a non-degenerate eigenvalue of $\hat{\mathcal{E}}$, since any change of the occupation configuration of the BdG quasi-particles will also change the eigenvalue~\footnote{For the clarity of key points, here we focus on the case where $N$ is full-rank. The discussions on the general cases can be found in Sec.~III of SM.}. Thus, $|\Psi_\text{SC}(\theta)\rangle$ is a ground-state of $\hat{H}_\text{int}$, and it hosts off-diagonal-long-range-order (ODLRO)~\cite{RevModPhys.34.694} with a superconducting phase $\theta$ in the sense that
\begin{align}\label{eq: ODLRO}
    \langle \Psi_\text{SC}(\theta) | \hat{O} |\Psi_\text{SC}(\theta)\rangle = \frac{\Phi_0}{2\mathsf{V}} \mathrm{e}^{-\mathrm{i}\theta} \equiv  \frac{\phi_0}{2} \mathrm{e}^{-\mathrm{i}\theta}
\end{align}

Note that by doing a global gauge transformation $\hat{\gamma}\rightarrow \hat{\gamma}\mathrm{e}^{-\mathrm{i}\theta/2}$, we are able to transform $\hat{\mathcal{E}}(\theta) \rightarrow \hat{\mathcal{E}}(\theta =0)$ and thus $|\Psi_\text{SC}(\theta)\rangle \rightarrow |\Psi_\text{SC}(\theta=0)\rangle$, while keeping $\hat{S}^{(I)}_{\bm{R}}$ unchanged. Thus, all $|\Psi_\text{SC}(\theta)\rangle$ with different $\theta$ are simultaneous eigenstate of $\hat{S}^{(I)}_{\bm{R}}$ with the same eigenvalue. The corresponding ground state of $\hat{H}_\text{int}$ with an {\it arbitrary} fixed (even) particle number $\mathsf{N}_{e}$ can be constructed as
\begin{align}
    |\Psi_\text{SC}(\mathsf{N}_{e})\rangle \equiv 
    \int_0^{2\pi} \mathrm{e}^{-\mathrm{i}\theta \mathsf{N}_{e} /2} |\Psi_\text{SC}(\theta)\rangle \ .
\end{align}

In the p-h case, the ground state of $\hat{H}_\text{int}$ can be similarly obtained by occupying any integer number of pseudo-bands of $\hat{\mathcal{E}}$ that are separated from other bands by finite gaps, since this gives rise to a {\it non-degenerate }~\cite{Note3} and generically non-zero eigenvalue $\Phi_0$ within the corresponding particle number sector. This implies that the system is solvable only at certain rational filling fractions of the flat bands with specific values determined by the OP. In contrast to the p-p case, here $\theta$ only cycles the pseudo-energies, permuting the equivalent translation-symmetry-breaking groundstates. It should be noted that when $\bm{Q}\neq 0$, $\hat{\mathcal{E}}$ has a folded BZ since it breaks the translation symmetry of the original electronic structure (corresponding to DW). Diagonalizing this pseudo-Hamiltonian then gives us $\mathsf{M}\mathsf{N}_\text{flat}$ pseudo-bands, where $\mathsf{M}$ is the smallest positive integer that makes $\mathsf{M}\bm{Q} =0$ modulo the reciprocal lattice. It is, in principle, possible to have $\varpi^{\bm{Q}}_0\approx 0$ in a continuous range of $\bm{Q}$, such that an incommensurate DW order is suggested. In such a case, the properties of each incommensurate $\bm{Q}$ should be approached to a good approximation from a nearby commensurate value.

While the four-fermion term in $H_\text{int}$ is translationally invariant, the single particle term is only translationally invariant when all $\langle \hat{S}^{(I)}_{\bm{R}}\rangle$ are.  For any order with $\bm{Q}=\bm{0}$, the translational invariance of $\langle \hat{S}^{(I)}_{\bm{R}} \rangle$ is obvious.  Even for $\bm{Q}\neq \bm{0}$, $\langle \hat{S}^{(I)}_{\bm{R}} \rangle$ can still be translationally invariant. For example, because the p-p orders we have constructed are of the FF type, translation by $\bm{R}$ is equivalent to a phase change of the OP by $\theta \to \theta + \bm{Q}\cdot \delta \bm{R}$;  since $\hat{S}^{(I)}_{\bm{R}}$, by assumption, are gauge invariant, their expectation values remain constant under translation. Similar situations can arise in the p-h cases, for example, when the order involved is a generalized ``spiral order'' and the used $\hat{S}^{(I)}_{\bm{R}}$ only overlap with the magnitude of the OP. In general, however, certain choices of $\hat{H}_\text{int}$ may explicitly break translational symmetry, but still preserve other symmetries that are broken spontaneously by $|\Psi\rangle$.

It should be noted that despite the spiritual similarity to the mean field theories, the Slater determinant states solved by the above approach are the {\it exact} many-body ground states of the ideal systems instead of just proxy states capturing the qualitative properties. The perturbations away from those ideal limits can be accounted for systematically since not only the ground states but also many aspects of the excitation spectrum can be solved, as we discuss below.

\section{Excitations in the ideal models} 
\label{sec: excitations}

In the ideal models constructed in Sec.~\ref{sec: interactions}, aspects of the excitation spectrum can also be determined in a systematic fashion, which we discuss in this section. 

\subsection{Few-body excitations}\label{sec: few body}
The strategy to solve the excitations with simple compositions is to construct operators $\hat{\xi}^{(c)}_{\bm{P}}$ within each charge-$c$, momentum-$\bm{P}$ sector that satisfy:
\begin{align}
[\hat{H}_\text{int}, \hat{\xi}^{(c)}_{\bm{P}} ]|\Psi\rangle = E^{(c)}(\bm{P}) \hat{\xi}^{(c)}_{\bm{P}}|\Psi\rangle.
\end{align}

This is easier said than done for typical interacting Hamiltonians, where the commutator of a one-particle operator, $\gamma_a^\dagger$, with the Hamiltonian produces higher order terms such as $\gamma_b^\dagger \gamma_c^\dagger\gamma_d$, and the commutator of many-particle operators produces still higher order terms in a hierarchy that never closes.  However, because $|\Psi\rangle$ is a zero-mode of all $\left(\hat{S}^{(I)}_{\bm{R}}-\langle \hat{S}^{(I)}_{\bm{R}}\rangle \right)$ within $\mathcal{H}_\text{flat}$,  sectors with e.g. $\gamma_a^\dagger$ and $\gamma_b^\dagger \gamma_c^\dagger\gamma_d$ excitations cannot mix~\cite{PhysRevLett.122.246401, PhysRevB.103.205415, herzog2022many,PhysRevB.104.075143} and thus the composition does not changed after the commutation. For example, for two-particle (Cooper pair) excitations,
\begin{align}
    & [\hat{H}_\text{int}, \hat{\gamma}^\dagger_{\bm{k}^+,n}\hat{\gamma}^\dagger_{\bm{k}^-,m} ]|\Psi\rangle&\nonumber\\
    =& \sum_{\bm{k}',n'm'\le \mathsf{N}_\text{flat}}  \Gamma^{\text{p-p}, \bm{P}}_{ nm, \bm{k}; n'm', \bm{k}' } \hat{\gamma}^\dagger_{\bm{k}'^+,n'}\hat{\gamma}^\dagger_{\bm{k}'^-,m'}  
  |\Psi\rangle 
\end{align}
where $\bm{k}^\pm \equiv  \bm{P}/2 \pm \bm{k}$. The commutation relations for single-particle/hole, two-hole, and particle-hole (exciton) excitations can be similarly derived (see Sec.~IV~A-C of SM for computationally feasible expressions in terms of parameters in the Hamiltonian). By diagonalizing the scattering matrix $\Gamma$, one can compute the corresponding eigenstates and spectrum of an excitation~\cite{PhysRevB.103.205415, herzog2022many}. For example, the charge $+1$ excitation spectrum only involves diagonalizing an effective single-particle Hamiltonian.  These excitations can be computed with polynomial scaling in the system size, rather than exponentially as in most interacting systems, and in this sense are solvable. For certain cases when the form of interaction is simple (e.g. attractive Hubbard~\cite{herzog2022many}), the quasi-particle spectrum can be even {\it analytically} solvable.

\subsection{Pseudo-spin $SU(2)$ symmetry of the p-p case and the phase stiffness} 
\label{sec: pseudospin}

In general, different types of excitations do not have any relation. However, in any ideal model in the p-p case, some of the excitations with the same fermion parity but different charge can be related, since $\hat{O}^{\text{p-p},\bm{Q},\dagger}$ or $\hat{O}^{\text{p-p},\bm{Q}}$ adds charge $\pm 2$ and 
momentum $\pm \bm{Q}$ with zero energy cost, 
\begin{align}
\hat{\xi}^{(c\pm 2)}_{\bm{P}\pm \bm{Q} } &\equiv  [\hat{\xi}^{(c)}_{\bm{P} }, \hat{O}^{\text{p-p},\bm{Q},\dagger} \text{ or } \hat{O}^{\text{p-p},\bm{Q}}] 
\end{align}
creates a charge $c\pm 2$, momentum $\bm{P}\pm \bm{Q}$, energy $E^{(c\pm 2)}(\bm{P}\pm \bm{Q}) = E^{(c)}(\bm{P})$ excitation when acting on a ground state. Therefore, for the p-p case, the particle and hole excitations share the same spectrum, and certain particle-particle, particle-hole, and hole-hole pair excitations share the same spectra. Furthermore, lowest branch of $E^{(0,\pm 2)} (\bm{P})$ must hit zero at $\bm{P} = 0, \pm \bm{Q}$ since $\hat{O}^{\text{p-p},\bm{Q},\dagger}$ creates a zero-energy p-p excitation with momentum $\bm{Q}$. This is due to a pseudospin $SU(2)$ symmetry in the p-p case, which is generated by:
\begin{align}
    \hat{J}^{x,y} &\equiv (\hat{O}^{\text{p-p},\dagger} \pm  \hat{O}^{\text{p-p}})/2, \quad 
    \hat{J}^z \equiv [\hat{O}^{\text{p-p},\dagger},  \hat{O}^{\text{p-p}}]/4
\end{align}
where for clarity, we consider the single-component SC case; multicomponent generalizations are discussed in Sec.~IV~D of SM. Clearly, the SC ground state $|\Psi_\text{SC}\rangle$ is a form of pseudospin ferromagnetism. Thus, by comparing the excitation spectrum to that of a non-linear sigma model of broken $SU(2)$ symmetry~\cite{PhysRevX.4.031057}, we obtain the following estimate for the pseudospin stiffness tensor:
\begin{align}
    \kappa_{ij} = \phi_0 M^{-1}_{ij} 
\end{align}
where $M^{-1}_{ij} $ is the Cooper pair inverse mass tensor, extracted by a Taylor expansion of the lowest branch of the charge-$2$ spectrum $E^{(2)}(\bm{P} = \bm{Q} + \delta \bm{Q} ) \approx \frac{1}{2}\delta \bm{Q}_{i} M^{-1}_{ij} \delta \bm{Q}_j $, and $\phi_0$ is the ODLRO density defined in Eq.~\ref{eq: ODLRO}. 

Given the quadratic dispersion of the Nambu-Goldstone mode, the Mermin-Wagner-Coleman theorem implies that spontaneous breaking of the $SU(2)$ symmetry cannot occur at finite temperature when the space dimension is $d \le 2$. (At zero temperature, in contrast, ferromagnetism is allowed in low dimensions.) However, we note that the susceptibility $\chi(T) \sim T^3/|\kappa|^4 \exp(4\pi |\kappa|/ T)$ is large at low temperature $T\rightarrow 0$~\cite{zinn2021quantum}, where $|\kappa| \equiv \sqrt[d]{\det{\kappa}}$ is the geometric mean of the stiffness in different directions. Then, when $d=2$, certain explicit breaking of the symmetry down to charge $U(1)$ will result in a quasi-long-range ordering of the $U(1)$ phase with a Berezinskii–Kosterlitz–Thouless (BKT) transition temperature $T_c$ that can be estimated as
\begin{align}
\label{eq:TBKT}
    \delta \cdot \chi(T_c) \sim 1 \implies T_c \sim 4\pi |\kappa|/ \ln(|\kappa|/\delta)
\end{align}
which quickly becomes comparable to $|\kappa|$ for non-zero symmetry-lowering energy scale $\delta$. We note that multiple factors can lead to a deviation from the idealized setup we constructed, and thus a weak explicit lowering of the pseudospin $SU(2)$ symmetry, including (but not limited to) the finite band gap, imperfect QGN and flatness of the bands, and the deviation of the interactions from the ideal form. Depending on the nature of those perturbations, the superconducting state may remain phase-coherent up to a temperature comparable to $|\kappa|$ for all dimensions $d \ge 2$. Our approach provides a direct estimate of the superfluid stiffness of {\it nearly ideal} systems, providing complementary estimates to generic upper bounds~\cite{PhysRevX.9.031049, hofmann2022heuristic, doi:10.1073/pnas.2217816120, mao2023upper}. 

\section{Examples of electronic structures with perfect QGN}\label{sec: examples}

Here, we present several examples of QGN. More detailed discussions, as well as another constructed model for a current density wave, can be found in Sec. V of the SM.

\subsection{Spinful time-reversal symmetric flat bands}
\label{sec: time reversal}
We first note that QGN can be guaranteed by symmetries. The simplest example of this is the combination of spin $S^z$ conservation and time-reversal symmetry. In such a system, we recognize the Wannier degeneracy (separating out the spin index $s$ from the orbital index $\mu$):
\begin{align}
    U_{\mu s,n s' } (\bm{k})
    &= \delta_{s s'} U_{\mu n}^s(\bm{k}) \ , \ \ 
    U_{\mu n}^s (\bm{k}) = \left[U_{\mu n}^{\bar{s}}(-\bm{k})\right]^\star
\end{align}
Then one can verify $N^{\text{p-p}, \bm{Q}=\bm{0}}_{\mu s, \nu s'} \equiv \sigma^y_{s s'} \mathbbm{1}_{\mu\nu}$ satisfies the perfect QGN condition in the p-p channel with  $\bm{Q}=\bm{0}$. Solvable Hamiltonians can thus be constructed with $B^{(I)}_{\mu s,\nu s'}$ matrices in Eq.~\ref{eq: separable form} taking the form $\sigma^{\mathsf{i} = 0,x,y,z}_{ s s'} \tilde{B}^{(I)}_{\mu\nu}$, where  $\sigma^{\mathsf{i}}$ are spin Pauli matrices and $\tilde{B}^{(I)}$ are arbitrary symmetric (for $\mathsf{i} = x,y,z$) or antisymmetric (for $\mathsf{i} = 0$) hermitian matrices for the orbital indices. This includes a huge class of possible interactions, e.g. the attractive Hubbard interactions on each site and orbital, whose solvability was pointed out in Refs.~\cite{PhysRevB.94.245149,herzog2022many} and where the superfluid stiffness is shown to fundamentally depend on the flat-band quantum geometry and topology in several previous works~\cite{PhysRevLett.130.226001,PhysRevB.95.024515,2021PNAS..11806744V,2018arXiv181112428H,PhysRevLett.124.167002,PhysRevLett.128.087002,2022PhRvB.106a4518H,herzog2022many}. 

\subsection{Chiral symmetric flat bands}

A less obvious example is associated with chiral symmetry. Specifically, when the lattice is bipartite and there are degenerate flat bands, one can verify that $N^{\text{p-h}, \bm{Q}=\bm{0}}_{\mu s , \nu s'} = \delta_{\mu\nu} (-1)^{\mu} \tilde{N}_{s s'}$ is a nesting matrix for the p-h channel at $\bm{Q}=\bm{0}$, where the parity of the orbital index $\mu$ represents the sublattice group, 
and $\tilde{N}$ is an arbitrary hermitian matrix associated with the {\it conserved} flavor index $s$ (e.g. spin and/or valley). The condensation of the corresponding OP then leads to symmetry breaking of the conserved flavor symmetry with a pattern specified by $\Tilde{N}$ (e.g. spin/valley polarization, inter-valley coherence, ...). This is exactly the case extensively discussed in the studies on magic-angle twisted bilayer graphene in its chiral-flat limit~\cite{PhysRevLett.122.246401, PhysRevB.104.075143,
PhysRevX.10.031034, PhysRevB.103.205414, PhysRevB.103.205415, ledwith2021strong}. In the language of QGN, it is now clear that these proposed solvable interacting Hamiltonians belong to a larger class, which is systematically constructible using the method in Sec.~\ref{sec: interactions} for each possible ordering. We also note that there is a large class of electronic structures that feature flat bands protected by chiral symmetry~\cite{cualuguaru2022general} and are approximately realized in materials~\cite{2022Natur.603..824R}.

It is of particular interest to identify and predict what interactions could lead to certain specific symmetry-breaking patterns in flat band materials. In this regard, the construction of solvable models for different OPs can provide important information by suggesting interactions that only favor certain types of symmetry-breaking patterns. We leave material-specific investigations to future studies. 

\subsection{Non-symmetry-related cases: An engineered model for pair-density-wave}

\label{sec: engineered model}

\begin{figure}[t]
    \centering
    \subfigure[]{\includegraphics[width = 0.98 \linewidth]{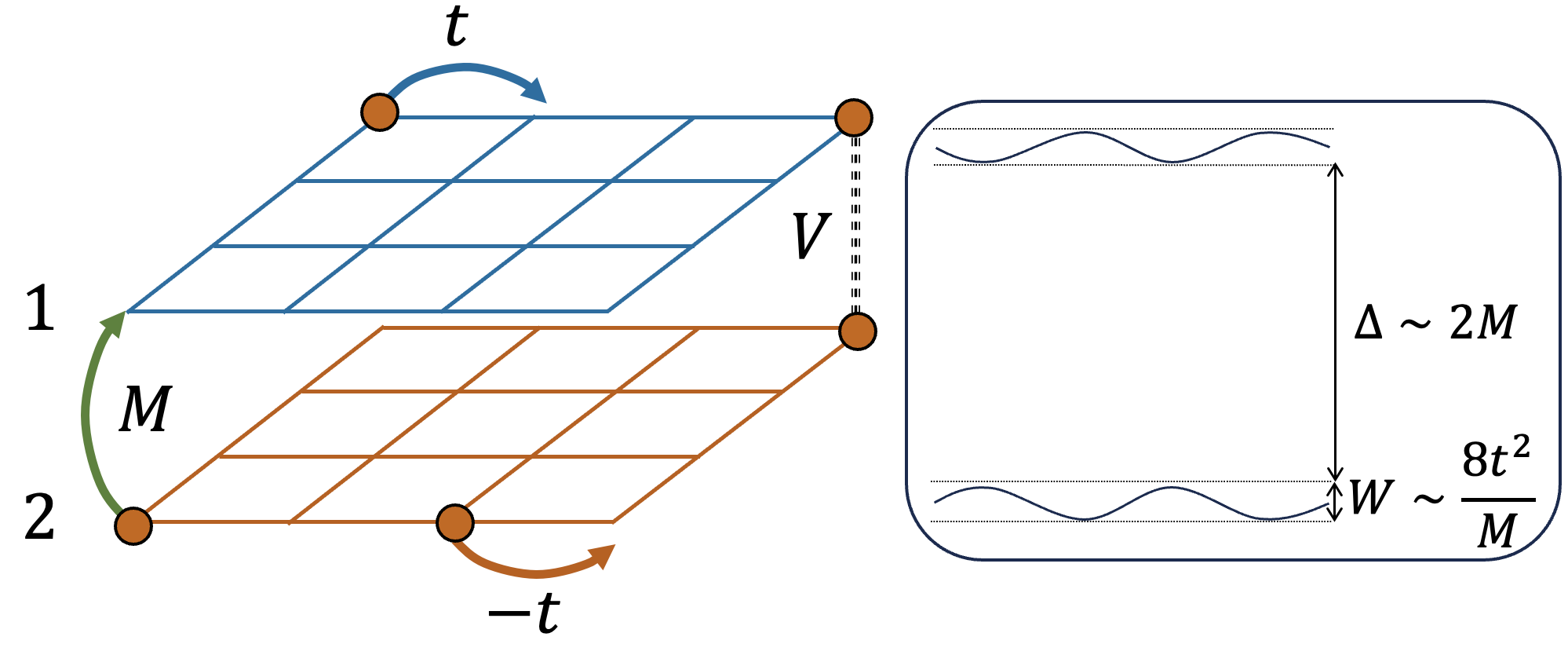} \label{fig: engineered model}}
    \subfigure[]{\includegraphics[width = 0.97 \linewidth]{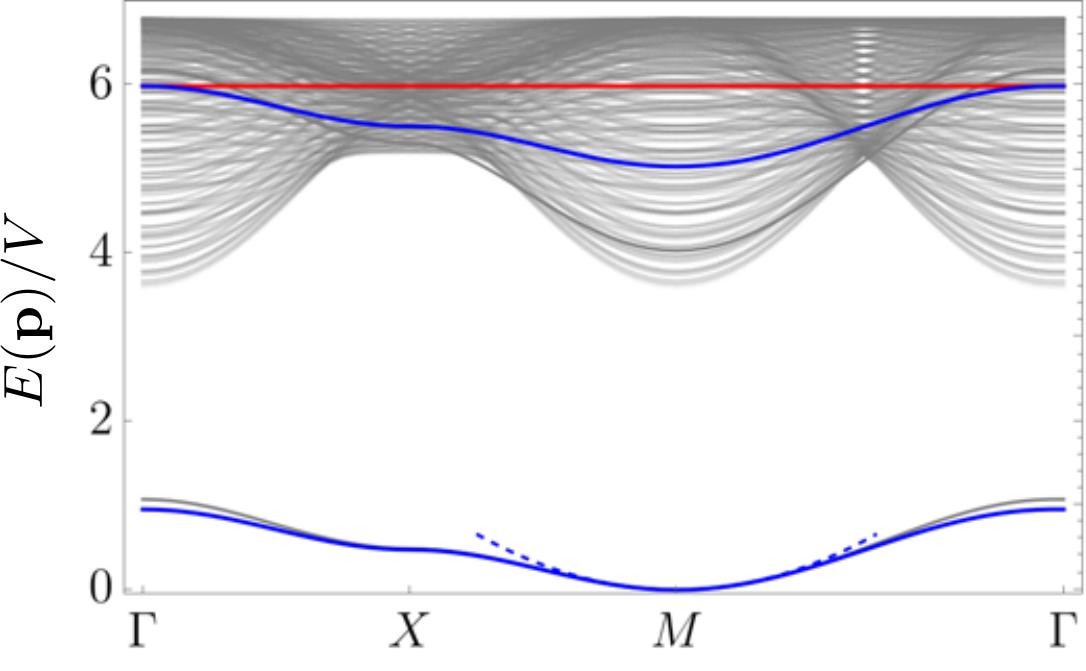} \label{fig: full}}
    \caption{(a) An illustration of the two-orbital model in Eqs.~\ref{eq: simple H} and the band structure. (b) Cooper pair excitation spectrum above the PDW groundstate for the onsite anti-ferromagnetic and inter-orbital attractive interactions in Eq.~\ref{eq: simple H int}, choosing $V=t=1, M = 5$. The analytical results for the artificial setup discussed in the text ($W=0,\Delta=\infty$ with quantum geometry fixed) are shown in blue (for the two bound states) and red (degenerate particle-particle continuum), whereas the numerical results that are accurate to the first order in $W/V$ are plotted in gray. The lowest, condensing Cooper pair band exhibits a zero mode at $\mbf{P} =  (\pi,\pi)$, agreeing with the fact that the PDW OP $\hat{O}^{\text{p-p},\bm{Q}=(\pi,\pi)}$ commutes with the projected Hamiltonian. It can be seen that this band is insensitive to the correction in $W/V$, while the continuum is significantly broadened, swallowing the upper bound state (Leggett mode) band. The condensing Cooper pair mass can be extracted from a quadratic fitting near $(\pi,\pi)$, which we plot with a dashed line; the value of the mass agrees with the result in the flat band limit Eq.~\ref{eq: charge 2 excitation}. }
\end{figure}

It is probably even more interesting to study cases in which perfect QGN is not associated with symmetry. To show such a possibility, we construct a simple model of a dimerized bilayer on a square lattice in which $\tau^{0,x,y,z}$ are Pauli matrices acting on orbital/layer indices, and $s$ is the spin index, 
\begin{align} \label{eq: simple H}
    \hat{H}_0 = \sum_{\bm{k},s, \mu\nu} \left[2t\left(\cos k_x +\cos k_y\right) \tau^z + M\tau^x\right]_{\mu \nu} \hat{c}^\dagger_{\bm{k}\mu s}  \hat{c}_{\bm{k}\nu s}
\end{align}
See Fig.~\ref{fig: engineered model} for an illustration. In the limit $M\gg |t|$, this electronic structure has two topologically trivial, spin-degenerate, flat bands with width $W\approx 8t^2/M$ separated by a band gap $\Delta \approx 2 M$. Due to time reversal and spin conservation, as discussed in Sec.~\ref{sec: time reversal}, QGN for uniform pairing (p-p channel at $\bm{Q}=\bm{0}$ with $N_{\mu s, \nu s'} = \tau^0_{\mu\nu} \sigma^y_{s s'}$) is automatically satisfied. Moreover, since the wavefunctions also obey $\tau^x U(\bm{k}) = U(\bm{k}+\bm{Q}) = U(\bm{Q}-\bm{k})$ with $\bm{Q} = (\pi,\pi)$, we find that QGN is also satisfied in both p-p and p-h channels at this momentum. This can be shown by computing the lowest two eigenvalues of $\Pi^{\bm{Q}}$ in both p-p and p-h channels (Eqs.~\ref{eq: p-p Pi}\&\ref{eq: p-h Pi}), $\varpi^{\bm{Q}}_{0,1} = 2 \pm (\cos Q_x + \cos Q_y)$ to leading order in $t/M$, so that one branch reaches zero at $\bm{Q} = (0,0)$ and the other at $\bm{Q} = (\pi,\pi)$. Solving the corresponding nesting matrices at $\bm{Q} = (\pi,\pi)$, we find $N_{\mu s, \nu s'} = \tau^x_{\mu\nu} \sigma^y_{s s'}$ in the p-p channel, and $N^{(\mathsf{i} = 0,x,y,z)}_{\mu s, \nu s'} = \tau^x_{\mu\nu} \sigma^{(\mathsf{i})}_{s s'}$ in the p-h case, where $\sigma^{(\mathsf{i}=0,x,y,z)}$ are Pauli matrices acting on spin indices. QGN thus suggests that, besides the uniform pairing order, the system is simultaneously prone to a singlet PDW, a charge DW, and a spin DW, all at $\bm{Q} = (\pi,\pi)$. [In Sec.~V~D of the SM, we generalize this construction to show that any single-particle Hamiltonian obeying $N h(\mbf{k}) N^\dag = h(\mbf{k}+\mbf{Q})$ with a unitary nesting matrix $N$ has perfect QGN in the p-h channel at $\bm{Q}$. (Analogous statements hold in the p-p channel.) For this class of systems satisfying QGN, we further show that the Fubini-Study quantum metric is also `nested', $g(\mbf{k}) = g(\mbf{k}+\mbf{Q})$, a property that may guide the search of electronic structures with QGN.]

For different suggested orders, we can construct different ideal interactions adopting our systematic scheme in Sec.~\ref{sec: interactions}. To give an example, below we focus on the PDW case with $\mbf{Q} = (\pi,\pi)$ where the nesting matrix is $\tau^x \sigma^y$ corresponding to inter-orbital singlet pairing. We pick a judicious choice of interaction,
\begin{align}
 \label{eq: simple H int}
\hat{H}_\text{int} = V \sum_{\bm{R}} (- 3\hat{n}_{\bm{R},1} \hat{n}_{\bm{R},2} + 4 \hat{\bm{S}}_{\bm{R},1} \cdot \hat{\bm{S}}_{\bm{R},2}),
\end{align}
that uniquely selects the PDW state as the ground state (see Sec.~V~C of the SM).  $\hat{\bm{S}}_{\bm{R},\mu} = \frac{1}{2}\sum_s \hat{c}^\dag_{\bm{R},\mu,s}\pmb{\sigma}_{ss'}\hat{c}_{\bm{R},\mu,s'}$ is the spin operator on layer-$\mu$ and site-$\bm{R}$. The exact ground states are generated by the p-p order parameter (Cooper pair creation operator)
\bea
\hat{O}^{\text{p-p},\dagger} = \sum_{\bm{k} \mu \nu} \hat{\bar{c}}^\dag_{\bm{k}+\bm{Q},\mu,\uparrow} \tau_{\mu \nu}^x \hat{\bar{c}}^\dag_{-\bm{k},\nu,\downarrow}
\eea
with $\hat{\bar{c}}^\dag_{\bm{k}+\bm{Q},\nu}= \sum_\nu \hat{\bar{c}}^\dag_{\bm{k}+\bm{Q},\mu} P_{\mu \nu}(\bm{k})$ the projection of the orbital operators onto the flat bands (see \Eq{eq: ph order parameter}). Projecting $\hat{H}_\text{int}$ in \Eq{eq: simple H int} to the flat band Hilbert space results in trivial one-body (Hartree-Fock-type) terms that act as a chemical potential on the flat bands. The triviality of the one-body terms generalizes the ``uniform pairing condition"~\cite{PhysRevB.94.245149,herzog2022many} utilized in attractive Hubbard models (see Sec.~II~D of the SM). 

Adopting the methods in Sec.~\ref{sec: excitations}, we can exactly solve the few-body excitations in an artificial setup by setting the $W\to 0$ and $\Delta \to \infty$ with the quantum geometry $U_{\mu n }(\bm{k})$ remaining fixed. First, we consider the charge $\pm1$ fermionic quasi-particles above the ground states. We find that these excitations are spin-degenerate, gapped, and flat, with energy $E^{(\pm 1)}(\mbf{p}) = 3V$. Next, we consider a Cooper pair excitation. This spectrum is also analytically solvable since the two-particle scattering matrix can be diagonalized by generalizing the approach of Ref.~\cite{herzog2022many} to extract a low-rank Hamiltonian for the pair binding energies. We find that there are two bound modes - the Goldstone mode of the PDW condensate, and a higher-energy Leggett mode - below the flat and macroscopically degenerate two-particle continuum at $6V$. The Goldstone mode has dispersion
\bea\label{eq: charge 2 excitation}
E^{(2)}_0(\mbf{P}) &= 6V g (2 + \cos P_x + \cos P_y) = 6V g \frac{\delta \bm{P}^2}{2} + \dots
\eea
to leading order in $t/M$, where the coefficient $g$ is the integrated Fubini-Study metric
\bea
\label{eq:gint}
g &= \int \frac{d^2k}{(2\pi)^2} \frac{1}{2} \text{Tr }(\partial_i P)^2 = \frac{t^2}{M^2} + \dots 
\eea
In the second equality of Eq.~\ref{eq: charge 2 excitation}, we expanded in $\delta \bm{P} = \mbf{P}-(\pi,\pi)$ to find the inverse Cooper pair mass $\approx \frac{6 V t^2}{M^2}$. We thus find the quantum geometric origin of PDW in this model is similar to the known case of uniform pairing~\cite{2015NatCo...6.8944P}; however, we note that for general cases, the Cooper pair mass may not directly relate to the value of the metric $g$. The dispersion of the higher bound state (Leggett mode) is identical to \Eq{eq: charge 2 excitation} up to a constant shift.

One may be worried about the effects of the actual small but non-zero bandwidth $W \sim t^2/M$ in our model. To address this issue, we first note that the quasi-particle excitations are relatively unaffected due to their large gap $\sim V \gg W$. For the two-particle excitations, we check the robustness of the key results by numerically calculating the excitation spectrum (shown in Fig.~\ref{fig: full}), which is obtained by diagonalizing the perturbed scattering matrix in the presence of a small bandwidth $W$ (compared to the interaction strength $V$) and is accurate to the leading order of $W/V$. We indeed see that although the particle-particle continuum around $6V$ is split by the finite bandwidth and the upper bound state branch merges into the continuum, the dispersion of the lowest branch is relatively unchanged. Especially, it retains its minimum at $(\pi,\pi)$ and the Cooper pair mass does not change significantly from its value in the idealized flat band setup. 

After verifying the robustness of the two-particle excitation results in Eq.~\ref{eq: charge 2 excitation}, we finally turn to the estimate of the superfluid stiffness and thus the BKT transition temperature $T_c$ of the system in the presence of symmetry lowering terms (see Sec.~\ref{sec: pseudospin}). Since for generic fillings, the ODLRO density of the system, $\phi_0$, is $\mathcal{O}(1)$, $T_c$ is expected to scale as 
\bea
T_c \sim  V \frac{t^2}{M^2} 
\eea
which applies as long as $\Delta \sim M \gg V \gg W \sim t^2/M$. In the further limit $V\gg t$, we see that $T_c\gg W^2/V$, the conventional strong coupling scaling for single-band systems. This observation further corroborates that the phase stiffness originates from the quantum geometry of the band instead of the (single-particle) kinetic energies. This example (along with similar efforts in Refs.~\cite{PhysRevB.96.064511,PhysRevLett.131.016002,2023SCPMA..6687212C}) is based on the geometric structure of the flat bands and is distinguished by its analytically controlled, microscopic approach. QGN thus points to a fresh direction in the pursuit of PDW phases,  intrinsically distinct from the existing literature~\cite{PhysRevB.105.L100509,PhysRevLett.130.126001,PhysRevB.109.L121101,PhysRevLett.130.026001,PhysRevLett.129.167001,PhysRevLett.125.167001,PhysRevB.108.174506,haldane2023d}.

\section{Possible Extensions} 

\label{sec: extensions}

We note that the form of solvable models presented in Sec.~\ref{sec: interactions} is not the only possibility for systems with perfect QGN. For instance, consider the one-dimensional SSH chain 
\bea
\hat{H}_0 = \sum_{k \al \be s} t [\tau^0 - \tau^z \cos k + \tau^y \sin k]_{\al \be} \hat{c}^\dag_{k,\al,s} \hat{c}_{k,\be,s}
\eea
which realizes an obstructed atomic limit \cite{2020Sci...367..794S} with lower-bounded quantum geometry \cite{PhysRevLett.128.087002,2023arXiv230302126E}. As discussed in the previous section, the time-reversal symmetry and spin conservation imply perfect QGN in the p-p channel, suggesting a singlet $s$-wave superconducting groundstate~\cite{herzog2022many}. However, we find that this result can be promoted to a singlet $p$-wave OP
 \bea
 \hat{O}^{\text{p-p},\dagger} = \sum_{k} \sin k \, \gamma^\dag_{k,1,\uparrow} \gamma^\dag_{-k,1,\downarrow}
 \eea
by choosing the interaction taking the form $H_\text{int} = U \sum_R \hat{S}^\dag_{R} \hat{S}_{R} \geq 0$ where $\hat{S}_{R}$ again satisfies $[\hat{P}\hat{S}_{R}\hat{P},\hat{O}^{\text{p-p},\dagger}] = 0$ but is {\it non-hermitian}. Explicitly, we choose $\hat{S}_{R}$ to take the form
\bea
\hat{S}_{R} &=  \sum_{\al \be,\sigma}  \sum_{\Delta R =-2}^2 s^z_\sigma \tau_{\al \be}(\Delta R) \hat{c}^\dag_{R+\Delta R,\al ,\sigma} \hat{c}_{R,\be ,\sigma}
\eea
where $s^z_{\uparrow/\downarrow} = \pm 1$ and $\tau(0)= \tau_2/4, \tau(\pm1) = \mp i \tau_0/4, \tau(\pm 2) = (-\tau_2 \pm i \tau_3)/8$. This generalizes the form assumed in Eq.~\ref{eq: interaction} (see Sec.~VI of SM for the explicit construction which guarantees trivial one-body terms). Remarkably, zero-energy ground states can still be exactly constructed according to $(\hat{O}^{\text{p-p},\dagger})^n\ket{0}$ and proven to be unique using a corollary of Lieb's well-known theorem~\cite{PhysRevLett.62.1201}. Although the excitations in this construction are not exactly solvable, we can prove (see Sec.~VI of SM) that the charge $\pm1$ excitation spectrum is gapless, as expected for a $p$-wave ground state. This example demonstrates that the possible construction schemes of ``ideal'' interacting Hamiltonians for systems with QGN have not been exhausted, which we leave for future explorations. 

Moreover, it remains to be investigated whether QGN is the {\it necessary} condition for a set of flat bands to admit solvable strongly coupled interacting Hamiltonians. It is interesting to explore whether new notions in quantum geometry~\cite{bouhon2023quantum} can help discover more general conditions of QGN. Our work advances the exploration of many-body phases enabled by topology and quantum geometry making use of the strongly coupled flat band limit, but it also important to explore how QGN can be combined with Fermi surface nesting for dispersive systems featuring significant quantum geometry.

{\bf Acknowledgement: } We would like to thank Erez Berg, Daniel Parker, Junkai Dong, Ashvin Vishwanath, Chaitanya Murthy, Biao Lian, Ben Feldman, Oskar Vafek, Jiabin Yu, Zhi-Da Song, and P\"aivi T\"orm\"a for helpful discussions. ZH was funded, in part, by a QuantEmX grant from ICAM and the Gordon and Betty Moore Foundation through Grant GBMF9616. {BAB was primarily supported by the the Simons Investigator Grant No. 404513. JHA was supported by a
Hertz Fellowship. BAB and JHA received additional support from the Gordon and Betty Moore Foundation through Grant No. GBMF8685 towards the Princeton theory program, Office of Naval Research (ONR Grant No. N00014-20-1-2303), BSF Israel US foundation No. 2018226 and NSF-MERSEC DMR-2011750, Princeton Global Scholar and the European Union's Horizon 2020 research and innovation program under Grant Agreement No 101017733 and from the European Research Council (ERC). SAK and ZH were supported, in part, by the Department of Energy, Office of Basic Energy Sciences, under contract DE-AC0276SF00515 at Stanford. }

\bibliography{ref}

\onecolumngrid
\clearpage

\end{document}


\title{``Quantum Geometric Nesting'' and Solvable Model  Flat-Band Systems: Supplemental Materials}

\author{Zhaoyu~Han}
\affiliation{Department of Physics, Stanford University, Stanford, CA 94305, USA}

\author{Jonah~Herzog-Arbeitman}
\affiliation{Department of Physics, Princeton University, Princeton, NJ 08544, USA}

\author{B.~Andrei~Bernevig}
\affiliation{Department of Physics, Princeton University, Princeton, NJ 08544, USA}
\affiliation{Donostia International Physics Center, P. Manuel de Lardizabal 4, 20018 Donostia-San Sebastian, Spain}
\affiliation{IKERBASQUE, Basque Foundation for Science, Bilbao, Spain}

\author{Steven~A.~Kivelson}
\affiliation{Department of Physics, Stanford University, Stanford, CA 94305, USA}
\maketitle

\tableofcontents

\section{Formalism}

\subsection{Preliminaries}

We consider electronic structure encoded in a tight-binding model with translation symmetry:
\begin{align}
\hat{H}_0 = \sum_{\bm{R}\bm{R}', \alpha\beta} t_{\alpha\beta, \bm{R}-\bm{R}'} \hat{c}^\dagger_{\bm{R},\alpha} \hat{c}_{\bm{R}',\beta}
\end{align}
where $\bm{R}$ indicate the unit cell center position, Greek letters $\alpha,\beta,\dots$ denote the orbital index in each unit cell. For simplicity, we combine the orbital and spin indices unless otherwise specified. In momentum space, we define basis $\hat{c}_{\bm{k},\alpha}\equiv\sum_{\bm{R}}\mathrm{e}^{\mathrm{i} \bm{k}\cdot (\bm{R}+\bm{r}_\alpha)} \hat{c}_{\bm{R},\alpha}/\sqrt{\mathsf{V}}$ and Fourier transform $t_{\alpha\beta} (\bm{k}) \equiv\sum_{\bm{R}}\mathrm{e}^{\mathrm{i} \bm{k}\cdot (\bm{R}+\bm{r}_\alpha-\bm{r}_\beta)} t_{\alpha\beta,\bm{R}}$ ($\mathsf{V}$ is the number of unit cells in the system, $\bm{r}_\alpha$ is the intra-unit-cell position of orbital $\alpha$) to rewrite
\begin{align}
\hat{H}_0 = \sum_{\bm{k}, \alpha\beta} t_{\alpha\beta}(\bm{k}) \hat{c}^\dagger_{\bm{k},\alpha} \hat{c}_{\bm{k},\beta}
\end{align}
Finally, we can diagonalize the Hamiltonian into the form
\begin{align}
\hat{H}_0 &= \sum_{\bm{k}, n} \epsilon_n(\bm{k}) \hat{\gamma}^\dagger_{\bm{k},n} \hat{\gamma}_{\bm{k},n}, \\
\hat{\gamma}_{\bm{k},n} &= U_{n\alpha}^\dagger(\bm{k}) \hat{c}_{\bm{k},\alpha},
\end{align}
where we use Latin letters ($n,m,\dots$) to label the bands. We assume there are $\mathsf{N}$ orbitals, and thus $\mathsf{N}$ bands in total.

The dispersion $\epsilon_n(\bm{k})$ and the unitary transformation matrix (the wavefunctions) $U_{\alpha n}(\bm{k})$ together define the electronic structure of the system. Conventional Fermi surface nesting conditions only use the information of the dispersion, and lose meaning when the bands are flat in comparison to the interaction strength. In this work, we aim to propose complementary concepts for flat band systems, utilizing the information stored in $U_{\alpha n}(\bm{k})$.

To do so, we consider an ideal setup where there are $\mathsf{N}_\text{flat}$ nearly degenerate and flat bands isolated from all the other bands by a large gap $\Delta$, and the relevant interaction strength $V$ is small compared to $\Delta$ but large compared to the flat bands' bare bandwidth $W$. Without loss of generality, we can relabel all the bands and assume the $n=1\dots \mathsf{N}_\text{flat}$ bands are the degenerate flat bands, and $n=\mathsf{N}_\text{flat}+1\dots \mathsf{N}$ bands are the others. For the single particle basis, we can thus define a projector matrix:
\begin{align}\label{eq: projector}
    P_{\alpha \beta} (\bm{k})\equiv  \sum_{n\le \mathsf{N}_\text{flat}} U_{\alpha n }(\bm{k}) U_{n\beta}^\dagger(\bm{k}),
\end{align}
For later convenience, we also define a projector matrix onto the remaining bands, $Q_{\alpha \beta} (\bm{k})\equiv \delta_{\alpha\beta} - P_{\alpha \beta} (\bm{k})$.

For the ideal scenario we are assuming, we define the ``flat band subspace'', $\mathcal{H}_\text{flat}$, as the Fock space of the electron modes $\gamma_{\bm{k},n=1\dots \mathsf{N}_\text{flat}}$ tensoring with the ``vacuum'' state of the other modes $\gamma_{\bm{k},n=\mathsf{N}_\text{flat}+1\dots \mathsf{N}}$ (i.e. all the modes with energy higher/lower than the flat bands are empty/occupied). We formally define the projector onto $\mathcal{H}_\text{flat}$ as $\hat{P}$. To the leading order of a perturbation series treating both $V/\Delta$ and $W/\Delta$ as small parameters, we can set $\epsilon_{n=1\dots \mathsf{N}_\text{flat}}(\bm{k}) =0$ and project the interacting Hamiltonian onto $\mathcal{H}_\text{flat}$. Formally, this should be achieved by $\hat{P}\hat{H}_\text{int} \hat{P}$, but operationally this could be more conveniently done by retaining terms that only involve $\hat{\gamma}_{\bm{k}, n\le \mathsf{N}_\text{flat}}$, or equivalently, by replacing all $\hat{c}_{\bm{k}\alpha}\rightarrow P_{\alpha \beta}(\bm{k}) \hat{c}_{\bm{k} \beta}$ in the interacting Hamiltonian. We note that this scheme in general will result in additional quadratic terms when compared to $\hat{P} \hat{H}_\text{int}\hat{P}$. We will discuss the subtlety led by this in Sec.~\ref{sec: caveat}.

We emphasize that the main results of this paper do not rely on symmetries other than translation symmetry and charge conservation, and they apply to flat-band systems in arbitrary dimensions.

\subsection{Quantum Geometric Nesting (QGN): definition}

Now we define QGN, which are based on $U_{\alpha n}(\bm{k})$ instead of the dispersion $\epsilon_n(\bm{k})$. Similar to conventional nesting conditions, to define a condition of QGN of an electronic system, one needs to specify the wavevector $\bm{Q}$ and whether it is in the particle-particle (p-p) channel or in the particle-hole (p-h) channel. In order to define the concept of QGN, we need to first define a linear operator for each case:
\begin{align}
    \Pi^{\text{p-p},\bm{Q}}_{\mu'\nu';\mu\nu} =\frac{1}{\mathsf{V}}\sum_{\bm{k}}\left[P^\star_{\mu'\mu}(\frac{\bm{Q}}{2}+\bm{k}) Q_{\nu \nu' } (\frac{\bm{Q}}{2}-\bm{k})+ Q^\star_{\mu'\mu}(\frac{\bm{Q}}{2}+\bm{k}) P_{\nu \nu' } (\frac{\bm{Q}}{2}-\bm{k})  \right], \label{eq: p-p Pi}\\
    \Pi^{\text{p-h},\bm{Q}}_{\mu'\nu';\mu\nu} =\frac{1}{\mathsf{V}}\sum_{\bm{k}}\left[P_{\mu'\mu}(\bm{k}+\frac{\bm{Q}}{2}) Q_{\nu \nu' } (\bm{k}-\frac{\bm{Q}}{2}) + Q_{\mu'\mu}(\bm{k}+\frac{\bm{Q}}{2}) P_{\nu \nu' } (\bm{k}-\frac{\bm{Q}}{2}) \right]. \label{eq: p-h Pi}
\end{align}
In Sec.~\ref{sec: general properties} we prove that the hyper-operator is hermitian and positive semi-definite. Therefore, it can be diagonalized, and all the eigenvalues are real and non-negative.

Then, we call an electronic structure satisfying QGN at a wavevector $\bm{Q}$ in the particle-particle (p-p) channel or the particle-hole (p-h) channel, if the smallest eigenvalue of the corresponding $\Pi^{\bm{Q}}$ is $0$. In other words, QGN is defined by the existence of a $\mathsf{N}\times\mathsf{N}$ matrix $N_{\mu\nu}$, which we call the ``nesting matrix'', satisfying (for any $\mu'\nu'$)
\begin{align}\label{eq: QGN definition}
\sum_{\mu\nu}\Pi_{\mu'\nu';\mu\nu}N_{\mu\nu} =0
\end{align}
Naturally, the value of the smallest eigenvalue of $\Pi^{\bm{Q}}$, $\varpi_0$, defines a ``geometric nestability'' that measures the closeness of a system to perfect geometric nesting in the corresponding channel and $\bm{Q}$; the smaller $\varpi_0$ is, the more ``nestable'' the system is.

This nesting matrix of perfect QGN further suggests an order parameter operator within the flat band subspace:
\begin{align}
    \hat{O}_{\bm{Q}}^\text{p-p} &\equiv \frac{1}{\mathsf{V}} \sum_{\bm{k}; n,m\le \mathsf{N}_\text{flat}} F^{\text{p-p},\bm{Q}}_{nm} (\bm{k})\hat{\gamma}_{\bm{Q}/2+\bm{k}, n} \hat{\gamma}_{\bm{Q}/2-\bm{k},m}  \label{eq: pp order parameter} \\
  \hat{O}_{\bm{Q}}^\text{p-h} &\equiv \frac{1}{\mathsf{V}} \sum_{\bm{k};n,m\le \mathsf{N}_\text{flat}} F^{\text{p-h},\bm{Q}}_{nm} (\bm{k})\hat{\gamma}^\dagger_{\bm{k}+\bm{Q}/2, n} \hat{\gamma}_{\bm{k}-\bm{Q}/2,m} \label{eq: ph order parameter}
\end{align}
which is specified by the form factor
\begin{align}
 \ &F^{\text{p-p},\bm{Q}}_{nm}(\bm{k}) \equiv U_{\mu n } (\frac{\bm{Q}}{2}+\bm{k}) N_{\mu\nu} U_{\nu m} (\frac{\bm{Q}}{2}-\bm{k})\label{eq: p-p form factor} \\
  \text{or}\   \ &F^{\text{p-h},\bm{Q}}_{nm}(\bm{k}) \equiv U^\dagger_{ n\mu } (\bm{k}+\frac{\bm{Q}}{2}) N_{\mu\nu} U_{\nu m} (\bm{k}-\frac{\bm{Q}}{2}) \label{eq: p-h form factor}
\end{align}

For QGN to be non-trivial in p-p channel (i.e. the order parameter is non-zero), it is additionally required that $F^{\text{p-p},\bm{Q}}_{nm} (\bm{k}) = - F^{\text{p-p},\bm{Q}}_{mn} (-\bm{k})$ and so that $N$ must be anti-symmetric. Similarly, for p-n channel, if $2\bm{Q} = 0$ (mod reciprocal lattice vectors), for $\hat{O}^{\text{p-h},\bm{Q}}$ to be hermitian, it is additionally required that $F^{\text{p-h},\bm{Q}}_{nm} (\bm{k}) = \left[F^{\text{p-h},\bm{Q}}_{mn} (\bm{k})\right]^\star$ which restricts $N$ to be hermitian. In Sec.~\ref{sec: general properties}, we will prove that in the p-p or p-h cases, the eigenmatrix of $\Pi^{\bm{Q}}$, in general, satisfies the corresponding condition.

In principle, a system can satisfy multiple QGN conditions in different channels and/or at different $\bm{Q}$s. It is also possible that there can be more than one linearly independent nesting matrix $N$ for each channel and $\bm{Q}$, which signals a further internal symmetry breaking, e.g. triplet SC in spin invariant systems.

The main conclusion of our paper is that, whenever QGN is satisfied, there exists an infinite class of interaction terms, $\hat{H}_\text{int}$, that feature solvable ground states with a non-zero corresponding order parameter. The form of such interaction terms will be given in Sec.~\ref{sec: interactions}, and the reason that they give rise to solvable ground states will be analyzed in Sec.~\ref{sec: ground state}. For any such solvable models, we further show in Sec.~\ref{sec: excitations} that the one- or two-particle excitations as well as certain neutral excitations are also exactly solvable, which can be used to estimate the superfluid stiffness of the superconducting states. In Sec.~\ref{sec: examples} we derive several scenarios where QGN are satisfied, which include many known systems. {Lastly, in Sec.~\ref{sec: generalization} we will present a generalization of the construction scheme, where ground states with non-trivial pairing symmetry are solvable, at the price of losing the ability to exactly solve the excitations.}

\subsection{General Properties}
\label{sec: general properties}

Before we proceed, we analyze some important properties of the linear operator $\Pi^{\bm{Q}}$ and QGN, which will be used for the following sections. {From here on, the superscript $\bm{Q}$ and the channel label will be omitted when not ambiguous, for simplicity.}

\begin{enumerate}

\item {\bf $\Pi$ is hermitian} in the sense that:
\begin{align}
    \Pi_{\mu'\nu';\mu\nu} = \left(\Pi_{\mu\nu ; \mu'\nu'}\right)^\star
\end{align}
which could be proven straightforwardly using the hermicity of $P$ and $Q$ matrices.

\item \label{property: positivity} {\bf $\Pi$ is positive semi-definite}: Using the the definitions of $\Pi$ in Eqs.~\ref{eq: p-h Pi}\&\ref{eq: p-p Pi}, $F$ in Eqs.~\ref{eq: p-p form factor}\&\ref{eq: p-h form factor} and $P$, $Q$ in and below Eq.~\ref{eq: projector}, it follows that, for any matrix $N$:
\begin{align}
    &\sum_{\mu'\nu';\mu\nu} N^\star_{\mu\nu}\Pi_{\mu'\nu';\mu\nu}N_{\mu\nu} = \frac{1}{\mathsf{V}} \sum_{k} \left(\sum_{\substack{n\le\mathsf{N}_\text{flat} \\ \mathsf{N}_\text{flat}<m\le\mathsf{N}}} |F_{nm}(\bm{k})|^2 + \sum_{\substack{m\le\mathsf{N}_\text{flat} \\ \mathsf{N}_\text{flat}<n\le\mathsf{N}}} |F_{nm}(\bm{k})|^2\right) \ge 0 \label{eq: Pi positiveness}
\end{align}

    \item {\bf Equivalent definition of QGN.} Since the equal sign in Eq.~\ref{eq: Pi positiveness} can only be taken when the QGN is satisfied, and it can be taken if and only if every term in the summation equals zero, we can give an equivalent definition of QGN:

    {\it There exists a matrix $N$, such that for arbitrary $\bm{k}$, the form factor $F_{nm}(\bm{k})$ defined by $N$ (as in Eq.~\ref{eq: p-p form factor}/\ref{eq: p-h form factor} depending on the channel) is a block diagonal matrix that disconnects flat band indices from the others, i.e. $F_{nm}(\bm{k})=0$ as long as $n,m$ are not simultaneously $~\le \mathsf{N}_\text{flat}$ or $> \mathsf{N}_\text{flat}$. }

\item {\bf  QGN is gauge invariant.}  Consider a unitary transformation, $V_{n \tilde{n}}(\bm{k})$, which rotates the degenerate bands, $F \rightarrow V^\dagger F V$ will be transformed accordingly, but the block-diagonality will be preserved since we assume the flat bands are disconnected from other bands. An equivalent way of seeing this is to verify that all the ingredients in defining $\Pi$, the projection matrices, are gauge invariant.

\item {\bf  If $N$ is a nesting matrix, $(N N^\dagger)^{n}N$ for $n=1,2,\cdots$ are all nesting matrices.} This is most conveniently seen with the equivalent definition of QGN: If $N$ is a nesting matrix, then the $F$ matrix defined by $N$ is block diagonal, and thus $F' \equiv (FF^\dagger)^n F$ is also block-diagonal for any positive integer $n$; this $F'$ is nothing but the form factor defined by $(N N^\dagger)^{n}N$.

\item {\bf Any nesting matrix $N$ can be decomposed} as
\begin{align}
    N = \sum_{\mathsf{i}=1}^{\mathsf{N}_\text{sv}} a_{\mathsf{i}}N^{(\mathsf{i})}
\end{align}
where $\mathsf{N}_\text{sv}$ is the number of distinct non-zero singular values of $N$, $a_{\mathsf{i}}$ are the non-zero singular values, each $N^{(\mathsf{i})}$ is a nesting matrix with all non-zero singular values equal to $1$,  and different $N^{(\mathsf{i})}$ are orthogonal in the sense that
\begin{align}
\left(N^{(\mathsf{i})}\middle| N^{(\mathsf{j\neq i})}\right)
    \equiv \Tr [(N^{(\mathsf{i})})^\dagger N^{(\mathsf{j\neq i})} ] = 0
\end{align}

To see this, we first recognize that, any nesting matrix can be singular-value-decomposed as
\begin{align}
    N = V_1 D V_2
\end{align}
where $V_1,V_2$ are unitary and
\begin{align}
    D = \left[\begin{array}{ c c c  c c }
   a_{\mathsf{1}} \mathbbm{1}_{d_\mathsf{1}} & & & & \\
   & a_{\mathsf{2}} \mathbbm{1}_{d_\mathsf{2}} & & &  \\
    & & \ddots & &  \\
    & & &  a_{\mathsf{N}_\text{sv}} \mathbbm{1}_{d_{\mathsf{N}_\text{sv}}} & \\
    & & & & \mathbf{0}
  \end{array}\right]  \equiv \sum_{\mathsf{i}=1}^{\mathsf{N}_\text{sv}} a_{\mathsf{i}} D^{(\mathsf{i})}
\end{align}
is diagonal and is a linear combination of identity matrices on singular value blocks ($d_{\mathsf{i}}$ is the degeneracy of $\mathsf{i}$-th singular value).

Then, $(N N^\dagger)^{n}N = V_1 D^{2n+1} V_2$ for $n=1,2,\cdots$ generate a series of nesting matrices  $V_1 D^{1} V_2, V_1 D^{3} V_2, \dots$ (according to the property 5 above). Linearly recombining those nesting matrices yields, on can obtain $N^{(\mathsf{i})} \equiv V_1 D^{(\mathsf{i})} V_2$, which are still nesting matrices. Then, apparently, $\left(N^{(\mathsf{i})}\middle| N^{(\mathsf{j\neq i})}\right) = \Tr [(N^{(\mathsf{i})})^\dagger N^{(\mathsf{j\neq i})} ] = \Tr [V_2^\dagger (D^{(\mathsf{i})})^\dagger V_1^\dagger V_1 D^{(\mathsf{j\neq i})} V_2 ] =  \Tr [(D^{(\mathsf{i})})^\dagger D^{(\mathsf{j\neq i})} ] =0 $

\item {\bf For any $\bm{Q}$ in p-p case, all the eigen-matrices of $\Pi$ are either symmetric or anti-symmetric.} To see this, we note that:
\begin{align}
    \Pi^{\text{p-p},\bm{Q}}_{\mu'\nu';\mu\nu} =&\frac{1}{\mathsf{V}}\sum_{\bm{k}}\left[P^\star_{\mu'\mu}(\frac{\bm{Q}}{2}+\bm{k}) Q_{\nu \nu' } (\frac{\bm{Q}}{2}-\bm{k})+ Q^\star_{\mu'\mu}(\frac{\bm{Q}}{2}+\bm{k}) P_{\nu \nu' } (\frac{\bm{Q}}{2}-\bm{k})  \right] \\
    =&\frac{1}{\mathsf{V}}\sum_{\bm{k}}\left[P^\star_{\mu'\mu}(\frac{\bm{Q}}{2}-\bm{k}) Q_{\nu \nu' } (\frac{\bm{Q}}{2}+\bm{k})+ Q^\star_{\mu'\mu}(\frac{\bm{Q}}{2}-\bm{k}) P_{\nu \nu' } (\frac{\bm{Q}}{2}+\bm{k})  \right] \\
    =&\frac{1}{\mathsf{V}}\sum_{\bm{k}}\left[P_{\mu\mu'}(\frac{\bm{Q}}{2}-\bm{k}) Q^\star_{\nu' \nu } (\frac{\bm{Q}}{2}+\bm{k})+ Q_{\mu\mu'}(\frac{\bm{Q}}{2}-\bm{k}) P^\star_{\nu' \nu } (\frac{\bm{Q}}{2}+\bm{k})  \right] \\
    =& \Pi^{\text{p-p},\bm{Q}}_{\nu'\mu';\nu\mu}
\end{align}
In the second equality, we changed the dummy variable $\bm{k}\rightarrow -\bm{k}$. In the third equality, we used the hermicity of $P$ and $Q$.

Therefore, if $\Pi_{\mu'\nu';\mu\nu} M_{\mu\nu} = \la M_{\mu\nu} $, $\Pi_{\mu'\nu';\mu\nu} M^T_{\mu\nu} = \Pi_{\nu'\mu';\nu\mu} M_{\nu\mu} = \la M^T_{\mu\nu} $, i.e. $M$ and $M^T$ are both eigenmatices of $\Pi$ with the same eigenvalue, then $(M\pm M^T)/2$ construct the symmetric or anti-symmetric eigenmatrix with eigenvalue $\la$.

\item {\bf For $2\bm{Q} =0 $ (mod reciprocal lattice vectors) case in p-h channel, all the eigen-matrices of $\Pi$ are hermitian} (up to an overall prefactor). To see this, we note that:
\begin{align}
    \Pi^{\text{p-h},\bm{Q}}_{\mu'\nu';\mu\nu} =&\frac{1}{\mathsf{V}}\sum_{\bm{k}}\left[P_{\mu'\mu}(\bm{k}+\frac{\bm{Q}}{2}) Q_{\nu \nu' } (\bm{k}-\frac{\bm{Q}}{2}) + Q_{\mu'\mu}(\bm{k}+\frac{\bm{Q}}{2}) P_{\nu \nu' } (\bm{k}-\frac{\bm{Q}}{2}) \right]  \\
    =&\frac{1}{\mathsf{V}}\sum_{\bm{k}}\left[P^\star_{\mu\mu'}(\bm{k}+\frac{\bm{Q}}{2}) Q^\star_{\nu' \nu } (\bm{k}-\frac{\bm{Q}}{2}) + Q^\star_{\mu\mu'}(\bm{k}+\frac{\bm{Q}}{2}) P^\star_{\nu' \nu } (\bm{k}-\frac{\bm{Q}}{2}) \right]  \\
    =&\frac{1}{\mathsf{V}}\sum_{\bm{k}}\left[P^\star_{\mu\mu'}(\bm{k}+\frac{3\bm{Q}}{2}) Q^\star_{\nu' \nu } (\bm{k}+\frac{\bm{Q}}{2}) + Q^\star_{\mu\mu'}(\bm{k}+\frac{3\bm{Q}}{2}) P^\star_{\nu' \nu } (\bm{k}+\frac{\bm{Q}}{2}) \right]  \\
    =&\frac{1}{\mathsf{V}}\sum_{\bm{k}}\left[P^\star_{\mu\mu'}(\bm{k}-\frac{\bm{Q}}{2}) Q^\star_{\nu' \nu } (\bm{k}+\frac{\bm{Q}}{2}) + Q^\star_{\mu\mu'}(\bm{k}-\frac{\bm{Q}}{2}) P^\star_{\nu' \nu } (\bm{k}+\frac{\bm{Q}}{2}) \right]  \\
    =&\left[\Pi^{\text{p-h},\bm{Q}}_{\nu'\mu';\nu\mu}\right]^\star
\end{align}
In the second equality, we used the hermicity of $P$ and $Q$. In the third equality, we shifted the dummy variable $\bm{k} \rightarrow \bm{k}+\bm{Q}$. In the fourth equality, we took into account the fact that $2\bm{Q}=0$.

Therefore, if $\Pi_{\mu'\nu';\mu\nu} M_{\mu\nu} = \la M_{\mu\nu} $, $\Pi_{\mu'\nu';\mu\nu} M^\dagger_{\mu\nu} = \Pi^\star_{\nu'\mu';\nu\mu} M^\star_{\nu\mu} = \la M^\dagger_{\mu\nu} $, i.e. $M$ and $M^\dagger$ are both eigenmatices of $\Pi$ with the same eigenvalue. Then $(M+ M^\dagger)/2$ and $(M- M^\dagger)/(2\mathrm{i})$ construct the hermitian eigenmatrices with eigenvalue $\la$.

\item  {\bf The susceptibility of the order parameters suggested by QGN originating from the flat bands saturate a theoretical upper bound.} To see this bound, we consider an arbitrary order parameter defined in the orbital basis
\begin{align}
        \hat{O}^{\text{p-h},\bm{Q}} \equiv \sum_{\alpha \beta} O_{\mu\nu}(\bm{k}) \hat{c}^\dagger_{\bm{k}+\bm{Q}/2,\mu} \hat{c}_{\bm{k}-\bm{Q}/2,\nu} \\
        \hat{O}^{\text{p-p},\bm{Q}} \equiv \sum_{\alpha \beta} O_{\mu\nu}(\bm{k}) \hat{c}_{\bm{Q}/2+\bm{k},\mu} \hat{c}_{\bm{Q}/2-\bm{k},\nu}
\end{align}
the static susceptibility originating from a set of flat bands at temperature $T$ (the contributions involving the remote bands are neglected) can be derived as (taking the p-h case for example):
\begin{align}
\chi_O &= \int_0^{1/T} d \tau \langle \hat{O}^\dagger(\tau ) \hat{O}\rangle  - \langle \hat{O}^\dagger\rangle\langle \hat{O}\rangle \\
&= \frac{1}{T}
    \sum_{\bm{k}} \tr\left[ P(\bm{k}+\bm{Q}/2)O(\bm{k}) P(\bm{k}-\bm{Q}/2)O^\dagger(\bm{k})  \right]
\end{align}
The trace in the above expression is a geometric factor that depends on the structure of the flat band subspace, and it can be rigorously upper-bounded by
\begin{align}
 &\tr\left[ P(\bm{k}+\bm{Q}/2)O(\bm{k}) P(\bm{k}-\bm{Q}/2)O^\dagger(\bm{k})  \right]  \nonumber\\
\leq &   \sqrt{\tr\left[ O(\bm{k}) P(\bm{k}-\bm{Q}/2)O^\dagger(\bm{k}) \right] \tr\left[O^\dagger(\bm{k}) P(\bm{k}+\bm{Q}/2)O(\bm{k}) \right]} \\
\leq &  \frac{\tr\left[ \mathbbm{1} O(\bm{k}) P(\bm{k}-\bm{Q}/2)O^\dagger(\bm{k}) \right] + \tr\left[  P(\bm{k}+\bm{Q}/2)O(\bm{k}) \mathbbm{1} O^\dagger(\bm{k})\right]}{2}\label{eq: upper bound}
\end{align}
where the first inequality is Cauchy–Schwarz inequality, and the second inequality is the inequality of arithmetic and geometric means.

Then, suppose the flat bands satisfy QGN at $\bm{Q}$ with the nesting matrix $N$, then the suggested order parameter is defined by $O(\bm{k}) = N$, and by definition,
\begin{align}
     \tr\left[P(\bm{k}+\bm{Q}/2) N Q(\bm{k}-\bm{Q}/2) N^\dagger\right] + (P \leftrightarrow Q) = 0
\end{align}
With simple algebraic manipulations, this immediately implies that the upper bound in Eq.~\ref{eq: upper bound} is saturated for every $\bm{k}$. In this sense, QGN of a flat-band system suggests a `best' order parameter with the largest possible susceptibility at low temperatures.

A similar derivation can be done for p-p case, where the susceptibility can be evaluated and bounded as follows ($\nu \equiv \tanh \frac{\mu}{2T} \in (-1,1)$ is the filling fraction of the flat bands and $\mu$ is the chemical potential, both measured from charge neutrality):
\begin{align}
\chi_O &= \int_0^{1/T} d \tau \langle \hat{O}^\dagger(\tau ) \hat{O}\rangle  - \langle \hat{O}^\dagger\rangle\langle \hat{O}\rangle \\
&= \frac{1}{4T} \frac{|\nu|}{\text{arctanh}|\nu|}
    \sum_{\bm{k}} \tr\left[ P^\star(\bm{Q}/2+\bm{k})O(\bm{k}) P(\bm{Q}/2-\bm{k})O^\dagger(\bm{k})  \right] \\
    &\leq \frac{1}{4T} \frac{|\nu|}{\text{arctanh}|\nu|}  \sum_{\bm{k}} \frac{ \tr\left[ \mathbbm{1} O(\bm{k}) P(\bm{Q}/2-\bm{k})O^\dagger(\bm{k})  \right]+\tr\left[ \mathbbm{1} P^\star(\bm{Q}/2+\bm{k})O(\bm{k}) \mathbbm{1} O^\dagger(\bm{k})  \right]}{2}
\end{align}
This bound is similarly saturated when the corresponding QGN is satisfied.

\item ({\bf The issue of embedding}) For the suggested OP to be physical, it is necessary that $F(\bm{k}+\bm{G}) = F(\bm{k})$ for any reciprocal lattice vector $\bm{G}$; this amounts to the restriction $N_{\mu\nu} \mathrm{e}^{\mathrm{i}\bm{G}\cdot (\bm{r}_\mu-\bm{r}_\nu)} = N_{\mu\nu}$ in both p-p or p-h cases. This implies that if $N$ is a nesting matrix at some embedding, it is also a nesting matrix at the periodic embedding $\bm{r}_\mu = \bm{0}$. Therefore, it suffices to consider only the periodic embedding throughout the discussion. Nonetheless, in the supplemental material, we will explicitly keep the embedding dependence in many places. The reduction to the periodic embedding is straightforward.

\end{enumerate}

\subsection{Generalized quantum geometry}
\label{sec: generalized quantum geometry}

The quantum geometry~\cite{provost1980riemannian,2011EPJB...79..121R} of a set of flat bands is a Riemann geometry defined by a notion of distance in the momentum space $d(\bm{k},\bm{k}')\equiv \sqrt{\mathsf{N}_{\text{flat}} - \tr \left[P(\bm{k}) P(\bm{k}')\right]}$ and correspondingly a Fubini-Study metric tensor $g_{ij}(\bm{k}) = \partial_{i}\partial_j d^2(\bm{k},\bm{k}') |_{\bm{k}'=\bm{k}}/2 $. In this section, we show that each satisfied QGN (at momentum $\bm{Q}$ with nesting matrix $N^{\bm{Q}}$ in either channel) defines a {\it generalized} quantum distance in the momentum space, $\mathcal{D}(\bm{k},\bm{k}')$:
\begin{align}
    2\left[\mathcal{D}^{\text{p-p},\bm{Q}}(\bm{k},\bm{k}')\right]^2 &\equiv  \tr\left[P^\star(\frac{\bm{Q}}{2}+\bm{k})N Q(\frac{\bm{Q}}{2}-\bm{k}') N^\dagger\right] + (P \leftrightarrow Q) \\
     2\left[\mathcal{D}^{\text{p-h},\bm{Q}}(\bm{k},\bm{k}')\right]^2 &\equiv  \tr\left[P(\bm{k}+\frac{\bm{Q}}{2})N Q(\bm{k}'-\frac{\bm{Q}}{2}) N^\dagger\right] + (P \leftrightarrow Q)
\end{align}
which satisfies all the rules for a distance (2-4 hold for at least for close enough generic momenta pairs):
\begin{enumerate}
    \item  The distance between a point and itself is always zero $\mathcal{D}(\bm{k},\bm{k})=0$. This is directly implied by the definition of QGN:
    \begin{align}
        \sum_{\mu'\nu';\mu\nu} N^\star_{\mu\nu}\Pi_{\mu'\nu';\mu\nu}N_{\mu\nu} =  \sum_{\bm{k} } 2\left[\mathcal{D}(\bm{k},\bm{k})\right]^2 = 0
    \end{align}
    since each term in the $\bm{k}$-sum is non-negative (Property \ref{property: positivity} of QGN), all the terms must be zero.

    \item The distance is positive between distinct points $\mathcal{D}(\bm{k},\bm{k}'\neq \bm{k})> 0$
    \item The distance is symmetric $\mathcal{D}(\bm{k},\bm{k}')=\mathcal{D}(\bm{k}',\bm{k})$
    \item Triangle inequality: $\mathcal{D}(\bm{k},\bm{k}') + \mathcal{D}(\bm{k}',\bm{k}'')\geq \mathcal{D}(\bm{k},\bm{k}'')$
\end{enumerate}
It also reduces to the usual distance $d(\bm{k},\bm{k}')$ when one consider the `trivial QGN' with $\bm{Q}=\bm{0}$ and $N=\mathbbm{1}$ in the p-h channel.

To see rules 2-4 above are generically approximately satisfied at least for small separations of $\bm{k}$, we note that $\mathcal{D}^2(\bm{k},\bm{k}')$ can be expanded for $\bm{k}'\approx \bm{k}$:
\begin{align}
    \mathcal{D}^2(\bm{k}+\delta \bm{k},\bm{k}) \approx \sum_{ij} \mathcal{G}_{ij}(\bm{k}) \delta \bm{k}_i \delta \bm{k}_j
\end{align}
which defines a {\it generalized} Fubini-study metric:
\begin{align}
    \mathcal{G}_{ij}(\bm{k}) \equiv &\frac{1}{2}\partial_{i}\partial_j \mathcal{D}^2(\bm{k},\bm{k}') |_{\bm{k}'=\bm{k}}\\
    = & - \frac{1}{2}\partial_{i}\partial'_j \mathcal{D}^2(\bm{k},\bm{k}') |_{\bm{k}'=\bm{k}} \\
    =  &\frac{1}{2}\tr\left[\partial_i P(\bm{k}+\bm{Q}/2)N \partial_j P(\bm{k}-\bm{Q}/2) N^\dagger \right]
\end{align}
Generically, $\mathcal{G}$ should be a positive tensor since $\mathcal{D}^2$ is always non-negative. After proper rescaling and rotation, it reduces to an identity tensor that gives the ordinary definition of distance and thus make rules 2-4 satisfied.

We will see below that this notion of generalized metric will determine the superfluid stiffness in a toy model for a superconducting state with non-zero ordering $\bm{Q}$ (Sec.~\ref{sec: engineered model}), similar to the familiar case where the usual quantum metric determines the superfluid stiffness in attractive Hubbard model (Sec.~\ref{sec: time reversal}).

Lastly, we prove that the generalized Fubini-Study metric closely relates to the second moment of the bare correlation functions of the order parameters from in the flat-band subspace, specifically:
\begin{align}
   4\sum_{\bm{k}} \mathcal{G}_{ij}(\bm{k}) = \sum_{\bm{R}\bm{R}'}\left\langle \left(\hat{O}^\dagger_{\bm{R}}-\langle \hat{O}^\dagger_{\bm{R}}\rangle  \right)\left(\bm{R}-\bm{R}'\right)_i\left(\bm{R}-\bm{R}'\right)_j\left(\hat{O}_{\bm{R}'} -\langle \hat{O}_{\bm{R}'}\rangle  \right)\right\rangle
\end{align}
where $\hat{O}_{\bm{R}} \equiv \sum_{\mu\nu} \mathrm{e}^{\mathrm{i}\bm{Q}\cdot \bm{R}} \hat{c}_{\bm{R}\mu} N_{\mu\nu} \hat{c}_{\bm{R}\nu} $ (for p-p case) or $\hat{O}_{\bm{R}} \equiv \sum_{\mu\nu} \mathrm{e}^{\mathrm{i}\bm{Q}\cdot \bm{R}} \hat{c}^{\dagger}_{\bm{R}\mu}  N_{\mu\nu} \hat{c}_{\bm{R}\nu}$ (for p-h case) is the unprojected order parameter density in real space. The expectation value is taken in an empty state in the flat-band subspace. To see this, consider p-h case for example:
\begin{align}
   &\sum_{\bm{R}\bm{R}'}\left\langle \left(\hat{O}^\dagger_{\bm{R}}-\langle \hat{O}^\dagger_{\bm{R}}\rangle  \right)\left(\bm{R}-\bm{R}'\right)_i\left(\bm{R}-\bm{R}'\right)_j\left(\hat{O}_{\bm{R}'} -\langle \hat{O}_{\bm{R}'}\rangle  \right)\right\rangle\nonumber\\
   =&\sum_{\bm{R}\bm{R'},\mu\nu,\mu'\nu'}  \mathrm{e}^{-\mathrm{i} \bm{Q} \cdot (\bm{R}-\bm{R}')} \left(\bm{R}-\bm{R}'\right)_i\left(\bm{R}-\bm{R}'\right)_j     N^\star_{\nu\mu} \langle \hat{c}_{\bm{R}\mu }   \hat{c}^\dagger_{\bm{R}'\mu' } \rangle  N_{\mu'\nu'} \langle \hat{c}_{\bm{R}' \nu'}\hat{c}^\dagger_{\bm{R} \nu}\rangle \\
   =& \sum_{\bm{R}\bm{R'},\mu\nu,\mu'\nu'}  \mathrm{e}^{-\mathrm{i} \bm{Q} \cdot (\bm{R}-\bm{R}')} \left(\bm{R}-\bm{R}'\right)_i\left(\bm{R}-\bm{R}'\right)_j   N^\star_{\nu\mu} P_{\mu\mu'} (\bm{R}-\bm{R}')   N_{\mu'\nu'} P_{\nu'\nu} (\bm{R}'-\bm{R})\\
   =&  2 \sum_{\bm{R}\bm{R'},\mu\nu,\mu'\nu'}  \mathrm{e}^{-\mathrm{i} \bm{Q} \cdot (\bm{R}-\bm{R}')}  N^\star_{\nu\mu} \bm{R}_i P_{\mu\mu'}  (\bm{R}-\bm{R}')   N_{\mu'\nu'} \bm{R}'_j  Q_{\nu'\nu} (\bm{R}'-\bm{R})  \\
   =& 2\sum_{\bm{k}} \tr\left[\partial_i P(\bm{k}+\bm{Q}/2)N \partial_j P(\bm{k}-\bm{Q}/2) N^\dagger \right] = 4\sum_{\bm{k}} \mathcal{G}_{ij}(\bm{k})
\end{align}
where in the first equality we used Wick's theorem, in the second equality we defined $P_{\mu\nu}(\bm{r}) \equiv \frac{1}{\mathsf{V}} \sum_{\bm{k}} \mathrm{e}^{-\mathrm{i}\bm{k}\cdot \bm{r}} P_{\mu\nu}(\bm{k})$ and similarly defined $Q(\bm{r})$, in the third equality we rearranged the terms and relabeled $\bm{R}$ and $\bm{R}'$ in some of them, and in the second last equality we used the general formula $ \bm{r}_i P(\bm{r}) =\sum_{\bm{k}}\left[  \mathrm{i}\partial_{\bm{k}_i} P(\bm{k}) \right] \mathrm{e}^{\mathrm{i}\bm{k}\cdot \bm{r}}$. This relation reduces to the usual case for the conventional quantum metric and electronic density correlations when one considers the `trivial QGN' with $\bm{Q}=\bm{0}$ and $N=\mathbbm{1}$ in the p-h channel. We note that, with a cumulant expansion method, higher-order geometric objects of this geometry (e.g. the Cristoffel symbol and the Riemann curvature) may be further related to the higher-order correlation functions~\cite{PhysRevA.108.032218}.

\section{Suggested solvable interactions}
\label{sec: interactions}

\subsection{General form of the interactions}

To construct the interactions that can give rise to the ground states with the promised orders $\langle \hat{O}\rangle \neq 0$, we first try to construct a set of local, hermitian, fermion bilinear operators $\{\hat{S}^{(I)}_{\bm{R}}\}$ centered at each unit cell and labeled by Latin letters in uppercase ($I,J,\dots$). The operators of interests are those that commute with the order parameter within the $\mathcal{H}_\text{flat}$
\begin{align}
    [\hat{O}, \hat{S}^{(I)}_{\bm{R}}] = 0 \ \ \text{within } \ \mathcal{H}_\text{flat}
\end{align}
{It should be noted that these operators are defined in the entire Hilbert space, whereas the above commutation relation only holds after projection.}

Then, any interaction terms that are built from those operators commute with the order parameter as well. In general, those interactions take the form
\begin{align}\label{eq: interaction}
    \hat{H}_\text{int} = \sum_{\bm{R},\bm{R}'; IJ} V_{IJ, \bm{R}-\bm{R}'}  \left(\hat{S}^{(I)}_{\bm{R}}-\langle \hat{S}^{(I)}_{\bm{R}}\rangle\right) \left(\hat{S}^{(J)}_{\bm{R}'}-\langle\hat{S}^{(J)}_{\bm{R}'}\rangle\right),
\end{align}
where $\langle \hat{S}_{\bm{R}}^{(I)}\rangle$ is simply a number for each $\bm{R}$ and $I$. [It should be noted that those ``interactions'' contain quadratic terms, and are not normal-ordered four-fermion operators. We will discuss the nomenclature subtlety led by these in Sec.~\ref{sec: caveat}.] Note that $\hat{S}^{(I)}_{\bm{R}}$ are not necessarily mutually commuting, although it could be possible to construct a set of $\{\hat{S}^{(I)}_{\bm{R}}\}$ that is mutually commuting.

For later convenience, we further rewrite the interaction as ($\delta \hat{S} \equiv \hat{S} - \langle \hat{S}\rangle$)
\begin{align}
    \hat{H}_\text{int} =\mathsf{V} \sum_{\bm{q}, IJ} V_{IJ}(\bm{q}) \delta \hat{S}^{(I)}(\bm{q}) \delta \hat{S}^{(J)}(- \bm{q})
\end{align}
where
\begin{align}
    V_{IJ}(\bm{q}) \equiv &  \sum_{\bm{R}} \mathrm{e}^{\mathrm{i} \bm{q}\cdot \bm{R}} \ V_{IJ,\bm{R}} \\
    \hat{S}^{(I)}(\bm{q}) \equiv &  \frac{1}{\mathsf{V}} \sum_{\bm{R}} \mathrm{e}^{- \mathrm{i} \bm{q}\cdot \bm{R}} \hat{S}^{(I)}_{\bm{R}} \label{eq: delta S}
\end{align}
For the hermicity of the Hamiltonian, we need $V_{IJ,\bm{R}-\bm{R}'} = [V_{JI,\bm{R}'-\bm{R}}]^\star$ and thus $V_{IJ}(\bm{q})$ is hermitian on all $\bm{q}$.

Since the kinetic energy is zero within the subspace, the order parameter further commutes with the whole Hamiltonian $\hat{H} = \hat{H}_0+\hat{H}_\text{int}$
\begin{align}
    [\hat{O},  \hat{H}_0+\hat{H}_\text{int}] = 0 \ \ \text{within } \ \mathcal{H}_\text{flat}
\end{align}

We will show in Sec.~\ref{sec: ground state} that, one can obtain a series of simultaneous exact eigenstates for all the $\{\hat{S}^{(I)}_{\bm{R}}\}$ operators. For any such an eigenstate, $|\Psi\rangle$, choosing $\Bar{S}^{(I)} = \langle \Psi| \hat{S}^{(I)}_{\bm{R}}|\Psi \rangle$ in Eq.~\ref{eq: interaction} will immediately result in $\hat{H}|\Psi\rangle=0$. Then, as long as $V_{IJ}(\bm{q})$ is postive semi-definite for all $\bm{q}$, $|\Psi\rangle$ must be a ground state. There are infinite choices in choosing $V_{IJ}(\bm{q})$ even if we require the interaction to be short-ranged. Actually, as we will show below, there are also many choices in defining $\hat{S}^{(I)}_{\bm{R}}$, even if we require them to be locally supported near $\bm{R}$. Thus we see that once QGN is satisfied, there exists an infinite class of interactions that feature solvable ground states with nonzero order parameters.

Below we will construct the specific form of the desired hermitian operators. In general, such an operator can be parameterized as:
\begin{align}
    \hat{S}^{(I)}_{\bm{R}} = \sum_{\mu\nu,\bm{R}_1\bm{R}_2} S^{(I)}_{\mu\nu}(\bm{R}_1,\bm{R}_2) \hat{c}^{\dagger}_{\bm{R}+\bm{R}_1,\mu} \hat{c}_{\bm{R}+\bm{R}_2,\nu}
\end{align}
which can be equivalently written with momentum bases as
\begin{align}
    \hat{S}^{(I)}_{\bm{R}} &= \frac{1}{\mathsf{V}}\sum_{\mu\nu,\bm{p}\bm{q}}  \mathrm{e}^{\mathrm{i}\bm{R}\cdot (\bm{p}-\bm{q})}S^{(I)}_{\mu\nu}(\bm{p},\bm{q}) \hat{c}^{\dagger}_{\bm{p},\mu} \hat{c}_{\bm{q},\nu} \\
    &= \frac{1}{\mathsf{V}} \sum_{nm,\bm{p}\bm{q}}  \mathrm{e}^{\mathrm{i}\bm{R}\cdot (\bm{p}-\bm{q})} S^{(I)}_{nm}(\bm{p},\bm{q}) \hat{\gamma}^{\dagger}_{\bm{p},n} \hat{\gamma}_{\bm{q},m} \label{eq: S eigenbasis}
\end{align}
where
\begin{align}
    S^{(I)}_{\mu\nu}(\bm{p},\bm{q}) &\equiv  \sum_{\bm{R}_1\bm{R}_2} S^{(I)}_{\mu\nu}(\bm{R}_1,\bm{R}_2) \mathrm{e}^{\mathrm{i}\left[(\bm{R}_1+\bm{r}_\mu)\cdot \bm{p} -(\bm{R}_2+\bm{r}_\nu) \cdot \bm{q}\right]} \\
    S^{(I)}_{nm}(\bm{p},\bm{q}) &\equiv U^\dagger_{ n \mu}(\bm{p}) S^{(I)}_{\mu\nu}(\bm{p},\bm{q})  U_{\nu m}(\bm{q}).
\end{align}
When projected onto $\mathcal{H}_\text{flat}$, one can simply restrict $n,m \le \mathsf{N}_\text{flat}$ in Eq.~\ref{eq: S eigenbasis}.

The hermicity of $\hat{S}^{(I)}_{\bm{R}}$ ensures that
\begin{align}
    S^{(I)}_{\mu\nu}(\bm{R}_1,\bm{R}_2) &= \left[S^{(I)}_{\nu\mu}(\bm{R}_2,\bm{R}_1)\right]^\star \\
    S^{(I)}_{nm}(\bm{p},\bm{q}) &= \left[S^{(I)}_{mn}(\bm{q},\bm{p})\right]^\star
\end{align}
The locality further implies that $S^{(I)}_{\mu\nu}(\bm{R}_1,\bm{R}_2)$ is a decaying function of both $\bm{R}_1,\bm{R}_2$.

Now we analyze the constraints on the form of $\hat{S}^{(I)}_{\bm{R}}$ to make them commute with the order parameter suggested by the QGN. For p-p and p-h cases, we need to discuss them separately.

\subsection{Particle-particle (p-p) case}

Within $\mathcal{H}_\text{flat}$, we compute
\begin{align}
    &[\hat{O}^\text{p-p}_{\bm{Q}}, \hat{S}^{(I)}_{\bm{R}}] \\
    =& \frac{1}{\mathsf{V}^2} \sum_{\substack{ nmkl \le \mathsf{N}_\text{flat} \\ \bm{k}\bm{p}\bm{q}}} \mathrm{e}^{\mathrm{i} \bm{R}\cdot (\bm{q}-\bm{p})} F^{\text{p-p},\bm{Q}}_{nm}(\bm{k}) S^{(I)}_{kl}(\bm{p},\bm{q}) \left[\hat{\gamma}_{\bm{Q}/2+\bm{k}, n} \hat{\gamma}_{\bm{Q}/2-\bm{k},m}, \hat{\gamma}^{\dagger}_{\bm{p},k} \hat{\gamma}_{\bm{q},l} \right] \\
    =& \frac{1}{\mathsf{V}^2} \sum_{\substack{ nmkl \le \mathsf{N}_\text{flat} \\ \bm{k}\bm{p}\bm{q}}} \mathrm{e}^{\mathrm{i} \bm{R}\cdot (\bm{q}-\bm{p})} F^{\text{p-p},\bm{Q}}_{nm}(\bm{k}) S^{(I)}_{kl}(\bm{p},\bm{q}) \left[\delta_{\bm{p},\frac{\bm{Q}}{2}-\bm{k}} \delta_{mk} \hat{\gamma}_{\frac{\bm{Q}}{2}+\bm{k},n} \hat{\gamma}_{\bm{q},l}  - \delta_{\bm{p},\frac{\bm{Q}}{2}+\bm{k}}  \delta_{nk} \hat{\gamma}_{\frac{\bm{Q}}{2}-\bm{k},m} \hat{\gamma}_{\bm{q},l} \right] \\
    =& \frac{2}{\mathsf{V}^2} \sum_{\substack{ nml \le \mathsf{N}_\text{flat} \\ \bm{k}\bm{q}}} \mathrm{e}^{\mathrm{i} \bm{R}\cdot (\bm{q}+\bm{k} - \frac{\bm{Q}}{2} )} F^{\text{p-p},\bm{Q}}_{nm}(\bm{k}) S^{(I)}_{ml}(\frac{\bm{Q}}{2}-\bm{k},\bm{q})  \hat{\gamma}_{\frac{\bm{Q}}{2}+\bm{k},n} \hat{\gamma}_{\bm{q},l} \\
    =& \frac{2}{\mathsf{V}^2} \sum_{\substack{ nl \le \mathsf{N}_\text{flat}, m\le \mathsf{N} \\ \bm{k}\bm{q}}} \mathrm{e}^{\mathrm{i} \bm{R}\cdot (\bm{q}+\bm{k} )} F^{\text{p-p},\bm{Q}}_{nm}(\bm{k}) S^{(I)}_{ml}(\frac{\bm{Q}}{2}-\bm{k},\frac{\bm{Q}}{2}+\bm{q})  \hat{\gamma}_{\frac{\bm{Q}}{2}+\bm{k},n} \hat{\gamma}_{\frac{\bm{Q}}{2}+\bm{q},l} \\
    =& \frac{1}{\mathsf{V}^2} \sum_{\substack{ nl \le \mathsf{N}_\text{flat}, m\le \mathsf{N} \\ \bm{k}\bm{q}}} \mathrm{e}^{\mathrm{i} \bm{R}\cdot (\bm{q}+\bm{k} )}
 \left[ F^{\text{p-p},\bm{Q}}_{nm}(\bm{k}) S^{(I)}_{ml}(\frac{\bm{Q}}{2}-\bm{k},\frac{\bm{Q}}{2}+\bm{q}) - S^{(I)}_{mn}(\frac{\bm{Q}}{2}-\bm{q},\frac{\bm{Q}}{2}+\bm{k}) F^{\text{p-p},\bm{Q}}_{lm}(\bm{q}) \right]  \hat{\gamma}_{\frac{\bm{Q}}{2}+\bm{k},n} \hat{\gamma}_{\frac{\bm{Q}}{2}+\bm{q},l} \\
     =& \frac{1}{\mathsf{V}^2} \sum_{\substack{ nl \le \mathsf{N}_\text{flat}\\ \bm{k}\bm{q}, \mu\nu\eta}} \mathrm{e}^{\mathrm{i} \bm{R}\cdot (\bm{q}+\bm{k} )} U_{\mu n } (\frac{\bm{Q}}{2}+\bm{k})\left[ N_{\mu\nu}  S^{(I)}_{\nu\eta}(\frac{\bm{Q}}{2}-\bm{k},\frac{\bm{Q}}{2}+\bm{q})  +  S^{(I)}_{\nu\mu}(\frac{\bm{Q}}{2}-\bm{q},\frac{\bm{Q}}{2}+\bm{k}) N_{\nu \eta}  \right] U_{\eta l}(\frac{\bm{Q}}{2}+\bm{q})   \hat{\gamma}_{\frac{\bm{Q}}{2}+\bm{k},n} \hat{\gamma}_{\frac{\bm{Q}}{2}+\bm{q},l}
\end{align}
In the third equality, we used that $F^\text{p-p}_{nm}(\bm{k}) = -F^\text{p-p}_{mn}(-\bm{k})$. In the fourth equality, we shift $\bm{q} \rightarrow \frac{\bm{Q}}{2}+\bm{q}$ and we use the equivalent definition of QGN (property 3 in Sec.~\ref{sec: general properties}) and recognize that the summation of the dummy variable $m$ can be over all possible bands, since the block diagonal structure of $F$ will ensure that the summation for $m>\mathsf{N}_F$ won't generate any results; this is the most crucial role played by QGN in our derivation. In the fifth equality, we recognize that $\bm{k},\bm{q}$ and $n,l$ are dummy variables that are exchangeable, and we are free two exchange them for half of the terms. In the last equality, we substitute in the definitions of $F$ and $S^{(I)}$ and recognize that $N$ is anti-symmetric for p-p case.

Therefore, the commutator vanishes if and only if
\begin{align}
   &\sum_\nu  N_{\mu\nu}  S^{(I)}_{\nu\eta}(\frac{\bm{Q}}{2}-\bm{k},\frac{\bm{Q}}{2}+\bm{q})  +  S^{(I)}_{\nu\mu}(\frac{\bm{Q}}{2}-\bm{q},\frac{\bm{Q}}{2}+\bm{k}) N_{\nu \eta}  = 0 \label{eq: pp primitive matrix condition}
\end{align}
for any $\bm{k},\bm{q}$ and $\mu,\nu$.

Then we choose $S^{(I)}_{\mu\nu}(\bm{p},\bm{q})$ to take a separable form:
\begin{align}\label{eq: separable form}
    S^{(I)}_{\mu\nu}(\bm{p},\bm{q}) = A^{(I)}(\bm{p},\bm{q}) B^{(I)}_{\mu\nu}
\end{align}
where $B$ is a hermitian matrix, and $A^{(I)}(\bm{p},\bm{q})  = \left[A^{(I)}(\bm{q},\bm{p})\right]^\star$ in order for $\hat{S}$ to be hermitian. In real space, this separation of orbital and momentum amounts to
\begin{align}\label{eq: separable form real space}
    \hat{S}^{(I)}_{\bm{R}} = \sum_{\bm{R}_1\bm{R}_2;\mu\nu} A^{(I)}(\bm{R}_1+\bm{r}_\mu,\bm{R}_2+\bm{r}_\nu) B_{\mu\nu} \hat{c}^\dagger_{\bm{R}+\bm{R}_1,\mu} \hat{c}_{\bm{R}+\bm{R}_2,\nu},
\end{align}
where $ A^{(I)}(\bm{r}_1,\bm{r}_2) \equiv \frac{1}{\mathsf{V}} \sum_{\bm{p}\bm{q}}\mathrm{e}^{-\mathrm{i}\left[\bm{r}_1\cdot \bm{p} - \bm{r}_2\cdot \bm{q}\right]} A^{(I)}(\bm{p},\bm{q}) $ is a spatial coefficient, and it satisfies $A^{(I)}(\bm{r}_1,\bm{r}_2)  = \left[A^{(I)}(\bm{r}_2,\bm{r}_1) \right]^\star$.

Substituting this orbital-momentum-separated form into the condition Eq.~\ref{eq: pp primitive matrix condition}, we find a set of {\it sufficient} (but probably not necessary) conditions on $A$ and $B$, respectively:
\begin{align}
    &A^{(I)}(\bm{p},\bm{q}) = A^{(I)} (\bm{Q}-\bm{q},\bm{Q}-\bm{p}) \label{eq: pp first condition} \\
    &  N \cdot B + B^T \cdot N = 0 \label{eq: pp second condition}
\end{align}

The first condition, Eq.~\ref{eq: pp first condition}, is quite loose; Specifically, in real space, it simply means
\begin{align}
     A^{(I)}(\bm{r}_1,\bm{r}_2) &  =\frac{1}{\mathsf{V}} \sum_{\bm{p}\bm{q}} \mathrm{e}^{-\mathrm{i}\bm{r}_1 \cdot \bm{p} + \mathrm{i} \bm{r}_2\cdot \bm{q}} A^{(I)}(\bm{p},\bm{q})  \\
     &  =\frac{1}{\mathsf{V}} \sum_{\bm{p}\bm{q}} \mathrm{e}^{-\mathrm{i}\bm{r}_1 \cdot \bm{p} + \mathrm{i} \bm{r}_2\cdot \bm{q}}  A^{(I)}(\bm{Q}-\bm{p},\bm{Q}-\bm{q})  \\
     &  =\frac{1}{\mathsf{V}} \sum_{\bm{p}\bm{q}} \mathrm{e}^{-\mathrm{i}\bm{r}_1 \cdot (\bm{Q}-\bm{p}) + \mathrm{i} \bm{r}_2\cdot (\bm{Q}-\bm{q})} A^{(I)}(\bm{p},\bm{q})  \\
     &=A^{(I)}(\bm{r}_2,\bm{r}_1)  \mathrm{e}^{-\mathrm{i}\bm{Q}\cdot (\bm{r}_1-\bm{r}_2)}\\
      &=\left[A^{(I)}(\bm{r}_1,\bm{r}_2) \right]^\star \mathrm{e}^{-\mathrm{i}\bm{Q}\cdot (\bm{r}_1-\bm{r}_2)}
\end{align}
which sets the phase of $A^{(I)}(\bm{r}_1,\bm{r}_2)$ to be $\mathrm{e}^{-\mathrm{i}\bm{Q}\cdot (\bm{r}_1-\bm{r}_2)/2}$, but not the amplitude. It is evident that there are many choices, including local ones, in choosing this coefficient.

The second condition, Eq.~\ref{eq: pp second condition}, is also easily satisfiable. In fact, one has at least $\sim 3\mathsf{N}_\text{flat}/2$ linear-independent choices of $B$ for any nesting matrix $N$. The procedure of finding those $B$ matrices is as follows. {First, we perform a spectral decomposition for $N$ with a unitary matrix $V$:
\begin{align}
    N = V G V^T
\end{align}
where $G$ is block diagonal, and each diagonal block is a $2\times 2$ matrix (if $\mathsf{N}_\text{flat}$ is odd, there will also be an extra zero on the diagonal). The $i$-th diagonal block is given by $a_{\mathsf{i}} (\mathrm{i} \sigma^y)$ where $a_{\mathsf{i}}$ is the ${\mathsf{i}}$-th singular value (which is non-negative) of $N$. Now, one can construct a hermitian matrix
\begin{align}
    B = V^\star G' V^T
\end{align}
where $G'$ is also block diagonal, and each  $2\times 2$ digonal block is either zero or $\sigma^{x,y,z}$. One can easily verify that such matrices are hermitian and satisfy Eq.~\ref{eq: pp second condition}. If a non-zero singular value $a_{\mathsf{i}}$ is $d_{\mathsf{i}}>2$-fold degenerate, we will have even more ways of choosing $G'$, since in this case one can also assign non-zero off-diagonal blocks to $G'$. Specifically, the $G$ matrix on the singular value block can now be written as $G = \mathrm{i}\sigma^y \otimes \mathbbm{1}_{d_{\mathsf{i}}/2}$ (note that $d_{\mathsf{i}}$ is even since $N$ is anti-symmetric). Then $G'$ can be either $\sigma^y \otimes \tilde{B}$ with $\tilde{B}$ an arbitrary $\frac{d_{\mathsf{i}}}{2}\times \frac{d_{\mathsf{i}}}{2}$ symmetric matrix, or $\sigma^{0,x,z} \otimes \tilde{B}$ with $\tilde{B}$ an arbitrary $\frac{d_{\mathsf{i}}}{2}\times \frac{d_{\mathsf{i}}}{2}$ anti-symmetric matrix. Therefore, overall, we have at least $\sim 3\mathsf{N}_\text{flat}/2$ linear-independent choices of $B$ in the p-p case.}

In summary, we have proven that, as long as $\hat{S}^{(I)}$ takes the separable form in Eq.~\ref{eq: separable form}, and the loose conditions Eqs.~\ref{eq: pp first condition}\&\ref{eq: pp second condition} are satisfied, $[\hat{O}^\text{p-p}_{\bm{Q}}, \hat{S}^{(I)}_{\bm{R}}] = 0$ within $\mathcal{H}_\text{flat}$. We further showed that Eqs.~\ref{eq: pp first condition}\&\ref{eq: pp second condition} are satisfiable for an infinite class of constructions.

We emphasize that, in principle, there may be other constructions of $\hat{S}^{(I)}_{\bm{R}}$ which make it commute with $\hat{H}$. Here we only prove one specific class of constructions with the specific form defined in Eq.~\ref{eq: separable form}.

\subsection{Particle-hole (p-h) case}

The calculation for p-h case is similar. Within $\mathcal{H}_\text{flat}$, we compute
\begin{align}
    &[\hat{O}^\text{p-h}_{\bm{Q}}, \hat{S}^{(I)}_{\bm{R}}] \\
    =& \frac{1}{\mathsf{V}^2} \sum_{\substack{ nmkl \le \mathsf{N}_\text{flat} \\ \bm{k}\bm{p}\bm{q}}} \mathrm{e}^{\mathrm{i} \bm{R}\cdot (\bm{q}-\bm{p})} F^{\text{p-h},\bm{Q}}_{nm}(\bm{k}) S^{(I)}_{kl}(\bm{p},\bm{q}) \left[\hat{\gamma}^\dagger_{\bm{k}+\bm{Q}/2, n} \hat{\gamma}_{\bm{k}-\bm{Q}/2,m}, \hat{\gamma}^{\dagger}_{\bm{p},k} \hat{\gamma}_{\bm{q},l} \right] \\
    =& \frac{1}{\mathsf{V}^2} \sum_{\substack{ nmkl \le \mathsf{N}_\text{flat} \\ \bm{k}\bm{p}\bm{q}}} \mathrm{e}^{\mathrm{i} \bm{R}\cdot (\bm{q}-\bm{p})} F^{\text{p-h},\bm{Q}}_{nm}(\bm{k}) S^{(I)}_{kl}(\bm{p},\bm{q}) \left[\delta_{\bm{p},\bm{k}-\frac{\bm{Q}}{2}} \delta_{mk} \hat{\gamma}^\dagger_{\bm{k}+ \frac{\bm{Q}}{2},n} \hat{\gamma}_{\bm{q},l}  - \delta_{\bm{q},\bm{k}+\frac{\bm{Q}}{2}}  \delta_{nl} \hat{\gamma}^\dagger_{\bm{p},k} \hat{\gamma}_{\bm{k}-\frac{\bm{Q}}{2},m} \right] \\
    =& \frac{1}{\mathsf{V}^2} \sum_{\substack{ nml \le \mathsf{N}_\text{flat} \\ \bm{k}\bm{q}}} \mathrm{e}^{\mathrm{i} \bm{R}\cdot (\bm{q}-\bm{k}+\frac{\bm{Q}}{2})} F^{\text{p-h},\bm{Q}}_{nm}(\bm{k}) S^{(I)}_{ml}(\bm{k}-\frac{\bm{Q}}{2},\bm{q})   \hat{\gamma}^\dagger_{\bm{k}+ \frac{\bm{Q}}{2},n} \hat{\gamma}_{\bm{q},l} \nonumber\\
    &\ \ \ \  - \frac{1}{\mathsf{V}^2} \sum_{\substack{ nmk \le \mathsf{N}_\text{flat} \\ \bm{k}\bm{p}}}
    \mathrm{e}^{\mathrm{i} \bm{R}\cdot (\bm{k}+\frac{\bm{Q}}{2}-\bm{p})} S^{(I)}_{kn} (\bm{p},\bm{k}+\frac{\bm{Q}}{2})  F^{\text{p-h},\bm{Q}}_{nm}(\bm{k}) \hat{\gamma}^\dagger_{\bm{p},k} \hat{\gamma}_{\bm{k}-\frac{\bm{Q}}{2},m} \\
     =& \frac{1}{\mathsf{V}^2} \sum_{\substack{ nl \le \mathsf{N}_\text{flat} , m\le \mathsf{N}\\ \bm{k}\bm{q}}} \mathrm{e}^{\mathrm{i} \bm{R}\cdot (\bm{q}-\bm{k})} \hat{\gamma}^\dagger_{\bm{k}+ \frac{\bm{Q}}{2},n}  \left[ F^{\text{p-h},\bm{Q}}_{nm}(\bm{k}) S^{(I)}_{ml}(\bm{k}-\frac{\bm{Q}}{2},\bm{q}-\frac{\bm{Q}}{2}) -  S^{(I)}_{nm}(\bm{k}+\frac{\bm{Q}}{2},\bm{q}+\frac{\bm{Q}}{2}) F^{\text{p-h},\bm{Q}}_{ml}(\bm{q})   \right] \hat{\gamma}_{\bm{q}-\frac{\bm{Q}}{2},l} \\
     =& \frac{1}{\mathsf{V}^2} \sum_{\substack{ nl \le \mathsf{N}_\text{flat}, \bm{k}\bm{q} \\ \mu\nu\eta}} \mathrm{e}^{\mathrm{i} \bm{R}\cdot (\bm{q}-\bm{k})} \hat{\gamma}^\dagger_{\bm{k}+ \frac{\bm{Q}}{2},n} U^\dagger_{n\mu}(\bm{k}+\frac{\bm{Q}}{2})   \left[ N_{\mu\nu} S^{(I)}_{\nu \eta}(\bm{k}-\frac{\bm{Q}}{2},\bm{q}-\frac{\bm{Q}}{2}) -  S^{(I)}_{\mu\nu}(\bm{k}+\frac{\bm{Q}}{2},\bm{q}+\frac{\bm{Q}}{2}) N_{\nu\eta}  \right]  U_{\eta l}(\bm{q}-\frac{\bm{Q}}{2}) \hat{\gamma}_{\bm{q}-\frac{\bm{Q}}{2},l}
\end{align}
In the second last equality we redefined $\bm{q}\rightarrow \bm{q}-\frac{\bm{Q}}{2}$ in the first summation, and $\bm{p}\rightarrow \bm{k}+\frac{\bm{Q}}{2}, \bm{k}\rightarrow \bm{q}$ in the second summation. Property 3 of QGN (see Sec.~\ref{sec: general properties}) is also used to allow the intermediate dummy index to sum over all bands. In the last equality, we substitute in the definitions of $F$ and $S^{(I)}$. Therefore, in order for the commutator to vanish, the matrix inside the square parenthesis in the last line must vanish for any $\bm{k}, \bm{q}$.

We again choose $\hat{S}^{(I)}_{\bm{R}}$ to take the separable form in Eq.~\ref{eq: separable form}, and we find the following conditions for the coefficients $A^{(I)}(\bm{p},\bm{q})$  and the hermitian matrix $B$ becomes:
\begin{align}
&A^{(I)}(\bm{p},\bm{q})  = A^{(I)}(\bm{p}+\bm{Q},\bm{q}+\bm{Q}) \label{eq: ph first condition} \\
   & [N,B] = 0 \label{eq: ph second condition}
\end{align}

Now we derive implication of the first condition, Eq.~\ref{eq: ph first condition}, on the form of $A^{(I)}(\bm{r}_1,\bm{r}_2)$ (See Eq.~\ref{eq: separable form real space} for definition):
\begin{align}
     A^{(I)}(\bm{r}_1,\bm{r}_2) &  =\frac{1}{\mathsf{V}} \sum_{\bm{p}\bm{q}} \mathrm{e}^{-\mathrm{i}\bm{r}_1 \cdot \bm{p} + \mathrm{i} \bm{r}_2\cdot \bm{q}} A^{(I)}(\bm{p},\bm{q})  \\
     &  =\frac{1}{\mathsf{V}} \sum_{\bm{p}\bm{q}} \mathrm{e}^{-\mathrm{i}\bm{r}_1 \cdot \bm{p} + \mathrm{i} \bm{r}_2\cdot \bm{q}}  A^{(I)}(\bm{p}+\bm{Q},\bm{q}+\bm{Q})  \\
     &  =\frac{1}{\mathsf{V}} \sum_{\bm{p}\bm{q}} \mathrm{e}^{-\mathrm{i}\bm{r}_1 \cdot (\bm{p}-\bm{Q}) + \mathrm{i} \bm{r}_2 \cdot (\bm{q}-\bm{Q})}  A^{(I)}(\bm{p},\bm{q})  \\
     &=A^{(I)}(\bm{r}_1,\bm{r}_2) \mathrm{e}^{\mathrm{i}\bm{Q}\cdot (\bm{r}_1-\bm{r}_2)}
\end{align}
which further suggests that $A^{(I)}(\bm{r}_1,\bm{r}_2) = 0$ if $\mathrm{e}^{\mathrm{i}\bm{Q}\cdot (\bm{R}_1-\bm{R}_2)}\neq 1$. We note that we are still left us with infinite choices. Especially, all the on-site density operators are always allowed.

To satisfy the second condition in Eq.~\ref{eq: ph second condition}, there are at least $\mathsf{N}_\text{flat}$ choices of such hermitian matrices $B$ if $N$ is hermitian, which is true when $2\bm{Q}=0$ (property 8, Sec.~\ref{sec: general properties}). Those $B$ matrices can be found by first diagonalizing $N$ with a unitary transformation
\begin{align}
    N = V D V^\dagger
\end{align}
where $V$ is unitary and $D$ is diagonal. Then $B$ can be constructed by $VD' V^\dagger$ with $D'$ any diagonal matrix; clearly, there are $\mathsf{N}_\text{flat}$ linearly independent choices. When there is degeneracy in the spectrum of $N$, one can even construct more $B$ matrices, since now any matrix defined on that block will commute with $D$.

For the general case where $N$ is not hermitian, one can repeat the above procedure in its diagonalizable subspace. At least, one can construct two $B$'s for Eq.~\ref{eq: ph second condition}, which is simply the identity matrix and $N$.


In summary, we have proven that, as long as $\hat{S}^{(I)}$ takes the separable form in Eq.~\ref{eq: separable form}, and the loose conditions Eqs.~\ref{eq: ph first condition}\&\ref{eq: ph second condition} are satisfied, $[\hat{O}^\text{p-h}_{\bm{Q}}, \hat{S}^{(I)}_{\bm{R}}] = 0$ within $\mathcal{H}_\text{flat}$. We further showed that Eqs.~\ref{eq: ph first condition}\&\ref{eq: ph second condition} are satisfiable for an infinite class of constructions.

\subsection{Caveat: definition of $\hat{H}_0$ and $\hat{H}_\text{int}$ } \label{sec: caveat}
{{
There were two subtle issues concerning the projection scheme and the definition of interactions we adopted, which we discuss in this section.

Recall that we have effectively enforced the projection by simply replacing $\hat{c}_{\bm{k}\alpha}\rightarrow P_{\alpha \beta} \hat{c}_{\bm{k}\beta}$, i.e. retaining the terms that only involves $\gamma_{\bm{k} n\le \mathsf{N}_\text{flat}}$. This is in general not equal to the projection with $\hat{P}$ for four fermion terms. To see this, we compose any $\hat{c}$ operator as $\hat{c} \equiv \Bar{c} +\tilde{c}$ where $\Bar{c}$ is the part that involves $\gamma_{\bm{k} n\le \mathsf{N}_\text{flat}}$ and $\Tilde{c}$ is the remaining part. Then, for any four fermion operator $\hat{c}_1^\dagger \hat{c}_2 \hat{c}_3^\dagger \hat{c}_4$, the projection with our scheme is:
\begin{align}
    \hat{c}_1^\dagger \hat{c}_2 \hat{c}_3^\dagger \hat{c}_4 \xrightarrow{\text{our scheme}} \bar{c}_1^\dagger \bar{c}_2 \bar{c}_3^\dagger \bar{c}_4
\end{align}
However, for the more rigorously defined projection, this will result in
\begin{align}
    &\hat{c}_1^\dagger \hat{c}_2 \hat{c}_3^\dagger \hat{c}_4 \xrightarrow{\hat{P}} \bar{c}_1^\dagger \bar{c}_2
    \bar{c}_3^\dagger \bar{c}_4 + (\text{quadratic terms}) + (\text{constant}) \\
    &\text{quadratic terms} = \langle \tilde{c}_1^\dagger \tilde{c}_2 \rangle \bar{c}_3^\dagger \bar{c}_4 +\langle \tilde{c}_3^\dagger \tilde{c}_4 \rangle \bar{c}_1^\dagger \bar{c}_2     + \langle \tilde{c}_2  \tilde{c}_3^\dagger  \rangle  \bar{c}_1^\dagger \bar{c}_4 + \langle \tilde{c}_1^\dagger  \tilde{c}_4\rangle \bar{c}_2 \bar{c}_3^\dagger
\end{align}
where $\langle \rangle$ is evaluated on the ``vacuum'' state of the other modes $\gamma_{\bm{k},n=\mathsf{N}_\text{flat}+1\dots \mathsf{N}}$ (i.e. all the modes with energy higher/lower than the flat bands are empty/occupied). Therefore, we see that these two schemes differ by some Hartree-Fock (HF) terms contributed by the electrons (or holes) in the remote bands. Specifically, for the form of interactions we constructed in Eq.~\ref{eq: interaction}, the difference is:
\begin{align}
    \hat{H}_\text{HF} = \sum_{\bm{R},\bm{R}'; IJ} V_{IJ, \bm{R}-\bm{R}'}  \left\{\hat{P}\hat{S}^{(I)}_{\bm{R}}\hat{S}^{(J)}_{\bm{R}'}\hat{P}- \hat{P}\hat{S}^{(I)}_{\bm{R}}\hat{P}\bar{S}^{(J)}_{\bm{R}'}\hat{P}\right\}
\end{align}



The bandwidth of the Hartree-Fock terms is generically $O(V)$, and hence can not be neglected in the ideal limit where the bare kinetic energy $H_0$ has nearly flat bands (i.e. bandwidth smaller than $V$). However, we note that the Hartree-Fock terms are of strength $\sim V \ll \Delta$ by assumption, so they do not alter the energy sequences between the flat bands and the remote bands, nor do they change the definition of the projector matrix $P$ (to the leading order of $V/\Delta$). Therefore, if one only regards the geometric nesting conditions as suggesting potential fermion bilinear orders in a strongly correlated system, one does not need to worry about these issues since there is no ambiguity in defining QGN with the projector matrix $P$.

When the construction of solvable models is concerned, a more careful discussion about the definitions of the definition of $\hat{H}_0$ and $\hat{H}_\text{int}$ is needed. A simple resolution to the subtlety is to regard $\hat{H}_\text{HF}$ as a part of the ``non-interacting Hamiltonian'' instead of part of the ``interaction''. Specifically, we can redefine
\begin{align}
    \hat{H}_0 \leftarrow \hat{H}_0 + \hat{H}_\text{HF} \ , \ \ \hat{H}_\text{int} \leftarrow  \hat{H}_\text{int} - \hat{H}_\text{HF}
\end{align}
in all the discussions in this paper.

On the other hand, if $\hat{H}_\text{HF}$ is just a chemical potential term for the flat bands, we can also avoid all the subtleties. Below we evaluate this possibility by explicitly deriving the condition for such a case. Without loss of generality, we consider an electronic structure with a flat band at the bottom of the spectrum, so that the vacuum state these HF terms are evaluated on is the empty state. In this case, there is only one nontrivial term given by the contraction $\braket{\tilde{c}_2 \tilde{c}_3^\dag}$. We will calculate this term explicitly to derive conditions on when it can be rigorously neglected, and the groundstates constructed in the Main text are asymptotically exact. In this case, the interactions we consider can be written
\begin{align}
    \hat{H}_\text{int} &= \sum_{\bm{q}, IJ} V_{IJ}(\bm{q}) \hat{S}^{(I)}_{\mbf{q}} \hat{S}^{(J)}_{-\mbf{q}}, \qquad \hat{S}^{(I)}_{\mbf{q}} = \frac{1}{\sqrt{\mathsf{V}}} \sum_{{\bm{k}} \mu \nu} S^{(I)}_{\mu \nu}({\bm{k}}+\mbf{q},{\bm{k}}) \hat{c}^\dag_{{\bm{k}}+\mbf{q},\mu} \hat{c}_{{\bm{k}},\nu} = \frac{1}{\sqrt{\mathsf{V}}} \sum_{{\bm{k}} mn} S^{(I)}_{mn}({\bm{k}}+\mbf{q},{\bm{k}}) \hat{\gamma}^\dag_{{\bm{k}}+\mbf{q},m} \hat{\gamma}_{{\bm{k}},n}
\end{align}
and the quadratic term arising from projection onto the flat band Hilbert space is
\bea
\hat{H}_\text{HF} &= \sum_{\mbf{q},IJ} V_{IJ}(\mbf{q}) \frac{1}{\mathsf{V}} \sum_{{\bm{k}} mn, {\bm{k}}' m'n'} S^{(I)}_{mn}({\bm{k}}+\mbf{q},{\bm{k}}) \hat{\gamma}^\dag_{{\bm{k}}+\mbf{q},m} \delta_{{\bm{k}},{\bm{k}}'-\mbf{q}} \delta_{n,m' > \mathsf{N}} S^{(J)}_{m' n'}({\bm{k}}'-\mbf{q},{\bm{k}}') \hat{\gamma}_{{\bm{k}}',n'} \\
&=  \sum_{{\bm{k}} mn} \hat{\gamma}^\dag_{{\bm{k}},m} \lp   \frac{1}{\mathsf{V}}\sum_{\mbf{q},IJ} V_{IJ}(\mbf{q}) [S^{(I)}({\bm{k}},{\bm{k}}-\mbf{q}) (1 - \mathbbm{1}_{\text{flat}}) S^{(J)}({\bm{k}}-\mbf{q},{\bm{k}})]_{mn} \rp \hat{\gamma}_{{\bm{k}},n} \\
\eea
where $(1 - \mathbbm{1}_{\text{flat}})$ is a projector onto the dispersive bands, e.g. $U({\bm{k}})(1 - \mathbbm{1}_{\text{flat}}) U^\dag({\bm{k}}) = Q({\bm{k}})$. $\hat{H}_\text{HF}$ can be neglected if it acts as a trivial chemical potential in the flat band Hilbert space, meaning the quantity in parentheses is independent of ${\bm{k}}$. 

We give two explicit examples. First, we consider the attractive Hubbard case with $V_{IJ}(\mbf{q}) = \frac{U}{2} \delta_{IJ}$ and $S^{(I)}_{\al s, \be s'}({\bm{k}}+\mbf{q},{\bm{k}}) = \delta_{\al I} \delta_{\be I} \sigma^z_{ss'}$ writing out $\mu = \al,s$ for orbital and spin respectively. Then the one-body term is
\bea
\hat{H}_\text{HF} &= \frac{U}{2} \sum_{{\bm{k}} mn} \hat{\gamma}^\dag_{{\bm{k}} m} \left[ \frac{1}{\mathsf{V}} \sum_{\al, \mbf{q}} U^\dag({\bm{k}}) D_\al Q({\bm{k}}-\mbf{q}) D_\al U({\bm{k}}) \right]_{mn} \hat{\gamma}_{{\bm{k}} n}
\eea
where $[D_\al]_{\be s, \be' ,s'} = \delta_{\al \be} \delta_{\al \be'} \delta_{ss'}$ is a diagonal matrix. Ref. \cite{herzog2022many} showed that a symmetry condition refered to as ``uniform pairing" suffices to prove that the the matrix is brakets is proportional to the identity (with the proportionality constant given by the density of the fully occupied flat bands), and thus $\hat{H}_\text{HF}$ is a trivial chemical potential.

The second example concerns the toy model in Sec.~V~C of the Main Text where we choose $V_{IJ}(\mbf{q}) = \delta_{IJ} V_I$ to be diagonal and on-site with
\bea
\label{eq:toymodelapp}
V_0 &= 3V, \quad S_{\mu \nu}^{(0)}(\mbf{p},\mbf{q}) = \tau^z \sigma^0\\
V_i &= V, \quad S_{\mu \nu}^{(i)}(\mbf{p},\mbf{q}) = \tau^0 \sigma^i \\
\eea
for $i = x,y,z$.  This interaction results in the one-body term
\bea
\label{eq:onebodytoy}
H_1 &= V\sum_{{\bm{k}} mn} \gamma^\dag_{{\bm{k}} m} \left[U^\dag({\bm{k}}) \lp \frac{1}{\mathsf{V}} \sum_{\mbf{q}} 3 \tau^z Q({\bm{k}}-\mbf{q}) \tau^z + 3 Q({\bm{k}}-\mbf{q})  \rp U({\bm{k}}) \right]_{mn} \gamma_{{\bm{k}} n}  \\
&= 3 V\sum_{{\bm{k}} mn} \gamma^\dag_{{\bm{k}} m} \left[U^\dag({\bm{k}})\frac{1}{\mathsf{V}} \sum_{\mbf{q}}  \lp Q(\mbf{q}) + \tau^z Q(\mbf{q}) \tau^z  \rp U({\bm{k}}) \right]_{mn} \gamma_{{\bm{k}} n} \\
&= 3 V\sum_{{\bm{k}} mn} \gamma^\dag_{{\bm{k}} m} \left[U^\dag({\bm{k}}) \sigma_0 \tau_0 U({\bm{k}}) \right]_{mn} \gamma_{{\bm{k}} n} \\
&= 3 V \bar{N}
\eea
and thus is a trivial chemical potential on the flat band Hilbert space. Here we used $[Q(\mbf{q}),\sigma^i] = 0$ from $SU(2)$ spin, which reduces $\sum_{\mbf{q}} Q(\mbf{q}) + \tau^z Q(\mbf{q}) \tau^z $ to a diagonal matrix, and all its entries must be equal due to $\tau_x Q(\mbf{q}) \tau_x = Q(\mbf{q}-(\pi,\pi))$, and hence must be $1$ since $\Tr Q({\bm{k}}) = 2$. This is an example of how ``uniform pairing" can be generalized to the more complicated interactions considered here.
}}

\section{Suggested ground states}\label{sec: ground state}

We have proven that once QGN is satisfied, there are several local hermitian operators $\{\hat{S}^{(I)}_{\bm{R}}\}$ within the flat band subspace that commutes with the order parameter
\begin{align}
    [\hat{O}, \hat{S}^{(I)}_{\bm{R}}] = [\hat{O}^\dagger, \hat{S}^{(I)}_{\bm{R}}] = 0 \ \ \text{within} \ \mathcal{H}_F
\end{align}
This motivates us to define a ``pseudo-Hamiltonian''
\begin{align}
\hat{\mathcal{E}}  =\mathsf{V} (\mathrm{e}^{\mathrm{i}\theta} \hat{O} + \mathrm{e}^{-\mathrm{i}\theta} \hat{O}^\dagger),
\end{align}
where $\theta$ is a free $U(1)$ phase. It is easy to see that $\hat{\mathcal{E}}$ also commutes with $\hat{S}^{(I)}_{\bm{R}}$. In this section, we will show that $\hat{\mathcal{E}}$ can be viewed as a certain `trial' Hamiltonian of the system, which provides exact solutions to ordered ground states of the interacting system.

Specifically, since $\hat{S}^{(I)}_{\bm{R}}$ and $\hat{\mathcal{E}}$ commute, we can deduce that any {\it non-degenerate} eigenstate of $\hat{\mathcal{E}}$ must also be an eigenstate of $\hat{S}^{(I)}_{\bm{R}}$. According to the analysis in Sec.~\ref{sec: interactions}, we know that once such an eigenstate is found, we can easily make it a ground state of $\hat{H}$ by adjusting constants. Therefore, the task is to find the unique eigenstates of $\hat{\mathcal{E}}$ and show that they are ordered, ground states of $\hat{H}_\text{int}$.
{
$\hat{\mathcal{E}}$ is Hermitian and quadratic in fermion operators within the flat band subspace, so its spectrum and eigenmodes can be exactly solved. Before proceeding to more detailed discussions, we note that the spectrum of $\hat{\mathcal{E}}$ always consists of {\it flat} pseudo-bands, i.e. the eigenmodes of $\hat{\mathcal{E}}$ on each $\bm{k}$ have identical pseudo-energies. To prove this, we define an operator
\begin{align}
    \hat{\mathcal{E}}' \equiv & \begin{cases}
        \sum_{\bm{k};  n,m\le \mathsf{N}} F^{\text{p-p},\bm{Q}}_{nm} (\bm{k}) \mathrm{e}^{\mathrm{i}\theta} \hat{\gamma}_{\bm{Q}/2+\bm{k}, n} \hat{\gamma}_{\bm{Q}/2-\bm{k},m} + \text{h.c.} & \text{ for p-p case} \\
        \sum_{\bm{k};  n,m\le \mathsf{N}} F^{\text{p-h},\bm{Q}}_{nm} (\bm{k}) \mathrm{e}^{\mathrm{i}\theta} \hat{\gamma}^\dagger_{\bm{k}+\bm{Q}/2, n} \hat{\gamma}_{\bm{k}-\bm{Q}/2,m} + \text{h.c.}  & \text{ for p-h case}
    \end{cases}
\end{align}
which differs from $\hat{\mathcal{E}}$ only in the range of summation of band indices (remember that $n,m \le \mathsf{N}_\text{flat}$ in $\hat{\mathcal{E}}$).  Clearly, since $F_{nm}$ is always block-diagonal, $\hat{\mathcal{E}}'$ does not mix the flat and the remote bands, thus the spectrum of $\hat{\mathcal{E}}$ must be a subset of that of $\hat{\mathcal{E}}'$. On the other hand, from the definition of $F$ (Eqs.~\ref{eq: p-p form factor}\&\ref{eq: p-h form factor}), we know that
\begin{align}
    \hat{\mathcal{E}}' = & \begin{cases}
        \sum_{\bm{k}; \alpha \beta} \mathrm{e}^{\mathrm{i}\theta}  N_{\alpha\beta} \hat{c}_{\bm{Q}/2+\bm{k}, \alpha} \hat{c}_{\bm{Q}/2-\bm{k}, \beta} + \text{h.c.} &  \text{ for p-p case} \\
        \sum_{\bm{k}; \alpha \beta} \mathrm{e}^{\mathrm{i}\theta}  N_{\alpha\beta} \hat{c}^\dagger_{\bm{k}+\bm{Q}/2, \alpha} \hat{c}_{\bm{k}-\bm{Q}/2, \beta} + \text{h.c.} & \text{ for p-h case}
    \end{cases}
\end{align}
which has $k$-independent spectrum. Therefore, we conclude that $\hat{\mathcal{E}}$ also only have flat pseudo-bands.

Now we discuss the specific cases. If the QGN is in the p-p channel (such that $N$ is antisymmetric),  $\hat{\mathcal{E}}$ can be written as
\begin{align}
    \hat{\mathcal{E}} = \frac{1}{2} \sum_{\bm{k}} \psi^\dagger_{\bm{k}} h^\text{BdG}(\bm{k}) \psi_{\bm{k}}
\end{align}
where
\begin{align}
    \psi_{\bm{k}} \equiv (\hat{\gamma}^\dagger_{\bm{Q}/2+\bm{k},1} , \dots , \hat{\gamma}^\dagger_{\bm{Q}/2+\bm{k},\mathsf{N}_\text{flat}}; \hat{\gamma}_{\bm{Q}/2-\bm{k},1} , \dots , \hat{\gamma}_{\bm{Q}/2-\bm{k},\mathsf{N}_\text{flat}} )
\end{align}
is the Nambu spinor on $\bm{k}$ and
\begin{align}
    h^\text{BdG}(\bm{k}) =
   \left[\begin{array}{ c c }
   & \mathrm{e}^{\mathrm{i} \theta} F(\bm{k}) \\
  \mathrm{e}^{-\mathrm{i} \theta}F^\dagger(\bm{k}) &
  \end{array}\right]
\end{align}
is the Bogoliubov-de-Gennes (BdG) pseudo-Hamiltonian on $\bm{k}$. Diagonlizing $h^\text{BdG}(\bm{k})$ will result in $2\mathsf{N}_\text{flat}$ quasiparticle  modes for each pair of $\pm \bm{k}$, and the eigenvalues come in pairs $\pm \la_i$, where $\la^2_i$ is the $i$-th eigenvalue of $F^\dagger(\bm{k})F(\bm{k})$. Further denoting the $i$-th (ortho-normal) eigenvector of $F^\dagger(\bm{k})F(\bm{k})$ as $\phi^{(i)}_{\bm{k}}$, the $(\pm i)$-th eigenvectors of $h^\text{BdG}(\bm{k})$ take the form $\left(\mathrm{e}^{\mathrm{i} \theta}F(\bm{k})\phi^{(i)}_{\bm{k}}, \pm \la_i \phi^{(i)}_{\bm{k}} \right)/(\sqrt{2}\lambda_i)$.

Therefore, the spectrum of $\hat{\mathcal{E}}$ consists of $\mathsf{N}_\text{flat}$ BdG quasiparticle bands, appearing in pairs. Fully occupying all the upper (or lower) bands gives rise to a state $|\Psi_\text{SC}(\theta)\rangle$ with a real eigenvalue $\bar{\mathcal{E}} \equiv  \bar{\varepsilon} \mathsf{V} $, where $\epsilon = \sum_{i} \lambda_i$ is a sum of all the pseudo-energies of the occupied bands and thus non-zero. It is clear that $\bar{\mathcal{E}}$ is a non-degenerate eigenvalue of $\hat{\mathcal{E}}$ and $|\Psi_\text{SC}(\theta)\rangle$ is its unique eigenstate since any change of the occupation configuration of the BdG quasi-particles will also change the eigenvalue. Thus we see that $|\Psi_\text{SC}(\theta)\rangle$ is an eigenstate of $\hat{S}^{(I)}_{\bm{R}}$ and thus $\hat{H}$.

By explicitly writing out the expression of the many-body wavefunction:
\begin{align}
    |\Psi_\text{SC}(\theta)\rangle \propto \prod_{\bm{k},i} \left\{\la_i+\sum_{n,m}\mathrm{e}^{-\mathrm{i}\theta}\phi^{(i)}_{\bm{k},n} \hat{\gamma}^\dagger_{\bm{Q}/2-\bm{k},n} \left[F(\bm{k})\phi^{(i)}_{\bm{k}}\right]^\star_m\hat{\gamma}^\dagger_{\bm{Q}/2+\bm{k},m}\right\}|\text{vac}\rangle,
\end{align}
we readily verify that it hosts an off-diagonal-long-range-order (ODLRO) with a superconducting phase $\theta$ in the sense that
\begin{align}\label{eq: ODLRO}
     \langle \Psi_\text{SC}(\theta)|\hat{O}|\Psi_\text{SC}(\theta)\rangle = \frac{\bar{\varepsilon}}{2} \mathrm{e}^{-\mathrm{i}\theta}
\end{align}

We note that by doing a global gauge transformation $\hat{\gamma}\rightarrow \hat{\gamma}\mathrm{e}^{-\mathrm{i}\theta/2}$, we are able to transform $\hat{\mathcal{E}}(\theta) \rightarrow \hat{\mathcal{E}}(\theta =0)$ and thus $|\Psi_\text{SC}(\theta)\rangle \rightarrow |\Psi_\text{SC}(\theta=0)\rangle$,while keeping $\hat{S}^{(I)}_{\bm{R}}$ unchanged (since it conserves charge). Thus, all $|\Psi_\text{SC}(\theta)\rangle$ with different $\theta$ are simultaneous eigenstate of $\hat{S}^{(I)}_{\bm{R}}$ with the same eigenvalue. Therefore, if one prefers to work with fixed particle numbers, we note that $\theta$ is arbitrary and
\begin{align}
    |\Psi_\text{SC}(\mathsf{N}_{e})\rangle \equiv
    \int_0^{2\pi} \mathrm{e}^{-\mathrm{i}\theta \mathsf{N}_{e} /2} |\Psi_\text{SC}(\theta)\rangle
\end{align}
defines a superconducting state with a fixed electron number $\mathsf{N}_{e}$.

In the above analysis, we have implicitly assumed that $\hat{\mathcal{E}}$ is full-rank, i.e. there are no zero modes. If this is not true, the uniqueness of $\Psi_\text{SC}(\theta)$ needs to be discussed with more care. In this case, we may consider a subspace of $\mathcal{H}_\text{flat}$, $\mathcal{H}_\text{flat}'$, in which the electron modes relevant to the BdG zero modes are all empty. Then, repeating the analysis above proves the uniqueness of $\Psi_\text{SC}(\theta)$ in this subspace. Since $\hat{S}^{(I)}_{\bm{R}}$ preserves $\mathcal{H}_\text{flat}'$, this still proves that $\Psi_\text{SC}(\theta)$ is an eigenstate of $\hat{S}^{(I)}_{\bm{R}}$.

On the other hand, if the QGN is in p-h channel, $\hat{\mathcal{E}}$ breaks the original translation symmetry of the problem for $\bm{Q}\neq 0$, and the BZ of $\hat{\mathcal{E}}$ will be folded. The folding number, $\mathsf{M}$, is the smallest positive integer that makes $\mathsf{M}\bm{Q} =0$ modulo the reciprocal lattice. Then, diagonalizing the pseudo-Hamiltonian gives us $\mathsf{M}\mathsf{N}_\text{flat}$ electron bands within the folded BZ. Again, occupying any integer number of bands that are separated from other bands by finite ``pseudo-gaps'' gives rise to a density-wave state $|\Psi_\text{DW}(\theta)\rangle$ with a real, {\it non-degenerate } and non-zero eigenvalue $\bar{\mathcal{E}}$ within the fixed particle number sector. 
In contrast to the p-p case, here $\theta$ only cycles the pseudo-energies of different pseudo-bands which corresponds to different translation symmetry-breaking patterns, but they do not have direct physical meaning.

}

\section{Excitations in the ideal Hamiltonians}
\label{sec: excitations}
In this section, we solve certain excitations for a system satisfying perfect QGN with an ideal Hamiltonian of the kind we constructed in Sec.~\ref{sec: interactions}. The strategy is to find operator $\xi$ that satisfies:
\begin{align}
    [\hat{H},\hat{\xi}]|\Psi\rangle = E\hat{\xi} |\Psi\rangle
\end{align}
for any ground state $|\Psi\rangle$. Since we have assumed momentum and charge conservation, we are able to seek such excitations within each charge and momentum sector. This method was proposed in Ref.~\cite{PhysRevB.103.205415}.

\subsection{Single particle/hole excitations}

We first compute the following commutators within $\mathcal{H}_\text{flat}$
\begin{align}
    [\hat{S}^{(I)}_{\bm{R}}, \hat{\gamma}_{\bm{k},l}] = &\frac{1}{\mathsf{V}} \sum_{nm\le\mathsf{N}_\text{flat},\bm{p}\bm{q}}  \mathrm{e}^{\mathrm{i}\bm{R}\cdot (\bm{p}-\bm{q})} S^{(I)}_{nm}(\bm{p},\bm{q}) [\hat{\gamma}^{\dagger}_{\bm{p},n} \hat{\gamma}_{\bm{q},m}, \hat{\gamma}_{\bm{k},l}]\nonumber \\
    =& - \frac{1}{\mathsf{V}} \sum_{m\le\mathsf{N}_\text{flat},\bm{q}}  \mathrm{e}^{\mathrm{i}\bm{R}\cdot (\bm{k}-\bm{q})} S^{(I)}_{lm}(\bm{k},\bm{q}) \hat{\gamma}_{\bm{q},m}
\end{align}
\begin{align}
[\hat{S}^{(I)}_{\bm{R}}\hat{S}^{(J)}_{\bm{R}'}, \hat{\gamma}_{\bm{k},l}] = & \hat{S}^{(I)}_{\bm{R}}[\hat{S}^{(J)}_{\bm{R}'}, \hat{\gamma}_{\bm{k},l}]+ [\hat{S}^{(I)}_{\bm{R}}, \hat{\gamma}_{\bm{k},l}]\hat{S}^{(J)}_{\bm{R}'} \nonumber \\
=& - \frac{1}{\mathsf{V}} \sum_{m\le\mathsf{N}_\text{flat},\bm{q}} \left[ \mathrm{e}^{\mathrm{i}\bm{R}'\cdot (\bm{k}-\bm{q})} S^{(J)}_{lm}(\bm{k},\bm{q}) \hat{S}^{(I)}_{\bm{R}}  \hat{\gamma}_{\bm{q},m} + \mathrm{e}^{\mathrm{i}\bm{R}\cdot (\bm{k}-\bm{q})} S^{(I)}_{lm}(\bm{k},\bm{q})   \hat{\gamma}_{\bm{q},m} \hat{S}^{(I)}_{\bm{R}} \right] \nonumber\\
=& \frac{1}{\mathsf{V}^2} \sum_{nm\le\mathsf{N}_\text{flat},\bm{p}\bm{q}} \left[ \mathrm{e}^{\mathrm{i}\bm{R}'\cdot (\bm{k}-\bm{q})} \mathrm{e}^{\mathrm{i}\bm{R}\cdot (\bm{k}-\bm{q})} S^{(J)}_{lm}(\bm{k},\bm{q})S^{(I)}_{mn}(\bm{q},\bm{p}) \hat{\gamma}_{\bm{p},n} \right]
 \nonumber\\
& - \frac{1}{\mathsf{V}} \sum_{m\le \mathsf{N}_\text{flat},\bm{q}} \left[ \mathrm{e}^{\mathrm{i}\bm{R}'\cdot (\bm{k}-\bm{q})} S^{(J)}_{lm}(\bm{k},\bm{q}) \hat{\gamma}_{\bm{q},m} \hat{S}^{(I)}_{\bm{R}}   + \mathrm{e}^{\mathrm{i}\bm{R}\cdot (\bm{k}-\bm{q})} S^{(I)}_{lm}(\bm{k},\bm{q})   \hat{\gamma}_{\bm{q},m} \hat{S}^{(I)}_{\bm{R}} \right]
\end{align}
\begin{align} \label{eq: gamma commutation}
    [\hat{H}_\text{int}, \hat{\gamma}_{\bm{k},l}] =& \frac{1}{\mathsf{V}} \sum_{IJ,nm \le \mathsf{N}_\text{flat},\bm{q}} V_{IJ}(\bm{q}) S^{(J)}_{lm}(\bm{k},\bm{q}+\bm{k}) S^{(I)}_{mn}(\bm{q}+\bm{k},\bm{k}) \hat{\gamma}_{\bm{k},n} \nonumber\\
    & - \sum_{IJ, m \le \mathsf{N}_\text{flat},\bm{q}}[V_{IJ}(\bm{q})+V_{JI}(-\bm{q})]S^{(J)}_{lm}(\bm{k},\bm{q}+\bm{k})\hat{\gamma}_{\bm{q}+\bm{k},m} \delta \hat{S}^{(I)}(\bm{q})
\end{align}

Now, for any ground state $|\Psi\rangle$, since $\delta \hat{S}^{(I)}(\bm{q}) |\Psi\rangle = 0$ (recall $\delta\hat{S}\equiv \hat{S}-\langle \hat{S}\rangle $), it follows that
\begin{align}
    [\hat{H}_\text{int}, \hat{\gamma}_{\bm{P},l}]|\Psi\rangle = \sum_{n \le \mathsf{N}_\text{flat}} \Gamma^{\text{h},\bm{P}}_{ln}\hat{\gamma}_{\bm{P},n} |\Psi\rangle
\end{align}
where $\Gamma^{\text{h},\bm{P}}$ is the transition matrix defined by
\begin{align}
    \Gamma^{\text{h},\bm{P}}_{ln}\equiv \frac{1}{\mathsf{V}} \sum_{IJ,m \le \mathsf{N}_\text{flat},\bm{q}} V_{IJ}(\bm{q}) S^{(J)}_{lm}(\bm{P},\bm{q}+\bm{P}) S^{(I)}_{mn}(\bm{q}+\bm{P},\bm{P})
\end{align}
Therefore, the eigenstates and eigenvalues of $\Gamma^{\text{h},\bm{P}}$ on each $\bm{P}$ describe the charge $-1$, single-hole excitations of the system.

Similarly, we obtain,
\begin{align} \label{eq: gamma dagger commutation}
    [\hat{H}_\text{int}, \hat{\gamma}^\dagger_{\bm{k},l}] =& \frac{1}{\mathsf{V}} \sum_{IJ,nm \le \mathsf{N}_\text{flat},\bm{q}} V_{IJ}  (-\bm{q}) \hat{\gamma}_{\bm{k},n}^\dagger S^{(I)}_{nm}(\bm{k},\bm{q}+\bm{k}) S^{(J)}_{ml}(\bm{q}+\bm{k},\bm{k})  \nonumber\\
    & + \sum_{IJ, m \le \mathsf{N}_\text{flat},\bm{q}}[V_{JI}(\bm{q})+V_{IJ}(-\bm{q})]\hat{\gamma}^\dagger_{\bm{q}+\bm{k},m} S^{(J)}_{ml}(\bm{k}+\bm{q},\bm{k}) \delta \hat{S}^{(I)}(-\bm{q})
\end{align}
\begin{align}
    [\hat{H}_\text{int}, \hat{\gamma}^\dagger_{\bm{P},l}]|\Psi\rangle = \sum_{n \le \mathsf{N}_\text{flat}} \Gamma^{\text{p},\bm{P}}_{ln} \hat{\gamma}^\dagger_{\bm{P},n}   |\Psi\rangle
\end{align}
where $\Gamma^{\text{p},\bm{P}}$ is the transition matrix defined by
\begin{align}
\label{eq:charge1app}
    \Gamma^{\text{p},\bm{P}}_{ln}\equiv \frac{1}{\mathsf{V}} \sum_{IJ,m \le \mathsf{N}_\text{flat},\bm{q}} V_{IJ}(-\bm{q}) S^{(I)}_{nm}(\bm{P},\bm{q}+\bm{P}) S^{(J)}_{ml}(\bm{q}+\bm{P},\bm{P})
\end{align}
Then, the eigenstates and eigenvalues of $\Gamma^{\text{p},\bm{P}}$ on each $\bm{P}$ describe the charge $+1$, single-electron excitations of the system.

We note that $\Gamma^{\text{p/h}}$, in general, do not share the same spectrum since there is no obvious particle-hole symmetry in the system. However, for the p-p case, there is an emergent ``particle-hole'' symmetry that allows us to relate the spectrum $\Gamma^{\text{p}}$ at $\bm{P}$ with that of $\Gamma^{\text{h}}$ at $\bm{P}-\bm{Q}$ ($\bm{Q}$ is the wavevector of the QGN). See Sec.~\ref{sec: excitation relations} for details.

\subsection{Particle-hole (exciton) excitations}

Defining $\bm{k}_\pm \equiv \bm{k}\pm \bm{P}/2$, using Eqs.~\ref{eq: gamma commutation}\&\ref{eq: gamma dagger commutation}, we compute within $\mathcal{H}_\text{flat}$
\begin{align}\label{eq: neutral commutator}
    [\hat{H}_\text{int}, \hat{\gamma}^\dagger_{\bm{k}_+,n}\hat{\gamma}_{\bm{k}_-,m}] =&\frac{1}{\mathsf{V}} \sum_{IJ,n'l \le \mathsf{N}_\text{flat},\bm{q}} V_{IJ}  (-\bm{q}) \hat{\gamma}_{\bm{k}_+,n'}^\dagger S^{(I)}_{n'l}(\bm{k}_+,\bm{q}+\bm{k}_+) S^{(J)}_{ln}(\bm{q}+\bm{k}_+,\bm{k}_+)  \hat{\gamma}_{\bm{k}_-,m} \nonumber\\
    & -\frac{1}{\mathsf{V}} \sum_{IJ, n'm' \le \mathsf{N}_\text{flat},\bm{q}}[V_{JI}(\bm{q})+V_{IJ}(-\bm{q})]\hat{\gamma}^\dagger_{\bm{q}+\bm{k}_+,n'} S^{(J)}_{n'n}( \bm{q}+\bm{k}_+,\bm{k}_+)  S^{(I)}_{mm'}( \bm{k}_-,\bm{q}+\bm{k}_-) \hat{\gamma}_{\bm{q}+\bm{k}_-,m'} \nonumber\\
    & + \frac{1}{\mathsf{V}} \sum_{IJ,m'l \le \mathsf{N}_\text{flat},\bm{q}} V_{IJ} (\bm{q})  \hat{\gamma}^\dagger_{\bm{k}_+,n} S^{(J)}_{ml}(\bm{k}_-,\bm{q}+\bm{k}_-) S^{(I)}_{lm'}(\bm{q}+\bm{k}_-,\bm{k}_-) \hat{\gamma}_{\bm{k}_-,m'} \nonumber\\
    & + \sum_{IJ, n' \le \mathsf{N}_\text{flat},\bm{q}}[V_{JI}(\bm{q})+V_{IJ}(-\bm{q})]\hat{\gamma}^\dagger_{\bm{q}+\bm{k}_+ ,n'} S^{(J)}_{n'n}( \bm{q}+\bm{k}_+,\bm{k}_+) \hat{\gamma}_{\bm{k}_-,m}\delta \hat{S}^{(I)}(-\bm{q})\nonumber\\
    & - \sum_{IJ, m'\le \mathsf{N}_\text{flat},\bm{q}}[V_{IJ}(\bm{q})+V_{JI}(-\bm{q})] \hat{\gamma}^\dagger_{\bm{k}_+,n} S^{(J)}_{mm'}(\bm{k}_-,\bm{q}+\bm{k}_-)\hat{\gamma}_{\bm{q}+\bm{k}_-,m'} \delta \hat{S}^{(I)}(\bm{q})
\end{align}
Then we find that starting from any ground state $|\Psi\rangle$ (recall $\delta \hat{S} \equiv \hat{S}-\langle\hat{S}\rangle$), since $\delta \hat{S}^{(I)}(\bm{q}) |\Psi\rangle = 0$, we have
\begin{align}
     [\hat{H}_\text{int}, \hat{\gamma}^\dagger_{\bm{k}_+,n}\hat{\gamma}_{\bm{k}_-,m}]|\Psi\rangle = \sum_{\bm{k}',n'm'\le \mathsf{N}_\text{flat}}  \Gamma^{\text{p-h}, \bm{P}}_{ nm, \bm{k} ; n'm', \bm{k}'} \hat{\gamma}^\dagger_{\bm{k}'_+,n'}\hat{\gamma}_{\bm{k}'_-,m'}  |\Psi\rangle
\end{align}
where $\Gamma^{\text{p-h},\bm{P}}$ is the scattering matrix for a particle and a hole with total momentum $\bm{P}$
\begin{align}
   \Gamma^{\text{p-h}, \bm{P}}_{ nm, \bm{k} ; n'm', \bm{k}'}  = &  \frac{1}{\mathsf{V}} \sum_{IJ,l \le \mathsf{N}_\text{flat},\bm{q}} V_{IJ}  (-\bm{q}) S^{(I)}_{n'l}(\bm{k}_+,\bm{q}+\bm{k}_+) S^{(J)}_{ln}(\bm{q}+\bm{k}_+,\bm{k}_+)  \delta_{mm'} \delta_{\bm{k},\bm{k}'}  \nonumber\\
    & - \frac{1}{\mathsf{V}}\sum_{IJ}[V_{JI}(\bm{k}'-\bm{k})+V_{IJ}(\bm{k}-\bm{k}')] S^{(J)}_{n'n}( \bm{k}'_+,\bm{k}_+)  S^{(I)}_{mm'}( \bm{k}_-,\bm{k}'_-) \nonumber\\
    & + \frac{1}{\mathsf{V}}\sum_{IJ,l \le \mathsf{N}_\text{flat},\bm{q}} V_{IJ} (\bm{q})  \delta_{nn'}\delta_{\bm{k}\bm{k}'} S^{(J)}_{ml}(\bm{k}_-,\bm{q}+\bm{k}_-) S^{(I)}_{lm'}(\bm{q}+\bm{k}_-,\bm{k}_-)
\end{align}
Solving the spectrum of $\Gamma^{\text{p-h},\bm{P}}$ on $\bm{P}$ will give the charge-neutral particle-hole pair (exciton) excitations with total momentum $\bm{P}$.

\subsection{particle-particle/hole-hole (Cooper) pair excitations}

Defining $\bm{k}^\pm \equiv \bm{P}/2 \pm\bm{k} $ (do not confuse with $\bm{k}_\pm$ defined for the previous section), we compute within $\mathcal{H}_\text{flat}$
\begin{align}
    [\hat{H}_\text{int}, \hat{\gamma}_{\bm{k}^+,n}\hat{\gamma}_{\bm{k}^-,m}] =& \frac{1}{\mathsf{V}}\sum_{IJ,n' l \le \mathsf{N}_\text{flat},\bm{q}} V_{IJ}(\bm{q}) S^{(J)}_{nl}(\bm{k}^+,\bm{q}+\bm{k}^+) S^{(I)}_{ln'}(\bm{q}+\bm{k}^+,\bm{k}^+) \hat{\gamma}_{\bm{k}^+,n'}\hat{\gamma}_{\bm{k}^-,m} \nonumber\\
    & +\frac{1}{\mathsf{V}}\sum_{IJ, n' \le \mathsf{N}_\text{flat},\bm{q}}[V_{IJ}(\bm{q})+V_{JI}(-\bm{q})] S^{(J)}_{nn'}(\bm{k}^+,\bm{q}+\bm{k}^+)\hat{\gamma}_{\bm{q}+\bm{k}^+ ,n'} S^{(I)}_{mm'}(\bm{k}^-,-\bm{q}+\bm{k}^-) \hat{\gamma}_{-\bm{q}+\bm{k}^-,m'} \nonumber\\
    & +\frac{1}{\mathsf{V}}\sum_{IJ,m'l\le \mathsf{N}_\text{flat},\bm{q}} V_{IJ}(\bm{q})  \hat{\gamma}_{\bm{k}^+,n} S^{(J)}_{ml}(\bm{k}^-,\bm{q}+\bm{k}^-) S^{(I)}_{lm'}(\bm{q}+\bm{k}^-,\bm{k}^-) \hat{\gamma}_{\bm{k}^-,m'} \nonumber\\
    & - \sum_{IJ, m \le \mathsf{N}_\text{flat},\bm{q}}[V_{IJ}(\bm{q})+V_{JI}(-\bm{q})] \hat{\gamma}_{\bm{k}^+,n} S^{(J)}_{mm'}(\bm{k},\bm{q}+\bm{k}^-)\hat{\gamma}_{\bm{q}+\bm{k}^-,m'} \delta \hat{S}^{(I)}(\bm{q}) \nonumber\\
     & - \sum_{IJ, n' \le \mathsf{N}_\text{flat},\bm{q}}[V_{IJ}(\bm{q})+V_{JI}(-\bm{q})] S^{(J)}_{nn'}(\bm{k}^+,\bm{q}+\bm{k}^+)\hat{\gamma}_{\bm{q}+\bm{k}^+ ,n'} \delta \hat{S}^{(I)}(\bm{q}) \hat{\gamma}_{\bm{k}^-,m}
\end{align}
\begin{align}
    [\hat{H}_\text{int}, \hat{\gamma}^\dagger_{\bm{k}^-,m} \hat{\gamma}^\dagger_{\bm{k}^+,n}] =& \frac{1}{\mathsf{V}} \sum_{IJ,m'l \le \mathsf{N}_\text{flat},\bm{q}} V_{IJ}  (-\bm{q}) \hat{\gamma}_{\bm{k}^-,m'}^\dagger S^{(I)}_{m'l}(\bm{k}^-,\bm{q}+\bm{k}^-) S^{(J)}_{lm}(\bm{q}+\bm{k}^-,\bm{k}^-) \hat{\gamma}^\dagger_{\bm{k}^+,n} \nonumber\\
    & + \frac{1}{\mathsf{V}}\sum_{IJ, n'm' \le \mathsf{N}_\text{flat},\bm{q}}[V_{IJ}(\bm{q})+ V_{JI}(-\bm{q})]\hat{\gamma}^\dagger_{\bm{q}+\bm{k}^-,m'} S^{(I)}_{m'm}(\bm{k}^-+\bm{q},\bm{k}^-) \hat{\gamma}^\dagger_{-\bm{q}+\bm{k}^+,n'}  S^{(J)}_{n'n}(-\bm{q}+\bm{k}^+,\bm{k}^+) \nonumber\\
    & +\frac{1}{\mathsf{V}}\sum_{IJ,n'l \le \mathsf{N}_\text{flat},\bm{q}} V_{IJ}  (-\bm{q}) \hat{\gamma}^\dagger_{\bm{k}^-,m} \hat{\gamma}_{\bm{k}^+,n'}^\dagger S^{(I)}_{n'l}(\bm{k}^+,\bm{q}+\bm{k}^+) S^{(J)}_{ln}(\bm{q}+\bm{k}^+,\bm{k}^+)  \nonumber\\
    & + \sum_{IJ, m' \le \mathsf{N}_\text{flat},\bm{q}}[V_{JI}(\bm{q})+V_{IJ}(-\bm{q})]\hat{\gamma}^\dagger_{\bm{q}+\bm{k}^-,m'} S^{(J)}_{m'm}(\bm{k}^-+\bm{q},\bm{k}^-) \delta \hat{S}^{(I)}(-\bm{q}) \hat{\gamma}^\dagger_{\bm{k}^+,n} \nonumber\\
    & + \sum_{IJ, n' \le \mathsf{N}_\text{flat},\bm{q}}[V_{JI}(\bm{q})+V_{IJ}(-\bm{q})]\hat{\gamma}^\dagger_{\bm{k}^-,m}\hat{\gamma}^\dagger_{\bm{q}+\bm{k}^+,n'} S^{(J)}_{n'n}(\bm{k}^++\bm{q},\bm{k}^+) \delta \hat{S}^{(I)}(-\bm{q})
\end{align}

Then we find that from any ground state $|\Psi\rangle$, since $\delta \hat{S}^{(I)}(\bm{q}) |\Psi\rangle = 0$ (recall $\delta \hat{S} \equiv \hat{S}-\langle\hat{S}\rangle$), we have
\begin{align}
     [\hat{H}_\text{int}, \hat{\gamma}_{\bm{k}^+,n}\hat{\gamma}_{\bm{k}^-,m}]|\Psi\rangle = \sum_{\bm{k}',n'm'\le \mathsf{N}_\text{flat}}
 \Gamma^{\text{h-h}, -\bm{P}}_{nm, \bm{k} ;n'm', \bm{k}'}  \hat{\gamma}_{\bm{k}'^+,n'}\hat{\gamma}_{\bm{k}'^-,m'}|\Psi\rangle \\
    [\hat{H}_\text{int}, \hat{\gamma}^\dagger_{\bm{k}^+,n}\hat{\gamma}^\dagger_{\bm{k}^-,m} ]|\Psi\rangle =  \sum_{\bm{k}',n'm'\le \mathsf{N}_\text{flat}}  \Gamma^{\text{p-p}, \bm{P}}_{ nm, \bm{k}; n'm', \bm{k}' } \hat{\gamma}^\dagger_{\bm{k}'^+,n'} \hat{\gamma}^\dagger_{\bm{k}'^-,m'}
  |\Psi\rangle
\end{align}
where $\Gamma^{\text{p-p},\bm{P}}$ is the scattering matrix for two particles with total momentum $\bm{P}$, and  $\Gamma^{\text{h-h}, -\bm{P}}$ is the scattering matrix for two holes with total momentum $-\bm{P}$,
\begin{align}
\label{eq:twoparticleGamma}
    \Gamma^{\text{h-h},-\bm{P}}_{nm, \bm{k};n'm', \bm{k}'} = &\frac{1}{\mathsf{V}} \sum_{IJ,l \le \mathsf{N}_\text{flat},\bm{q}} V_{IJ}  (\bm{q}) S^{(J)}_{nl}(\bm{k}^+,\bm{q}+\bm{k}^+) S^{(I)}_{ln'}(\bm{q}+\bm{k}^+,\bm{k}^+)  \delta_{mm'} \delta_{\bm{k},\bm{k}'}  \nonumber\\
    & + \frac{1}{\mathsf{V}}\sum_{IJ}[V_{IJ}(\bm{k}'-\bm{k})+V_{JI}(\bm{k}-\bm{k}')] S^{(J)}_{nn'}( \bm{k}^+,\bm{k}'^+)  S^{(I)}_{mm'}( \bm{k}^-,\bm{k}'^-) \nonumber\\
    & +\frac{1}{\mathsf{V}} \sum_{IJ,l \le \mathsf{N}_\text{flat},\bm{q}} V_{IJ} (\bm{q})  \delta_{nn'}\delta_{\bm{k}\bm{k}'} S^{(J)}_{ml}(\bm{k}^-,\bm{q}+\bm{k}^-) S^{(I)}_{lm'}(\bm{q}+\bm{k}^-,\bm{k}^-) \\
    \Gamma^{\text{p-p},\bm{P}}_{nm, \bm{k};n'm', \bm{k}'}   = & \frac{1}{\mathsf{V}}\sum_{IJ,l \le \mathsf{N}_\text{flat},\bm{q}} V_{IJ}  (-\bm{q}) S^{(I)}_{m'l}(\bm{k}^-,\bm{q}+\bm{k}^-) S^{(J)}_{lm}(\bm{q}+\bm{k}^-,\bm{k}^-)  \delta_{nn'} \delta_{\bm{k},\bm{k}'}  \nonumber\\
    & + \frac{1}{\mathsf{V}} \sum_{IJ}[V_{JI}(\bm{k}-\bm{k}')+V_{IJ}(\bm{k}'-\bm{k})] S^{(I)}_{m'm}( \bm{k}'^-,\bm{k}^-)  S^{(J)}_{n'n}( \bm{k}'^+,\bm{k}^+) \nonumber\\
    & + \frac{1}{\mathsf{V}} \sum_{IJ,l \le \mathsf{N}_\text{flat},\bm{q}} V_{IJ} (\bm{q})  \delta_{mm'}\delta_{\bm{k}\bm{k}'} S^{(I)}_{n'l}(\bm{k}^+,\bm{q}+\bm{k}^+) S^{(J)}_{ln}(\bm{q}+\bm{k}^+,\bm{k}^+)
\end{align}
Solving the spectrum and eigenstates of the scattering matrices will give rise to the particle-particle pair or hole-hole pair excitations with charge $\pm 2$ and momentum $\pm \bm{P}$. {Note that the first and last lines of \Eq{eq:twoparticleGamma} have $\delta_{{\bm{k}},{\bm{k}}'}$ factors, and are exactly the charge $\pm1$ excitation energies, and the middle line has the interpretation of the scattering matrix. It is important to note that the basis on which $\Gamma$ acts is $\gamma^\dag_{{\bm{k}}_+,m}\gamma^\dag_{{\bm{k}}_-,n}$ which anti-symmetrizes the matrix due to the anti-commutation relations of the electron operators. Thus if $\psi_{{\bm{k}}mn}(\mbf{p})$ is an eigenvector of $\Gamma$, it only corresponds to a physical excitation of the model if
\bea
\sum_{{\bm{k}}mn} (\psi_{{\bm{k}} mn}(\mbf{p})-\psi_{-\mbf{p}-{\bm{k}} nm}(\mbf{p})) \gamma^\dag_{\mbf{p}+{\bm{k}},m}\gamma^\dag_{-{\bm{k}},n} = \sum_{{\bm{k}}mn} (\psi_{-{\bm{k}}_- mn}(\mbf{p})-\psi_{-{\bm{k}}_+ nm}(\mbf{p})) \gamma^\dag_{{\bm{k}}_+,m}\gamma^\dag_{{\bm{k}}_-,n}
\eea
is nonzero. For models with $S_z$ symmetry, we can decompose $\Gamma$ into spin 0 and spin $\pm 1$ sectors which are anti-symmetric/symmetric in spin, respectively. In the Hubbard case for instance, one can show the nontrivial scattering term vanishes on the spin $\pm1$ states \cite{herzog2022many}.

To be explicit, we will write up the spin-single branch of the Cooper pair scattering matrix assuming $SU(2)$ symmetry and time-reversal. We find
\bea
\label{eq:spinzero}
\Gamma^{\text{p-p} \u\d,\bm{P}}_{nm, \bm{k};n'm', \bm{k}'} &= (\Gamma^{\text{p.}}_{m \u,m' \u}({\bm{k}}^-) \delta_{nn'}+\delta_{mm'}\Gamma^{\text{p.}}_{n \d, n' \d}({\bm{k}}^+)) \delta_{\bm{k} \bm{k}'} \\
&\!\!\!\!\!\!\!\! + \frac{1}{\mathsf{V}} \sum_{IJ}[V_{JI}(\bm{k}-\bm{k}')+V_{IJ}(\bm{k}'-\bm{k})] \big( S^{(I)}_{m' \u,m \u}( \bm{k}'^-,\bm{k}^-)  S^{(J)}_{n' \d, n \d}( \bm{k}'^+,\bm{k}^+) - S^{(I)}_{m' \d,m \u}(-\bm{k}'^-,\bm{k}^-)  S^{(J)}_{n' \u, n \d}(-\bm{k}'^+,\bm{k}^+) \big)
\eea
where we took $m \to m,s$ to label the $m$th band in the spin $s$ sector. }

\subsection{Peculiarities of the p-p case}

\subsubsection{Relation between different types of excitations}\label{sec: excitation relations}

In general, the excitations derived in previous sections do not have any relation. However, in the p-p case, we find that certain excitations with the same fermion (and thus charge) parity could be related. To see this, suppose $\hat{\xi}^{(c)}_{\bm{P}}$ creates a charge $c$ excitation with total momentum $\bm{P}$ and energy $E^{(c)}(\bm{P})$, then
\begin{align}
\hat{\xi}^{(c+2)}_{\bm{P}+\bm{Q} } &\equiv  [\hat{\xi}^{(c)}_{\bm{P} }, (\hat{O}^{\text{p-p},\bm{Q}})^\dagger] \\
\hat{\xi}^{(c-2)}_{\bm{P}-\bm{Q} } &\equiv  [\hat{\xi}^{(c)}_{\bm{P} }, \hat{O}^{\text{p-p},\bm{Q}}]
\end{align}
define a charge $c\pm 2$ excitation with momentum $\bm{P}\pm \bm{Q}$ and energy $E^{(c\pm 2)}(\bm{P}\pm \bm{Q}) = E^{(c)}(\bm{P})$. This is because that
\begin{align}
     [\hat{H}_\text{int}, \hat{\xi}^{(c-2)}_{\bm{P}-\bm{Q} }]|\Psi\rangle =& \left\{[\hat{H}_\text{int},\hat{\xi}^{(c)}_{\bm{P} }]\hat{O}^{\text{p-p},\bm{Q}} + \hat{\xi}^{(c)}_{\bm{P} }[\hat{H}_\text{int},\hat{O}^{\text{p-p},\bm{Q}}] - [\hat{H}_\text{int},\hat{O}^{\text{p-p},\bm{Q}}]\hat{\xi}^{(c)}_{\bm{P} } - \hat{O}^{\text{p-p},\bm{Q}}[\hat{H}_\text{int},\hat{\xi}^{(c)}_{\bm{P} }]\right\}
     |\Psi\rangle \nonumber\\
     =&  \left[ E^{(c)}(\bm{P}) \hat{\xi}^{(c)}_{\bm{P} }\hat{O}^{\text{p-p},\bm{Q}} + 0-0-  E^{(c)}(\bm{P})\hat{O}^{\text{p-p},\bm{Q}}  \hat{\xi}^{(c)}_{\bm{P} } \right]|\Psi\rangle \nonumber \\
     =& E^{(c)}(\bm{P}) \hat{\xi}^{(c-2)}_{\bm{P}-\bm{Q} }|\Psi\rangle
\end{align}
and similarly for $\xi^{(c+2)}_{\bm{P}+\bm{Q} }$.

Applying this to the previous cases, we know that the particle and hole excitations share the same spectrum, and the particle-particle, particle-hole, and hole-hole pair excitations share the same spectrum.

\subsubsection{Pseudo-spin $SU(2)$ symmetry in the p-p case}
In this section, we will show that there are $\mathsf{N}_\text{sv}$ pseudo-spin $SU(2)$ symmetries in the solvable Hamiltonian we constructed for the p-p case, where $\mathsf{N}_\text{sv}$ is the number of different non-zero singular values of $N$.

First, we recall that each nesting matrix can be decomposed as  (property 6 in Sec.~\ref{sec: general properties})
\begin{align}
    N = \sum_{\mathsf{i}= 1,\dots,\mathsf{N}_\text{sv}}a_{\mathsf{i}}N^{(\mathsf{i})}
\end{align}
where $a_{\mathsf{i}}$ are the singular values of $N$,  and each $N^{(\mathsf{i})}$ is a nesting matrix with singular values equal to $0,1$. Different $N^{(\mathsf{i})}$ are orthogonal.

Then we recognize that there are $\mathsf{N}_\text{sv}$ linearly independent order parameters that commute with the Hamiltonian. They are given by
\begin{align}
    \hat{O}^{(\mathsf{i})} &\equiv \frac{1}{\mathsf{V}} \sum_{\bm{k}; n,m\le \mathsf{N}_\text{flat}} F^{(\mathsf{i})}_{nm} (\bm{k})\hat{\gamma}_{\bm{Q}/2+\bm{k}, n} \hat{\gamma}_{\bm{Q}/2-\bm{k},m}
\end{align}
where $F^{(\mathsf{i})}$ is given by $N^{(\mathsf{i})}$ according to Eq.~\ref{eq: p-p form factor}. To prove this, we repeat Eq.~\ref{eq: pp primitive matrix condition} here (but recast into a more concise form), which is the sufficient and necessary condition for a $\hat{S}_{\bm{R}}$ to commute with $\hat{O}$ (constructed by $N$):
\begin{align}
   & N \cdot  S(\bm{k},\bm{q})  +  S^T(\bm{Q}-\bm{q},\bm{Q}-\bm{k}) \cdot  N  = 0
\end{align}
Utilizing this we know that
\begin{align}
    N N^\dagger N S(\bm{k},\bm{q}) =&- N N^\dagger S^T(\bm{Q}-\bm{q},\bm{Q}-\bm{k})  N \nonumber\\
   =& -N \left[S^\star(\bm{Q}-\bm{q},\bm{Q}-\bm{k}) N \right]^\dagger  N \nonumber\\
    =& -N \left[S^T(\bm{Q}-\bm{k},\bm{Q}-\bm{q}) N \right]^\dagger  N \nonumber\\
   =& N \left[ N S(\bm{q},\bm{k}) \right]^\dagger   N \nonumber\\
   =& N S(\bm{k},\bm{q})  N^\dagger    N \nonumber\\
   =& -S^T(\bm{Q}-\bm{q},\bm{Q}-\bm{k}) N N^\dagger N
\end{align}
where in the third and the fifth equality we used the hermicity of $S$: $S(\bm{k},\bm{q}) = S^\dagger(\bm{q},\bm{k})$. Therefore, we know that the Hamiltonian we constructed also commutes with the order parameter $\hat{O}$ constructed by $NN^\dagger N$ and all even those corresponding to $(NN^\dagger)^{n=1,2,\dots} N$. Since linearly combining $(NN^\dagger)^{n=1,2,\dots} N$ can yield $\{N^{(\mathsf{i})}\}$ (see property 6 in Sec.~\ref{sec: general properties}), linearly combining the corresponding order parameters can also yield $\{\hat{O}^{(\mathsf{i})}\}$. We thus conclude that all those order parameters $\{\hat{O}^{(\mathsf{i})}\}$ commute with the Hamiltonian we constructed.

Finally, we recognize that, for each $\hat{O}^{(\mathsf{i})}$, we can construct three operators:
\begin{align}
    \hat{J}^{(\mathsf{i})}_x &\equiv \left[\hat{O}^{(\mathsf{i}),\dagger} + \hat{O}^{(\mathsf{i})}\right]/2 \\
    \hat{J}^{(\mathsf{i})}_y &\equiv \left[\hat{O}^{(\mathsf{i}),\dagger} - \hat{O}^{(\mathsf{i})}\right] /(2\mathrm{i}) \\
    \hat{J}^{(\mathsf{i})}_z &\equiv [\hat{O}^{(\mathsf{i}),\dagger},  \hat{O}^{(\mathsf{i})}]/4
\end{align}
all of which commute with the Hamiltonian, and form an $\mathfrak{su}(2)$ algebra (``$\eta$-pairing'') in the sense that:
\begin{align} [\hat{J}^{(\mathsf{i})}_a,\hat{J}^{(\mathsf{i})}_b] = \mathrm{i} 2 \epsilon^{abc} \hat{J}^{(\mathsf{i})}_c.
\end{align}
To check these commutation relations, we compute (defining $\bm{k}^\pm \equiv \bm{Q}/2 \pm\bm{k} $)
\begin{align}
    \hat{J}^{(\mathsf{i})}_z =  &
    \frac{1}{4}\sum_{\bm{k},nmn'm'\le \mathsf{N}_\text{flat} } F^{(\mathsf{i})}_{nm}(\bm{k})F^{(\mathsf{i}),\star}_{n'm'}(\bm{k})[\hat{\gamma}^\dagger_{\bm{k}^-,m'}\hat{\gamma}^\dagger_{\bm{k}^+, n'}  ,\hat{\gamma}_{\bm{k}^+, n} \hat{\gamma}_{\bm{k}^-,m} ] + F^{(\mathsf{i})}_{nm}(\bm{k})F^{(\mathsf{i}),\star}_{n'm'}(-\bm{k})[\hat{\gamma}^\dagger_{\bm{k}^+,m'}\hat{\gamma}^\dagger_{\bm{k}^-, n'}  ,\hat{\gamma}_{\bm{k}^+, n} \hat{\gamma}_{\bm{k}^-,m} ] \nonumber\\
     = & \frac{1}{4} \sum_{\bm{k},nmn'm'\le \mathsf{N}_\text{flat} } F^{(\mathsf{i})}_{nm}(\bm{k})F^{(\mathsf{i}),\star}_{n'm'}(\bm{k}) [\delta_{nn'} \hat{\gamma}^\dagger_{\bm{k}^-, m'}  \hat{\gamma}_{\bm{k}^-, m} - \delta_{mm'} \hat{\gamma}_{\bm{k}^+, n}\hat{\gamma}^\dagger_{\bm{k}^+, n'}  ]  \nonumber\\
     & + \frac{1}{4}\sum_{\bm{k},nmn'm'\le \mathsf{N}_\text{flat} } F^{(\mathsf{i})}_{mn}(\bm{k})F^{(\mathsf{i}),\star}_{n'm'}(-\bm{k}) [\delta_{n'm} \hat{\gamma}_{\bm{k}^+, n}\hat{\gamma}^\dagger_{\bm{k}^+, m'} - \delta_{nm'}
     \hat{\gamma}^\dagger_{\bm{k}^-, n'}  \hat{\gamma}_{\bm{k}^-, m}    ]  \nonumber\\
     = &  \sum_{\bm{k},nm\le \mathsf{N}_\text{flat} }  [F^{(\mathsf{i})}(\bm{k})F^{(\mathsf{i}),\dagger}(\bm{k})]_{nm}  \left(\frac{1}{2}\delta_{nm} -\hat{\gamma}_{\bm{k}^+, n}\hat{\gamma}^\dagger_{\bm{k}^+, m}\right).
\end{align}
Then for $\hat{J}^{(\mathsf{i})}_- \equiv \hat{J}^{(\mathsf{i})}_x - \mathrm{i} \hat{J}^{(\mathsf{i})}_y = \hat{O}^{(\mathsf{i})}$, it is straightforward to check
\begin{align}
    [\hat{J}^{(\mathsf{i})}_-, \hat{J}^{(\mathsf{i})}_z] = & 2\sum_{\bm{k},nm\le \mathsf{N}_\text{flat} }  [F^{(\mathsf{i})}(\bm{k})F^{(\mathsf{i}),\dagger}(\bm{k}) F^{(\mathsf{i})}(\bm{k})]_{nm}  \hat{\gamma}_{\bm{k}^+, n}\hat{\gamma}_{\bm{k}^-, m} \nonumber\\
    = & 2\sum_{\bm{k},nm\le \mathsf{N}_\text{flat} }  [F^{(\mathsf{i})}(\bm{k})]_{nm}  \hat{\gamma}_{\bm{k}^+, n}\hat{\gamma}_{\bm{k}^-, m} =2 \hat{J}^{(\mathsf{i})}_-
\end{align}
where the second equality used the fact that all the singular values of $N^{(\mathsf{i})}$ are either $0$ or $1$. Similarly, we can obtain
\begin{align}
    [\hat{J}^{(\mathsf{i})}_+, \hat{J}^{(\mathsf{i})}_z] = -2\hat{J}^{(\mathsf{i})}_+
\end{align}
for $\hat{J}^{(\mathsf{i})}_+ \equiv \hat{J}^{(\mathsf{i})}_x + \mathrm{i} \hat{J}^{(\mathsf{i})}_y = \hat{O}^{(\mathsf{i}),\dagger}$. Combining these two equality we thus verify that $\hat{J}^{(\mathsf{i})}_{x,y,z}$ form an $\mathfrak{su}(2)$ algebra. Since they all commute with the Hamiltonian, the system has an $SU(2)$ symmetry generated by them.

It is straightforward to check that the symmetry generators commute for different $(\mathsf{i})$. Therefore, we conclude that there are $\mathsf{N}_\text{sv}$ pseudo-spin $SU(2)$ symmetries in the solvable Hamiltonian we constructed for the p-p case. Intriguingly, for each of them, {the conservation of $\hat{J}^{(\mathsf{i})}_z$ implies charge conservation in a part of the flat bands}.

\subsubsection{The pseudospin stiffness}

We first review some general results for systems with $SU(2)$ symmetry. It is well known that its symmetry-broken phase is described by a non-linear sigma model (NLSM)~\cite{zinn2021quantum, PhysRevX.4.031057}:
\begin{align}
    \mathcal{L}_\text{eff} = \frac{1}{2} \sum_{a,b}  \left[ \rho^{ab} \pi_a \partial_t \pi_b + g \partial_t \pi_a \partial_t \pi_b - \sum_{ij} D_{ij} \delta^{ab} \nabla_i \pi_a \nabla_j \pi_b \right] + \dots
\end{align}
where $\pi_{a=x,y,z}$ are the Nambu–Goldstone (NG) fields of the pseudospin symmetry, $\rho^{ab} = \sum_{c} 2 \epsilon^{abc} \langle \hat{J}^{c}\rangle /\mathsf{V}$ is related to the pseudospin density evaluated in the symmetry broken phase, and $D_{ij}$, $g$ are undetermined effective parameters. We note that $D_{ij}$ is the pseudospin stiffness tensor.

From this NLSM, it can be derived that, when the pseudospin $\Vec{J}$ is ferromagnetically ordered, i.e. $ \bar{J} = |\langle\vec{J}\rangle| /\mathsf{V} \neq 0 $ in the thermodynamic limit, the NG fields in the orthogonal directions are mutual conjugate variables, and the dispersion relation of the NG mode at small wavevector $\delta \bm{Q}$ will be given by
\begin{align}\label{eq: NG dispersion}
    E_\text{NG}(\delta \bm{Q})  = \delta \bm{Q}_{i} \frac{D_{ij}}{4\bar{J}}\delta \bm{Q}_{j}
\end{align}
Given the quadratic dispersion of the NG mode, the Mermin-Wagner-Coleman theorem further states that spontaneous symmetry breaking of the pseudospin $SU(2)$ symmetry generically cannot occur at finite temperature when the space dimension is $d \le 2$ (at zero temperature, in contrast, it can happen at all positive dimensions). However, when $d=2$, the susceptibility $\chi(T) \sim T^3/|D|^4 \exp(4\pi |D|/ T)$ (where $|D| = \sqrt{\det{D}}$) is large at low temperature $T$~\cite{zinn2021quantum}. In this case, any explicit breaking of the symmetry with an energy scale $D'$ will result in the ordering of the pseudospin, with critical temperature $T_c$ estimated by
\begin{align}
    D' \chi(T_c) \sim 1 \implies T_c \sim 4\pi |D|/ \ln(|D|/D')
\end{align}
which quickly become comparable to $|D|$ as $D'$ becomes non-zero.

With these general understandings in mind, we now come back to our constructed, solvable systems. For simplicity, we will restrict to the case of $\mathsf{N}_\text{sv} =1$, but the generalization to other $\mathsf{N}_\text{sv}$ is straightforward.

From Eq.~\ref{eq: ODLRO} it can be directly inferred that the ground states $|\Psi_\text{SC}(\theta)\rangle$ have pseudospin order in its $xy$-plane with amplitude $\bar{J} = |\langle \Vec{J}\rangle|/2 = |\langle \cos\theta \hat{J}_x + \sin\theta \hat{J}_y \rangle|/2  = \bar{\varepsilon} \mathsf{V}/2$. On top of these ground states, the lowest branch of the Cooper pair or neutral spectrum, $E^{(0,\pm 2)}(\bm{P})$, must hit zero at $\bm{P} = \bm{Q}$. This is because $\hat{O}$ is a pair excitation operator with zero energy and momentum $\bm{Q}$. Then, the dispersion $E^{(0, \pm 2)}(\bm{P})$ near $\bm{Q}$ can be Taylor expanded:
\begin{align}\label{eq: pair dispersion}
    E^{(0,\pm 2)}(\bm{P} = \bm{Q} + \delta \bm{Q} ) = \frac{1}{2}\delta \bm{Q}_{i} M^{-1}_{ij} \delta \bm{Q}_j + \mathcal{O}(|\delta \bm{Q}|^3)
\end{align}
where $M^{-1}_{ij}$ is the inverse of the mass tensor of the cooper pairs. These modes are nothing but the NG mode corresponding to the spontaneous pseudospin symmetry breaking. Thus, comparing the dispersions in Eq.~\ref{eq: pair dispersion} and Eq.~\ref{eq: NG dispersion}, one can conclude that
\begin{align}
   D_{ij} = \bar{\varepsilon}M^{-1}_{ij}
\end{align}

We also note that there are multiple factors that can lead to a deviation from the idealized setup we assumed and thus a weak explicit breaking of the pseudospin $SU(2)$ symmetry, including (but not limited to) the finite band gap, the imperfect flatness of the bands or electronic structure, and the deviation of interaction from the obtained form. Depending on the nature of those perturbations, the superconducting state may remain phase-coherent up to a temperature comparable to $|D|$ in $d\ge 2$.

\section{Examples of electronic structures satisfying QGN}

\label{sec: examples}

In this section, we give several examples of electronic structures satisfying certain QGN.

\subsection{Time reversal symmetric and $S^z$ conserving systems}\label{sec: time reversal}

To give the simplest example of QGN, we consider any system with spin $S^z$ conservation and time-reversal symmetry. Then we separate out spin index from the band and orbital indices and recognize the implications of the symmetries:
\begin{align}
    U_{\alpha s,n s' } (\bm{k})
    &= \delta_{s s'} U_{\alpha n}^s(\bm{k})  \\
    U_{\alpha n}^s(\bm{k}) &= \left[U_{\alpha n}^{\bar{s}}(-\bm{k})\right]^\star
\end{align}
Then  $N_{\mu s_1, \nu s_2} \equiv \sigma^y_{s_1 s_2} \mathbbm{1}_{\mu\nu}$ satisfies QGN at $\bm{Q}=0$ in the p-p channel, since
\begin{align}
    F^{\text{p-p},\bm{Q}=0}_{n s_1, m s_2}(\bm{k}) &= U_{\mu s_1,n s_1' } (\bm{k}) \sigma^y_{s_1' s_2'} \mathbbm{1}_{\mu\nu} U_{\nu s_2,m s_2' } (-\bm{k}) \\
    & = U^{s_1}_{\mu n} (\bm{k})\sigma^y_{s_1 s_2} \mathbbm{1}_{\mu\nu} U^{s_2}_{\nu m } (-\bm{k}) \\
    & = U^{s_1}_{\mu n} (\bm{k})\sigma^y_{s_1 s_2} \mathbbm{1}_{\mu\nu} [U^{\bar{s}_2}_{\nu m } (\bm{k})]^\star \\
    & = \mathrm{i}\sigma^z_{\sigma_1\sigma_2} \mathbbm{1}_{n m}
\end{align}
is block diagonal (equivalent definition of QGN in property 3 of Sec.~\ref{sec: general properties}). The consequence is that, for this QGN, solvable Hamiltonians can be constructed with $B^{(I)}_{\mu\sigma_1,\nu\sigma_2} = \sigma^{\mathsf{i}=0,x,y,z}_{\sigma_1\sigma_2} \tilde{B}^{(I)}_{\mu\nu}$ where $\tilde{B}^{(I)}$  are {\it arbitrary} symmetric (for $\mathsf{i} = x,y,z$) or antisymmetric (for $\mathsf{i} = 0$) hermitian matrices for the orbital indices. This includes a huge class of possibilities, e.g. the attractive Hubbard interactions on each site, which were pointed out in Ref.~\cite{herzog2022many,PhysRevB.94.245149}.

We note that, given those symmetries, the conventional Fermi surface nesting is also satisfied for the dispersive cases. In fact, both cases are the manifestation of the Wannier degeneracy.

\subsection{Chiral systems with flat zero-mode bands}

When the electronic system is on a bipartite lattice, the single-particle Hamiltonian on each $\bm{k}$ can be written as
\begin{align}
    t(\bm{k}) = \begin{bmatrix}
        0 & T(\bm{k}) \\
        T^\dagger(\bm{k}) & 0
    \end{bmatrix}
\end{align}
where $T$ is a $\mathsf{N}_A \times \mathsf{N}_B$ matrix characterizing the hopping between the two sublattices $A$ and $B$ consisting of $\mathsf{N}_A$ and $\mathsf{N}_B$ orbitals, respectively. Without loss of generality, we assume $\mathsf{N}_A \le \mathsf{N}_B$. Then $T(\bm{k})$ can be singular value decomposed into
\begin{align}
    T(\bm{k}) = V_A^\dagger(\bm{k})D(\bm{k}) V_B(\bm{k})
\end{align}
where $D(\bm{k})$ is a $\mathsf{N}_B \times \mathsf{N}_B$ diagonal matrix with non-negative entries and rank $\mathcal{R} \le \mathsf{N}_A$, i.e. (assuming $0 \le \la_{1}(\bm{k}) \le \la_{2} (\bm{k})\le \la_{{\mathsf{N}_B}}(\bm{k})$)
\begin{align}
  D(\bm{k}) = \left[\begin{array}{ c | c c  c }
   \mathbf{0}_{B-A} & & & \\
      \hline
   & \la_{1}(\bm{k}) & &  \\
    & & \ddots &  \\
    & & &  \la_{{\mathsf{N}_B}}(\bm{k})
  \end{array}\right]
  \equiv \left[\begin{array}{c | c }
   \mathbf{0}_{B-A} & \\
    \hline
    & \tilde{D}(\bm{k})
  \end{array}\right],
\end{align}
and $V_A(\bm{k})$ and $V_B(\bm{k})$ are unitary matrices of sizes $\mathsf{N}_A \times \mathsf{N}_A$ and $\mathsf{N}_B \times \mathsf{N}_B$.

Then, $t(\bm{k})$ can be diagonalized into $t(\bm{k}) = U^\dagger(\bm{k}) \epsilon(\bm{k}) U(\bm{k})$ with the spectrum
\begin{align}
    \epsilon(\bm{k}) = \left[\begin{array}{ c | c }
    -\tilde{D}(\bm{k}) &  \\
    \hline
     & D(\bm{k})
  \end{array}\right]
\end{align}
and eigenbasis defined by
\begin{align}
    U(\bm{k}) = \frac{1}{\sqrt{2}} \left[\begin{array}{ c | c }
    -V_A(\bm{k}) & \tilde{V}_B(\bm{k}) \\
    \hline
    \tilde{V}_A(\bm{k}) & V_B(\bm{k})
  \end{array}\right]
\end{align}
where $\tilde{V}_A$ is a $\mathsf{N}_B\times \mathsf{N}_A$ matrix expanded from $V_A$ by adding
$\mathsf{N}_B-\mathsf{N}_A$ empty rows to the top, and $\tilde{V}_B$ is a $\mathsf{N}_A\times \mathsf{N}_B$ matrix trimmed from $V_B$ by deleting the first $\mathsf{N}_B-\mathsf{N}_A$ rows.

\begin{align}\label{eq: chiral nesting matrix}
    N_{\mu s; \nu s'} \equiv \mathbbm{1}_{\mu\nu} \sigma^y_{ss'}
\end{align}

Then we note that,
\begin{align}\label{eq: chiral nesting matrix}
    N \equiv \left[\begin{array}{ c | c }
    -\mathbbm{1}_A &  \\
    \hline
     & \mathbbm{1}_B
  \end{array}\right]
\end{align}
is a nesting matrix for the p-h channel at $\bm{Q}=0$ in the sense that the form factor matrix defined by
\begin{align}
    F(\bm{k}) \equiv U^\dagger(\bm{k}) N U(\bm{k}) =  \left[\begin{array}{ c | c | c }
     &  &  \mathbbm{1}_A \\
    \hline
      & \frac{\mathbbm{1}_{A-B}}{2} & \\
     \hline
     \mathbbm{1}_A & &
  \end{array}\right]
\end{align}
keep {\it all the zero modes} disconnected from other finite energy modes (which is easier to see if one re-indexes the bands and makes $n=1\dots \mathsf{N}_\text{flat}$ the flat bands). Therefore, according to the alternative defintion of QGN (property 3 in Sec.~\ref{sec: general properties}), when there are {\it flat bands at zero energy} in chiral systems, the system satisfies QGN in the p-h channel. This QGN is particularly useful when there are additional internal symmetries, e.g. spin and/or valley. In this case, the nesting matrix can be a tensor product of Eq.~\ref{eq: chiral nesting matrix} defined on purely the orbital basis and an {\it arbitrary} matrix on the internal symmetry basis. The condensation of the corresponding order parameter then leads to the symmetry breaking of the internal symmetry. This is exactly the case extensively discussed in the studies on magic-angle twisted bilayer graphene in its flat and chiral-flat limits~\cite{PhysRevX.10.031034, PhysRevLett.122.246401, PhysRevB.103.205414, PhysRevB.103.205415}.

It is of particular interest to identify or even predict what form of perturbations and interactions could lead to certain specific symmetry-breaking patterns in flat band materials, e.g. picking inter-valley coherence, valley polarization, etc, out of the symmetry-related manifold. We expect QGN to provide important information in this aspect since it may suggest interactions that are only favorable for one type of symmetry-breaking pattern but not the others. We leave the investigations to specific materials for further studies.

\subsection{An engineered model for PDW}

\label{sec: engineered model}

To show that QGN can be applicable to other systems, here we explicitly construct a toy, minimal system where QGN is satisfied in multiple channels. Specifically, we consider a two-band, spinful model on the square lattice, whose band structure is given by
\begin{align}
    \hat{H}_0 = \sum_{\bm{k},\sigma, \alpha\beta} \left[-2t\left(\cos k_x +\cos k_y\right) \tau^z + M\tau^x\right]_{\alpha \beta} \hat{c}^\dagger_{\bm{k}\alpha\sigma}  \hat{c}_{\bm{k}\beta\sigma}
\end{align}
where we have restored the spin index $\sigma$, and $\tau^{i=x,y,z}$ are the Pauli matrices acting on the orbital indices. In real space, this model can be viewed as a bilayer model with an interlayer hopping $M$ and an intralayer nearest neighbor hopping $t$ with an opposite sign but the same magnitude on two layers. It is easy to tune the system to a flat band limit, where the band gap is much greater than the bandwidths, by taking $M\gg |t|$.

This system is time-reversal invariant and spin rotation invariant so that the QGN for uniform pairing discussed in the previous subsection is automatically satisfied. However, there are actually more QGN satisfied in this system. To recognize these channels, we note that in this band structure, $U_{\alpha s,n s}(\bm{k}) = \delta_{ss'} U_{\alpha n } (\bm{k})$ and for $\bm{Q} = (\pi,\pi)$,
\begin{align}
    \tau^x_{\alpha \beta}U_{\beta n}(\bm{k}) = U^\dagger_{n\alpha}(\bm{k}+\bm{Q}) = U_{\alpha n} (\bm{Q}-\bm{k}).
\end{align}
This suggests QGN in p-p channel with nesting matrix $N^{\text{p-p},\bm{Q}}_{\alpha s, \beta s'} = \tau^x_{\alpha\beta} \sigma^y_{s s'}$ (and thus form factors $F^{\text{p-p},\bm{Q}}_{n s,m s'}(\bm{k}) = \delta_{nm}\sigma^y_{s s'}$), and a series of QGNs in p-h channel with nesting matrices $N^{\text{p-h},\bm{Q}}_{\alpha s, \beta s'} = \tau^x_{\alpha\beta} \sigma^{\mathsf{i} = 0,x,y,z}_{s s'}$ (and thus form factors $F^{\text{p-h},\bm{Q}}_{ns,ms'}(\bm{k}) = \delta_{nm}\sigma^{\mathsf{i}}_{s s'}$). These channels of perfect QGN suggest that, besides the uniform pairing order, the system is simultaneously prone to a singlet pair density wave, a charge density wave, and a spin density wave, all at $\bm{Q} = (\pi,\pi)$.

Applying the construction scheme in Sec.~\ref{sec: interactions}, we find that different possible $B^{(I)}_{\mu\sigma,\nu\sigma'}$ matrices (factorized by $\tau^{\mathsf{i}=0,x,y,z}_{\mu\nu}\sigma^{\mathsf{j}=0,x,y,z}_{s s'}$) can serve as building blocks of ideal interactions for different orders, as listed in Table.~\ref{tab: B}

\begin{table}[h]
    \centering
    \begin{tabular}{|c|c|c|c|c|}
       \hline
        & $\sigma^0$  &  $\sigma^x$ &  $\sigma^y$ &  $\sigma^z$\\
        \hline
        $\tau^0$ & CDW, SDW & uSC, PDW, CDW, xSDW & uSC, PDW, CDW, ySDW & uSC, PDW, CDW, zSDW \\
        \hline
         $\tau^x$ & PDW, SDW & uSC, xSDW & uSC,  ySDW & uSC, zSDW \\
         \hline
          $\tau^y$ & uSC, PDW, CDW & CDW, ySDW, zSDW & CDW, xSDW, zSDW & CDW, xSDW, ySDW \\
          \hline
          $\tau^z$ & PDW & uSC, ySDW, zSDW & uSC,  xSDW, zSDW & uSC, xSDW, ySDW \\
          \hline
    \end{tabular}
    \caption{The corresponding orders of the different possible $B^{(I=\mathsf{i}\mathsf{j})}_{\mu\sigma,\nu\sigma'}=\tau^{\mathsf{i}=0,x,y,z}_{\mu\nu}\sigma^{\mathsf{j}=0,x,y,z}_{s s'}$ matrices, which can be used to construct the ideal interactions. uSC stands for uniform superconductivity with nesting matrix $N^{\text{p-p},\bm{Q}=0}_{\mu s, \nu s'} = \tau^0_{\mu\nu}\sigma^y_{ss'}$. CDW stands for charge density wave with nesting matrix $N^{\text{p-h},\bm{Q}=(\pi,\pi)}_{\mu s, \nu s'} = \tau^x_{\mu\nu}\sigma^0_{ss'}$. PDW stands for pair density wave with nesting matrix $N^{\text{p-p},\bm{Q}=(\pi,\pi)}_{\mu s, \nu s'} = \tau^x_{\mu\nu}\sigma^y_{ss'}$. $x,y,z$-SDW stand for spin density wave polarized in the corresponding direction with nesting matrix $N^{\text{p-h},\bm{Q}=(\pi,\pi)}_{\mu s, \nu s'} = \tau^x_{\mu\nu}\sigma^{x,y,z}_{ss'}$.  }
    \label{tab: B}
\end{table}

Aiming to construct a simple ideal model for PDW in this quantum geometry, we consider the following local interaction made of four of the six possible $B^{(I)}$ matrices in Table.~\ref{tab: B}
\begin{align}
    \hat{H}_{\text{int}} &= V \sum_{\bm{R} } 3 (\hat{S}^{(I=z0)}_{\bm{R}})^2 + (\hat{S}^{(I=0x)}_{\bm{R}})^2 + (\hat{S}^{(I=0y)}_{\bm{R}})^2 + (\hat{S}^{(I=0z)}_{\bm{R}})^2 \\
    &= 6 V \sum_{\bm{R}} \hat{n}_{\bm{R}} + V \sum_{\bm{R}} (- 3\hat{n}_{\bm{R},1} \hat{n}_{\bm{R},2} + 4 \hat{\bm{S}}_{\bm{R},1} \cdot \hat{\bm{S}}_{\bm{R},2})
\end{align}
showing a combination of anti-ferromagnetism and inter-orbital attraction. This form of interaction is particularly simple and is spin invariant.

We now compute the charge $\pm 1$ (quasi-particle) and charge $2$ excitation (Cooper pair) excitations for this model. Note that the one-body terms generated by projection are all trivial, and thus the groundstate and excitations are exact in the strong coupling limit $V \gg t^2/M$.

The charge $\pm1$ excitations are also fully gapped and flat. The calculation is very similar to \Eq{eq:onebodytoy}. From \Eq{eq:charge1app}, the effective Hamiltonian of the charge $+1$ excitations is
\bea
\Gamma^{\text{p.},\mbf{P}}_{mn} &= \left[ U^\dag({\bm{k}}) \lp \frac{1}{\mathsf{V}} \sum_{\mbf{q}} 3 \tau^z P({\bm{k}}-\mbf{q}) \tau^z + 3 P({\bm{k}}-\mbf{q})  \rp U({\bm{k}}) \right]_{mn} \\
&= 3 V \left[U^\dag({\bm{k}})\frac{1}{\mathsf{V}} \sum_{\mbf{q}}  \lp P(\mbf{q}) + \tau^z P(\mbf{q}) \tau^z  \rp U({\bm{k}}) \right]_{mn}\\
&= 3 V [\sigma_0]_{mn}
\eea
where $m,n \in \{\u,\d\}$ since there is 1 flat band per spin in the projected Hilbert space. The flatness of the charge 1 bands leads to the solvability of the Cooper pair spectrum. We will focus on the spin 0 sector (see \Eq{eq:spinzero}) where
\bea
\Gamma^{\text{p.-p.} \u \d,\mathbf{p}}_{{\bm{k}},{\bm{k}}'} &= 2 (3 V) \delta_{{\bm{k}'} {\bm{k}}} + \frac{2}{\mathsf{V}} \sum_I V_I \big( S^{(I)}_{\u,\u}( \bm{k}'^-,\bm{k}^-)  S^{(J)}_{\d, \d}( \bm{k}'^+,\bm{k}^+) - S^{(I)}_{\d,\u}(-\bm{k}'^-,\bm{k}^-)  S^{(J)}_{\u, \d}(-\bm{k}'^+,\bm{k}^+) \big) \\
\eea
using the fact that the charge $1$ excitation energies are flat and equal to $3V$.

To diagonalize the Hamiltonian, it is sufficient to diagonalize the scattering matrix, which we are able to solve in some generality. We need the following identities
\bea
\frac{2}{\mathsf{V}} \sum_I V_I S_{\u \u}^{(I)}({\bm{k}}'^-,{\bm{k}}^-) S_{\d \d}^{(I)}({\bm{k}}'^+,{\bm{k}}^+) &= \frac{2}{\mathsf{V}} \sum_I V_I U_\u^\dag({\bm{k}}'^-) B^{(I)} U_\u({\bm{k}}^-) U_\d^\dag({\bm{k}}'^+) B^{(I)} U_\d({\bm{k}}^+) \\
&= \frac{2}{\mathsf{V}} \sum_{I,\al\be,\al' \be'} V_I U_{\al'}^*({\bm{k}}'^-) B_{\al' \u,\be' \u}^{(I)} U_{\be'}({\bm{k}}^-) U_{\al}^*({\bm{k}}'^+) B_{\al \d, \be \d}^{(I)} U_{\be}({\bm{k}}^+) \\
&= \frac{2}{\mathsf{V}} \sum_{I,\al\be,\al' \be'}  U_{\be'}({\bm{k}}^-)  U_{\be}({\bm{k}}^+) V_I  B_{\al' \u,\be' \u}^{(I)}  B_{\al \d, \be \d}^{(I)} U_{\al'}^*({\bm{k}}'^-) U_{\al}^*({\bm{k}}'^+) \\
\frac{2}{\mathsf{V}} \sum_I - V_I S_{\d \u}^{(I)}(-{\bm{k}}'^-,{\bm{k}}^-) S_{\u \d}^{(I)}(-{\bm{k}}'^+,{\bm{k}}^+) &= \frac{2}{\mathsf{V}} \sum_{I,\al\be,\al' \be'} - U_{\be'}({\bm{k}}^-)  U_{\be}({\bm{k}}^+) V_I  B_{\al' \d,\be' \u}^{(I)}  B_{\al \u, \be \d}^{(I)} U_{\al'}^*({\bm{k}}'^-) U_{\al}^*({\bm{k}}'^+)
\eea
where we dropped the spin index of the eigenvector using $SU(2)$ symmetry, and took $U({\bm{k}}) = U(-{\bm{k}})$ by spin-less time-reversal. We now write the scattering matrix in the outer product notation
\bea
\label{eq:Bethesalpeterapp}
\Gamma^{\text{p.-p.} \u \d,\mathbf{p}} &= 2\Gamma^{\text{p.}} \mathbbm{1} - \frac{2}{\mathsf{V}} \, \mathcal{U}(\mbf{p}) \mathcal{B} \mathcal{U}^\dag(\mbf{p})
\eea
where $\Gamma^{\text{p.}}$ is the flat charge 1 excitation energy, and we have defined the $\mathsf{V} \times N_{orb}^2$ matrix (recall ${\bm{k}}^\pm = \mbf{p}/2 \pm {\bm{k}}$)
\bea
\null [\mathcal{U}(\mbf{p})]_{{\bm{k}},\be \be'} &= U_{\be}({\bm{k}}^+) U_{\be'}({\bm{k}}^-)
\eea
and the $N_{orb}^2\times N_{orb}^2$ interaction matrix
\bea
\mathcal{B}_{\be \be',\al \al'} = -\sum_I V_I (B_{\al' \u,\be' \u}^{(I)}  B_{\al \d, \be \d}^{(I)} -  B_{\al' \d,\be' \u}^{(I)}  B_{\al \u, \be \d}^{(I)}) \ .
\eea
Here $ N_{orb}$ is the number of orbitals per spin. Since the rank of $\mathcal{U}(\mbf{p})$ and $\mathcal{B}$ can be no greater than $N_{orb}^2$, the scattering matrix, although thermodynamically large, has finite rank at most $N_{orb}^2$. Its kernel contains all eigenvectors annihilated by $\mathcal{U}^\dag(\mbf{p})$, which hence have total energy $6V$ (twice the charge $+1$ energy) and form the flat particle-particle continuum. To derive the nontrivial eigenvalues, we will assume that $\mathcal{B}$ is positive semi-definite. Then we can write $\mathcal{B} = \mathcal{A} \mathcal{A}^\dag$ via a singular-value decomposition, leading to
\bea
-\frac{2}{\mathsf{V}} \mathcal{U}(\mbf{p}) \mathcal{B} \mathcal{U}^\dag(\mbf{p}) &=  \frac{2}{\mathsf{V}} \mathcal{U}(\mbf{p}) \mathcal{A} \mathcal{A}^\dag \mathcal{U}^\dag(\mbf{p})
\eea
showing that the non-zero eigenvalues are those of
\bea
h(\mbf{p}) &= \frac{2}{\mathsf{V}}  \mathcal{A}^\dag \mathcal{U}^\dag(\mbf{p})\mathcal{U}(\mbf{p}) \mathcal{A} \\
\eea
whose eigenvalues are the binding energies of the Cooper pair branches below the particle-particle continuum. Written out in components, we find
\bea
\null [\frac{1}{\mathsf{V}} \mathcal{U}^\dag(\mbf{p})\mathcal{U}(\mbf{p})]_{\al \al',\be \be'} &= \frac{1}{\mathsf{V}} \sum_{\bm{k}} P^*_{\al \be}(\mbf{p}/2 + {\bm{k}}) P^*_{\al' \be'}(\mbf{p}/2 - {\bm{k}})
\eea
generalizing the form previously found in Ref. \cite{herzog2022many}. To leading order in $t/M$, we can compute the ${\bm{k}}$ integrals analytically to find
\bea
h(\mbf{p}) &= 3 V\left(
\begin{array}{cccc}
 0 &  &  &  \\
  &  1-\frac{2 t^2}{M^2}(\cos p_x + \cos p_y) &  1-\frac{4 t^2}{M^2} &  \\
  & 1-\frac{4 t^2}{M^2} &  1-\frac{2 t^2}{M^2}(\cos p_x + \cos p_y) &  \\
  &  &  & 0 \\
\end{array}
\right) \ .
\eea
Using the fact that $2 \Gamma^{\text{p}.} = 6V$ is twice the charge $+1$ excitation spectrum, \Eq{eq:Bethesalpeterapp} gives the many-body energies
\bea
E^{\text{p-p}}(\mbf{p}) &= \left\{ \frac{6 Vt^2}{M^2}(2 + \cos p_x + \cos p_y), \quad 6V + \frac{6 Vt^2}{M^2}(-2 + \cos p_x + \cos p_y) \right\}
\eea
neglecting the additional two modes with zero binding energy. The condensing PDW mode is the focus of this work, which behaves like
\bea
 \frac{6 Vt^2}{M^2}(2 + \cos p_x + \cos p_y) &= 0 + \frac{6 V t^2}{M^2} \frac{\delta \bf{p}^2}{2} + \dots
\eea
for $\mbf{p} = (\pi,\pi) + \delta \bf{p}$. Thus we determine the inverse mass of the Cooper pair to be $\frac{6 V t^2}{M^2}$, which is proportional to the interaction strength and the integrated Fubini-Study metric
\bea
g &= \int \frac{1}{2} \Tr (\del_i P)^2 \frac{d^2k}{(2\pi)^2} = \frac{t^2}{M^2} + O(t^4/M^4) \ .
\eea
In particular, the ratio of the inverse mass (in units where the lattice constant is 1) to the quasi-particle gap is $\frac{6 V t^2}{M^2} / 3V = 2g$ is identical to the attractive Hubbard model case with $s$-wave pairing \cite{herzog2022many}.

\subsection{A class of electronic structures with QGN}

Like how a single-band model has perfect Fermi-surface nesting if $E(\bm{k}+\bm{Q}) = E(\bm{k})$, we can engineer bands with perfect QGN obeying $N U(\bm{k}) = U(\bm{k}+\bm{Q})$ where $N$ is an $\mathsf{N} \times \mathsf{N}$ {\it unitary} matrix. Our method is to realize a single-particle Hamiltonian obeying
\bea
\label{eq:hkQ}
N h(\bm{k}) N^\dag = h(\bm{k}+\bm{Q})
\eea
where in this section we require a periodic embedding $h_{\mu \nu}(\bm{k}) = h_{\mu \nu}(\bm{k}+\mbf{G})$ for all reciprocal lattice vectors $\mbf{G}$. \Eq{eq:hkQ} is a multi-band generalization of Fermi surface nesting since \Eq{eq:hkQ} guarantees the spectrum also nests according to $E(\bm{k}+\bm{Q}) = E(\bm{k})$. We will consider the flat band limit where the Fermi surface is ill-defined and quantum geometry instead drives nesting. First we note that \Eq{eq:hkQ} can be written as a real-space condition on the hoppings
\bea
N t(\bm{R}) N^\dag = e^{i \mbf{Q} \cdot \mbf{R}} t(\bm{R})
\eea
where $t(\bm{R})$, the Fourier transform of $h(\mbf{k})$, is the hopping matrix between unit cells $\mbf{R}$ apart.

Given \Eq{eq:hkQ}, it is clear that $U(\mbf{k}+\mbf{Q})$ and $N U(\mbf{k})$ are eigenvectors of $h(\bm{k}+\bm{Q})$, and thus the projector onto the set of (quasi-)flat  bands obeys
\bea
\label{eq:PkQperiodicity}
N P(\bm{k}) N^\dag = P(\bm{k}+\bm{Q}) \ .
\eea
This condition is gauge-invariant, although it is restricted to the periodic embedding.

We now show that $N$ yields a nesting matrix as defined by the criteria in the Main Text, $\Pi_{\mu' \nu',\mu \nu} N_{\mu \nu} = 0$ for the p.-p. and p.-h. cases. To show this, we shift $\mbf{k}$ in the sum to yield (considering the p.-h. case) first
\bea
    \Pi^{\text{p-h},\bm{Q}}_{\mu'\nu';\mu\nu} &= \frac{1}{\mathsf{V}}\sum_{\bm{k}}P_{\mu'\mu}(\bm{k}+\bm{Q}) Q_{\nu \nu' } (\bm{k}) +  (P\leftrightarrow Q) \\
    &= \frac{1}{\mathsf{V}}\sum_{\bm{k}} [NP(\bm{k})N^\dag]_{\mu'\mu} Q_{\nu \nu' } (\bm{k}) +  (P\leftrightarrow Q)  \\
\eea
so that we find
\bea
\sum_{\mu \nu} \Pi_{\mu' \nu',\mu \nu} N_{\mu \nu} &= \frac{1}{\mathsf{V}}\sum_{\bm{k}} [NP(\bm{k})N^\dag]_{\mu'\mu} N_{\mu \nu} Q_{\nu \nu' } (\bm{k}) +  (P\leftrightarrow Q) \\
&= \sum_{\nu} \Pi_{\mu' \nu',\mu \nu} N_{\mu \nu} \\
&= \frac{1}{\mathsf{V}}\sum_{\bm{k}} [NP(\bm{k})]_{\mu'\nu}  Q_{\nu \nu' } (\bm{k}) +  (P\leftrightarrow Q) \\
&= 0
\eea
since $P(\mbf{k}) Q(\mbf{k}) = 0$. We now make the observation that the solution to the QGN criterion exhibited here manifests as a conventional ``nesting instability" of the Fubini-Study metric. This is because \Eq{eq:PkQperiodicity} implies
\bea
g(\mbf{k}+\mbf{Q}) &= \frac{1}{2} \text{Tr } \partial_i P(\mbf{k}+\mbf{Q}) \partial_i P(\mbf{k}+\mbf{Q}) \\
&= \frac{1}{2} \text{Tr } N \partial_i P(\mbf{k}) \partial_i P(\mbf{k}) N^\dag \\
&= g(\mbf{k}) \ .
\eea
If we consider $g(\mbf{k})$ as an effective dispersion $E(\mbf{k})$ on the BZ, then QGN can be visualized as a nesting of $g(\mbf{k})$.

In general, perfect QGN such that $\Pi N = 0$ implies $g(\mbf{k}+\mbf{Q}) = g(\mbf{k})$ if $N$ is unitary (up to an overall normalization, i.e. $N^\dag N \propto 1$). To see this, note that every term in the $\mbf{k}$ sum defining $\Pi$ is positive semi-definite, so $\Pi N = 0$ implies that
\bea
\label{eq:PNQ}
0 &= P(\bm{k}+\bm{Q}) N Q(\bm{k}) = \left( N^\dag P(\bm{k}+\bm{Q}) N \right) Q(\bm{k}) \\
\eea
so that $N^\dag P(\bm{k}+\bm{Q}) N$ is a Hermitian matrix in the null space of $\mbf{Q}(\mbf{k})$. If $N$ is unitary, then $N^\dag P(\bm{k}+\bm{Q}) N$ is a Hermitian projector with rank $\Tr N^\dag P(\bm{k}+\bm{Q}) N = P(\bm{k}+\bm{Q}) = \Tr P(\bm{k})$. This proves $N^\dag P(\bm{k}+\bm{Q}) N = P(\mbf{k})$ since the maximal-rank projector onto the nullspace of $\mbf{Q}(\mbf{k})$ is unique. If $N$ is not unitary, $N^\dag P(\bm{k}+\bm{Q}) N$ is not guaranteed to be a projector, and this argument does not hold.

For the p.-p. channel, we will assume there is also time-reversal symmetry $h^*(\mbf{k}) = h(-\mbf{k})$, in which case
\begin{align}
    \Pi^{\text{p-p},\bm{Q}}_{\mu'\nu';\mu\nu} &=\frac{1}{\mathsf{V}}\sum_{\bm{k}}P^\star_{\mu'\mu}(\bm{Q}+\bm{k}) Q_{\nu \nu' } (-\bm{k})+ (P\leftrightarrow Q) \\
    &=\frac{1}{\mathsf{V}}\sum_{\bm{k}}P^\star_{\mu'\mu}(\bm{Q}+\bm{k}) Q^*_{\nu \nu' } (\bm{k})+ (P\leftrightarrow Q) \\
    &=\frac{1}{\mathsf{V}}\sum_{\bm{k}}[N^* P^\star(\bm{k}) N^T]_{\mu'\mu} Q^*_{\nu \nu' } (\bm{k})+ (P\leftrightarrow Q) \\
\end{align}
and thus we find that $N^*$ is a nesting matrix obeying
\bea
\sum_{\mu \nu} \Pi^{\text{p-p},\bm{Q}}_{\mu'\nu';\mu\nu} N_{\mu \nu}^* &= \frac{1}{\mathsf{V} \mu \nu}\sum_{\bm{k}}[N^* P^*(\bm{k}) N^T]_{\mu'\mu} N^*_{\mu \nu} Q^*_{\nu \nu' } (\bm{k})+ (P\leftrightarrow Q) \\
&= \frac{1}{\mathsf{V}}\sum_{\bm{k} \nu}[N^* P^*(\bm{k})]_{\mu'\nu} Q^*_{\nu \nu' } (\bm{k})+ (P\leftrightarrow Q) \\
&= 0
\eea
which completes the proof. Note that a weaker requirement for the p.-p. case is $P(\mbf{k} + \mbf{Q}) = (N P(-\mbf{k}) N^\dag)^*$. Following the same arguments as the p.-h. case, it follows that $g(\mbf{k}+\mbf{Q}) = g(-\mbf{k})$.

\subsection{An engineered model for current density wave}

At first glance, the order parameter considered in our work seems orthogonal to any current operator when expressed on an orbital basis, thus excluding the possibility of a current (flux) density wave. However, in a multiband system with non-trivial quantum geometry, the projected order parameter may acquire overlap with the current operator and lead to the current density wave order. To see this, we construct a simple example below.

Consider a one-dimensional, two-leg (labeled by $\alpha =1,2$), spinless electronic system with electronic structure specified by:
\begin{align}
    \hat{H}_0 = \sum_{\alpha\beta, k} h\left(k\right)_{\alpha \beta} \hat{c}^\dagger_{k\alpha} \hat{c}_{k\beta} = \sum_{\alpha\beta, k} h_0 \left[\cos \theta(k) \tau^{z} + \sin \theta(k) \tau^{x}  \right]_{\alpha \beta} \hat{c}^\dagger_{k\alpha} \hat{c}_{k\beta}
\end{align}
where $\tau^{i}$ are Pauli matrices acting on the leg index, and we take a special choice
\begin{align}
    \theta(k) = k + \delta \cos(2k)
\end{align}
with $\delta$ an arbitrary real number. This electronic structure has two perfectly flat bands (labeled by $n=\pm$):
\begin{align}
    \hat{H}_0 &=\sum_{n=\pm, k} \pm h_0 \hat{\gamma}^\dagger_{k n} \hat{\gamma}_{k n} \\
    \hat{\gamma}_{kn} &= \sum_\alpha U^\dagger_{n\alpha}(k) \hat{c}_{k\alpha}  = \sum_\alpha \begin{bmatrix}
        \cos \frac{\theta(k)}{2} & \sin \frac{\theta(k)}{2} \\
        -\sin \frac{\theta(k)}{2} & \cos \frac{\theta(k)}{2}
    \end{bmatrix}_{n\alpha}\hat{c}_{k\alpha}
\end{align}

Then, when $Q=\pi$,
\begin{align}
    F^{\text{p-h},Q=\pi}_{nm}(k)  = \sum_{\alpha\beta} U^\dagger_{n\alpha}(k+Q/2) \mathrm{i} \tau^y_{\alpha\beta} U_{\beta m}(k-Q/2) = \delta_{nm}
\end{align}
Therefore, according to the alternative definition of QGN, any one of the two bands satisfies QGN in p-h channel with $Q=\pi$ with nesting matrix:
\begin{align}
    N^{\text{p-h},Q=\pi} = \mathrm{i} \tau^y.
\end{align}
Then, the ground states at $1/4$ or $3/4$ filling for any constructed ideal interacting models will have a certain density wave order with momentum $\pi$, the nature of which is to be determined. To be concrete, we focus on the $1/4$ filling where one of the ground states can be obtained by occupying the modes $(\gamma_{k,n=-} + \gamma_{k+\pi,n=-})/\sqrt{2}$ on all $k$. With this many-body wavefunction, we can compute the expectation value of the current (going from leg $\alpha$ to $\beta$) density operator at momentum $\pi$:
\begin{align}
    \hat{J}_{\alpha\beta} (q) \equiv & \sum_{k} \left[\partial_k h(k)\right]_{\alpha\beta} \hat{c}^\dagger_{k+q/2,\alpha} \hat{c}_{k-q/2,\beta}\\
    \langle \hat{J}_{11} (q=\pi) \rangle =& \frac{1}{2}\int \frac{\mathrm{d} k}{2\pi} \sin \frac{\theta(k+\pi/2)}{2} \cdot \partial_k  \cos \theta(k)\cdot \sin \frac{\theta(k-\pi/2)}{2} =-\frac{\delta}{8} = - \langle \hat{J}_{22} (q=\pi) \rangle \\
     \langle \hat{J}_{12} (q=\pi) \rangle =& \frac{1}{2} \int \frac{\mathrm{d} k}{2\pi}  -\sin \frac{\theta(k+\pi/2)}{2} \cdot \partial_k  \sin \theta(k)\cdot \cos \frac{\theta(k-\pi/2)}{2} =\frac{\delta}{8} = - \langle \hat{J}_{21} (q=\pi) \rangle
\end{align}
Meanwhile, the number density operators (on leg $\alpha$) have vanishing expectation values on this momentum:
\begin{align}
    \hat{\rho}_{\alpha} (q) \equiv & \sum_{k}  \hat{c}^\dagger_{k+q/2,\alpha} \hat{c}_{k-q/2,\alpha}\\
    \langle \hat{\rho}_{1} (q=\pi) \rangle =& \frac{1}{2} \int \frac{\mathrm{d} k}{2\pi} \sin \frac{\theta(k+\pi/2)}{2} \cdot \sin \frac{\theta(k-\pi/2)}{2} =0 = \langle \hat{\rho}_{2} (q=\pi) \rangle
\end{align}
Therefore, the ground state is purely a current density wave. The corresponding ideal interaction can be constructed with the general scheme in Sec.~\ref{sec: interactions}. In this case, the viable $B$ matrices are $\tau_y$ or $\tau_0=\mathbbm{1}$. Therefore, the ideal interaction can be as simple as a density repulsion
\begin{align}
    \hat{H}_{\text{int}} = V \sum_{\bm{R}} \left(\hat{n}_{\bm{R}1}+\hat{n}_{\bm{R}2} - \frac{1}{4}\right)^2.
\end{align}

\section{A Generalization of the Construction Scheme of Solvable Models}\label{sec: generalization}

The construction scheme for solvable models so far encompasses the known solvable cases of flat bands. However, it is not exhaustive for systems satisfying perfect QGN. We now show that relaxing the Hermiticity constraint of the spin operators in \Eq{eq: interaction} (while preserving Hermiticity of the Hamiltonian) enables us to obtain nontrivial pairings in flat-band superconductors. This generalizes the construction presented thus far, where nontrivial pairing symmetry requires additional global symmetries.

Below we will show a minimal example of this extension in the superconducting case by constructing a 1D $p$-wave eta pairing groundstate.

In this section, we construct a family of models with solvable eta-pairing ground states with nonzero angular momentum ($p$-wave, $d$-wave, and $f$-wave pairings). Our construction generalizes the $s$-wave groundstates obtained in Ref. \cite{herzog2022many} by allowing combined spin-hopping operators (denoted $S_\mbf{R}$ below) that are not Hermitian. This will allow nesting matrices with ${\bm{k}}$-dependence that realize nontrivial pairing symmetries. A key requirement will be that $S_\mbf{R}$ is strictly local (it has finite support), which we can satisfy in strictly local generalized bipartite flat band models that have been exhasutively constructed in Ref. \cite{2022NatPh..18..185C}.

We will consider a system $S_z$ spin U(1) symmetry, spin-ful time-reversal symmetry $\mathcal{T}$, and a single-particle Hamiltonian $H_0$ with a set of exactly flat bands. For convenience, we will assume all electron orbitals on which the flat band is supported are located at the origin of the unit cell, $\mbf{r}_\al = 0$, as can be easily implemented in the construction of Ref. \cite{2022NatPh..18..185C}.

We now define a generalized spin/hopping operator
\bea
\label{eq:Srnonherm}
S_{\mbf{R}} &= \sum_{\mbf{R}', \al \be,\sigma} s^z_\sigma \tau^\sigma_{\be \al}(\mbf{R}'-\mbf{R}) c^\dag_{\mbf{R}',\be,\sigma} c_{\mbf{R},\al,\sigma} \\
\eea
where $s^z_\sigma = \pm$ for spin $\u/\d$ and $\tau^\sigma_{\be \al}(\mbf{R}'-\mbf{R})$ is a finite range hopping whose form is determined by a solvability criterion we will derive momentarily. (We use the notation $\tau$ instead of $S$ to avoid confusion with the $s^z_\sigma$ spin.) From $S_{\mbf{R}}$ we obtain the positive semi-definite interaction
\bea
H_{int} &= U \sum_{\mbf{R}} S_\mbf{R}^\dag S_\mbf{R}
\eea
although it is possible to generalize our results to longer-ranged positive-semi definite interactions. For $S_{\mbf{R}}$ to be time-reversal symmetric, i.e. $\mathcal{T} S_{\mbf{R},\al} \mathcal{T}^{-1} = - S_{\mbf{R},\al}$ and $\mathcal{T} H_{int} \mathcal{T}^{-1}$, we take $\tau^\sigma_{\be \al}(\mbf{R}'-\mbf{R}) = \tau^{-\sigma}_{\be \al}(\mbf{R}'-\mbf{R})^*$. It will be useful to rewrite this Hamiltonian in momentum space as
\bea
H_{int} = U \sum_{\mbf{q}} S_\mbf{q}^\dag S_\mbf{q}, \qquad S_{\mbf{q}} &= \frac{1}{\sqrt{\mathsf{V}}} \sum_\mbf{R} e^{- i \mbf{q} \cdot \mbf{R}} S_{\mbf{R}} \ .
\eea
We now go to the band basis via the following Fourier transformations
\bea
c^\dag_{{\bm{k}}\al} &= \frac{1}{\sqrt{\mathsf{V}}} \sum_\mbf{R} e^{-i {\bm{k}} \cdot \mbf{R} } c^\dag_{\mbf{R},\al}, \qquad \tau^\sigma_{\be \al}(\mbf{R}'-\mbf{R}) = \frac{1}{\mathsf{V}} \sum_\mbf{q} e^{-i \mbf{q} \cdot (\mbf{R}'-\mbf{R})} \tau^\sigma_{\be \al}(\mbf{q})  \\
\eea
and $\gamma^\dag_{{\bm{k}},n,\sigma} = \sum_\al c^\dag_{{\bm{k}},\al,\sigma} U^\sigma_{\al,n}({\bm{k}})$ in terms of the single-particle eigenvectors $U^\sigma({\bm{k}})$ of $H_0$. We obtain
\bea
S_{\mbf{R}} &= \frac{1}{\mathsf{V}^2} \sum_{\mbf{R}',\mbf{q}{\bm{k}}{\bm{k}}', \al \be,\sigma}s^z_\sigma e^{i ({\bm{k}}'-\mbf{q}) \cdot \mbf{R}'-i ({\bm{k}}-\mbf{q}) \cdot \mbf{R}}  \tau^\sigma_{\be \al}(\mbf{q}) c^\dag_{{\bm{k}}',\be,\sigma} c_{{\bm{k}},\al,\sigma} = \frac{1}{\mathsf{V}} \sum_{{\bm{k}}{\bm{k}}', \al \be,\sigma}s^z_\sigma e^{-i ({\bm{k}}-{\bm{k}}') \cdot \mbf{R} }  \tau^\sigma_{\be \al}({\bm{k}}') c^\dag_{{\bm{k}}',\be,\sigma} c_{{\bm{k}},\al,\sigma} \\
\eea
leading to the momentum space expression
\bea
\label{eq:Sqband}
S_{\mbf{q}}&= \frac{1}{\sqrt{\mathsf{V}}} \sum_{{\bm{k}}, \al \be,\sigma}s^z_\sigma \tau^\sigma_{\be \al}({\bm{k}}+\mbf{q}) c^\dag_{{\bm{k}}+\mbf{q},\be,\sigma} c_{{\bm{k}},\al,\sigma} = \frac{1}{\sqrt{\mathsf{V}}} \sum_{{\bm{k}},\al \be,\sigma}s^z_\sigma \tau^\sigma_{\be \al}({\bm{k}}) c^\dag_{{\bm{k}},\be,\sigma} c_{{\bm{k}}-\mbf{q},\al,\sigma} \\
&= \frac{1}{\sqrt{\mathsf{V}}} \sum_{{\bm{k}}, mn,\sigma}s^z_\sigma [U^{\sigma\dag}({\bm{k}})  \tau^\sigma({\bm{k}})U^\sigma({\bm{k}}-\mbf{q})]_{nm} \gamma^\dag_{{\bm{k}},n,\sigma}  \gamma_{{\bm{k}}-\mbf{q},m,\sigma} \ . \\
\eea
We define a projected interaction
\bea
H_{int, flat} &= U \sum_{\mbf{q}} \bar{S}_\mbf{q}^\dag \bar{S}_\mbf{q}
\eea
where $\bar{S}_\mbf{q}$ is obtained restricting $m,n \in 1,\dots, \mathsf{N}_\text{flat}$ in \Eq{eq:Sqband} to sum only over the flat bands. There is also a one-body Hartree-Fock term given by
\bea
\label{eq:onebodylieb}
H_{1} &= U \sum_{{\bm{k}} mn,\sigma} \gamma^\dag_{{\bm{k}},m,\sigma} \lp  \frac{1}{\mathsf{V}}\sum_{\mbf{q}} [(U^{\sigma\dag}({\bm{k}}+\bm{q})  \tau^\sigma({\bm{k}}+\bm{q})U^\sigma({\bm{k}}) )^\dag(1 - \mathbbm{1}_{\text{flat}})U^{\sigma\dag}({\bm{k}}+\bm{q})  \tau^\sigma({\bm{k}}+\bm{q})U^\sigma({\bm{k}}) ]_{mn} \rp \gamma_{{\bm{k}},n,\sigma} \\
&= U \sum_{{\bm{k}} mn,\sigma} \gamma^\dag_{{\bm{k}},m,\sigma} \lp  \frac{1}{\mathsf{V}}\sum_{\mbf{q}} [U^{\sigma \dag}(\bm{k}) \tau^{\sigma \dag}(\bm{q}) Q^\sigma(\bm{q}) \tau^\sigma({\bm{q}})U^\sigma(\bm{k}) ]_{mn} \rp \gamma_{{\bm{k}},n,\sigma} \\
\eea
which can be neglected as long as it acts as a chemical potential in the flat band Hilbert space. In the following construction, we will show that this can be ensured.

This completes our construction of the projected model. Note that $H_{int,flat}$ is the full Hamiltonian after projection, since the projected kinetic term $\bar{H}_0$ is assumed to be a constant due to the flatness of the bands.

We now look for additional symmetries of the $\bar{S}_\mbf{q}$ spin operators. We make the generalized eta pair ansatz
\bea
\eta^\dag &= \sum_{{\bm{k}},mn} F_{mn}({\bm{k}}) \gamma^\dag_{{\bm{k}},m,\u} \gamma^\dag_{-{\bm{k}},n,\d}
\eea
where $F_{mn}({\bm{k}})$ is pairing matrix among the flat bands index by $mn$. (In the Main Text, we use the notation $\eta^\dag = \hat{O}^{\text{p.-p.}\dag}$. We change notation here to alert the reader to the generalization of the formalism developed, where now $S_\mbf{q}$ is not a Hermitian operator.) We compute the commutator:
\bea
\null [S_{\mbf{q}}, \eta^\dag] &= \frac{1}{\sqrt{\mathsf{V}}} \sum_{{\bm{k}} mn, {\bm{k}}' m'n',\sigma}s^z_\sigma [U^{\sigma \dag}({\bm{k}})  \tau^\sigma({\bm{k}}) U^\sigma({\bm{k}}-\mbf{q})]_{nm} F_{m'n'}({\bm{k}}') [\gamma^\dag_{{\bm{k}},n,\sigma}  \gamma_{{\bm{k}}-\mbf{q},m,\sigma} , \gamma^\dag_{{\bm{k}}',m',\u} \gamma^\dag_{-{\bm{k}}',n',\d}] \\
&= \frac{1}{\sqrt{\mathsf{V}}} \sum_{{\bm{k}} mn, {\bm{k}}' m'n',\sigma}s^z_\sigma [U^{\sigma \dag}({\bm{k}})  \tau^\sigma({\bm{k}}) U^\sigma({\bm{k}}-\mbf{q})]_{nm} F_{m'n'}({\bm{k}}')  \gamma^\dag_{{\bm{k}},n,\sigma} (\delta_{{\bm{k}}-\mbf{q}, {\bm{k}}'} \delta_{mm'} \delta_{\sigma \u} \gamma^\dag_{-{\bm{k}}',n',\d} - \delta_{{\bm{k}}-\mbf{q},- {\bm{k}}'} \delta_{mn'} \delta_{\sigma \d}  \gamma^\dag_{{\bm{k}}',m',\u}) \\
&= \frac{1}{\sqrt{\mathsf{V}}} \sum_{{\bm{k}} n n' m'} ([U^{\u \dag}({\bm{k}})  \tau^\u({\bm{k}}) U^\u({\bm{k}}-\mbf{q})]_{nm'} F_{m'n'}({\bm{k}}-\mbf{q}) \gamma^\dag_{{\bm{k}},n,\u} \gamma^\dag_{\mbf{q}-{\bm{k}},n',\d}\\
&\qquad\qquad\qquad + [U^{\d \dag}({\bm{k}})  \tau^\d({\bm{k}}) U^\d({\bm{k}}-\mbf{q})]_{nn'} F_{m'n'}(\mbf{q}-{\bm{k}})  \gamma^\dag_{{\bm{k}},n,\d} \gamma^\dag_{\mbf{q}-{\bm{k}},m',\u}) \\
&= \frac{1}{\sqrt{\mathsf{V}}} \sum_{{\bm{k}} n n' m'} ([U^{\u \dag}({\bm{k}})  \tau^\u({\bm{k}})  U^\u({\bm{k}}-\mbf{q})]_{nm'} F_{m'n'}({\bm{k}}-\mbf{q}) - [U^{\d \dag}(\mbf{q}-{\bm{k}})  \tau^\d(\mbf{q}-{\bm{k}})  U^\d(-{\bm{k}})]_{n'm'} F_{nm'}({\bm{k}})) \gamma^\dag_{{\bm{k}},n,\u} \gamma^\dag_{\mbf{q}-{\bm{k}},n',\d} \ . \\
\eea
Thus for the commutator to vanish, we have the condition (analogous to \Eq{eq: pp primitive matrix condition})
\bea
\label{eq:Ucomspin}
U^{\u \dag}({\bm{k}})  \tau^\u({\bm{k}}) U^\u({\bm{k}}-\mbf{q}) F({\bm{k}}-\mbf{q}) &= F({\bm{k}}) [U^{\d \dag}(\mbf{q}-{\bm{k}})  \tau^\d(\mbf{q}-{\bm{k}})  U^\d(-{\bm{k}})]^T \\
\eea
written in matrix notation. By $\mathcal{T}$ and $S_z$, we write $U^\u({\bm{k}}) = U({\bm{k}})$ and $U^\d({\bm{k}}) = U^*(-{\bm{k}})$, and $\tau({\bm{k}}) = \tau^\u({\bm{k}}) = \tau^\d(-{\bm{k}})^*$. Then \Eq{eq:Ucomspin} is equivalent to
\bea
\label{eq:Ucomspin2}
U^\dag({\bm{k}})  \tau({\bm{k}}) U({\bm{k}}-\mbf{q}) F({\bm{k}}-\mbf{q}) &= F({\bm{k}}) U^\dag({\bm{k}}) \tau^\dag({\bm{k}}-\mbf{q})  U({\bm{k}}-\mbf{q})\ . \\
\eea
We will now find a solution to this equation for $\tau({\bm{k}})$ in terms of $U({\bm{k}})$. Note that the usual s-wave case can be recovered by taking $\tau_{\al\be}({\bm{k}}) = \delta_{\al\be}$ and $F_{mn}({\bm{k}}) = \delta_{mn}$. However, we can find a nontrivial solution by identifying a different ansatz for $\tau({\bm{k}})$ in the form
\bea
\label{eq:taudef}
\tau({\bm{k}}) = U({\bm{k}}) F({\bm{k}}) U^\dag({\bm{k}})
\eea
which satisfies \Eq{eq:Ucomspin2} as long as $F({\bm{k}})$  is Hermitian:
\bea
U^\dag({\bm{k}})  U({\bm{k}}) F({\bm{k}}) U^\dag({\bm{k}})U({\bm{k}}-\mbf{q}) F({\bm{k}}-\mbf{q}) &= F({\bm{k}}) U^\dag({\bm{k}})U({\bm{k}}-\mbf{q}) F({\bm{k}}-\mbf{q}) U^\dag({\bm{k}}-\mbf{q}) U({\bm{k}}-\mbf{q}) \\
F({\bm{k}}) U^\dag({\bm{k}})U({\bm{k}}-\mbf{q}) F({\bm{k}}-\mbf{q}) &= F({\bm{k}}) U^\dag({\bm{k}})U({\bm{k}}-\mbf{q}) F({\bm{k}}-\mbf{q})\\
\eea
so we see that the form of the interaction $S_\mbf{q}$ determined by $\tau({\bm{k}})$ is directly related to the Cooper pair wavefunction $F({\bm{k}})$. Secondly, \Eq{eq:onebodylieb} automatically shows that the one-body term $H_1$ vanishes, since $\mbf{Q}(\mbf{q}) \tau(\mbf{q}) = 0$. Thus we see that
\bea
\ket{n} = \eta^{\dag n} \ket{0}
\eea
are (unnormalized) pairing groundstates at fixed particle number. This follows from $ [\bar{S}_{\mbf{q}}, \eta^\dag] = 0$ because
\bea
\braket{n| H_{int,flat} |n} &= \sum_\mbf{q}\big( \bar{S}_\mbf{q} \eta^{\dag n}\ket{0} \big)^2 = 0
\eea
and thus $\ket{n}$ must be zero energy eigenstates because of the positive semi-definite form of $H_{int,flat}$. Hence they are groundstates. This approach agrees with the particle-number polarized groundstates of the psuedo-Hamiltonian in \ref{sec: ground state}.

We now will show that $p$-wave, $d$-wave, and $f$-wave Cooper pairing groundstates can be constructed from this more general ansatz, and then we will give a minimal example.

Using \Eq{eq:taudef}, we see that the Cooper pair can be written
\bea
\eta^\dag &= \sum_{{\bm{k}},mn} \gamma^\dag_{{\bm{k}},m,\u} F_{mn}({\bm{k}})\gamma^\dag_{-{\bm{k}},n,\d} \\
&= \sum_{{\bm{k}},mn} \bar{c}^\dag_{{\bm{k}},\al,\u} U^\u_{\al,m}({\bm{k}}) F_{mn}({\bm{k}})U^\d_{\al,n}(-{\bm{k}}) \bar{c}^\dag_{-{\bm{k}},\be,\d} \\
&= \sum_{{\bm{k}},mn} \bar{c}^\dag_{{\bm{k}},\al,\u} U_{\al,m}({\bm{k}}) F_{mn}({\bm{k}})U^*_{\al,n}({\bm{k}}) \bar{c}^\dag_{-{\bm{k}},\be,\d} \\
&= \sum_{{\bm{k}},mn} \bar{c}^\dag_{{\bm{k}},\al,\u}\tau_{\al \be}({\bm{k}}) \bar{c}^\dag_{-{\bm{k}},\be,\d} \\
\eea
so that $\tau({\bm{k}})$ is the order parameter/Cooper pair wavefunction. To have $\eta^\dag$ transform nontrivially, we check its symmetry transformation properties:
\bea
\label{eq:CPgtrans}
g \eta^\dag g^\dag &= \sum_{{\bm{k}},mn} \bar{c}^\dag_{{\bm{k}},\al,\u}[D[g]\tau(g^{-1}{\bm{k}})D^\dag[g]]_{\al \be} \bar{c}^\dag_{-{\bm{k}},\be,\d} \\
\eea
where $g \bar{c}^\dag_{{\bm{k}},\al,\sigma}g^\dag = \sum_{\al'}\bar{c}^\dag_{g{\bm{k}},\al',\sigma}D^\sigma_{\al' \al}[g]$ and $D^\sigma[g]$ is the representation of the symmetry $g$ on the spin $\sigma$ electrons, which obey $D^\u[g] = D^\d[g]^*$ by $\mathcal{T}$. Because we required $\tau({\bm{k}})$ to be Hermitian in order to satisfy \Eq{eq:Ucomspin2}, the only allowed transformations of $\tau({\bm{k}})$ are
\bea
D[g]\tau(g^{-1}{\bm{k}})D^\dag[g] = \pm \tau({\bm{k}})
\eea
which from \Eq{eq:CPgtrans} yields
\bea
g \eta^\dag g^\dag &= \sum_{{\bm{k}},mn} \bar{c}^\dag_{{\bm{k}},\al,\u} (\pm\tau_{\al \be}({\bm{k}})) \bar{c}^\dag_{-{\bm{k}},\be,\d} = \pm \eta^\dag \ .
\eea
We see that only $\pm1$ eigenvalues of $g = C_n$ are allowed, which is expected given that we have enforced spin-ful time-reversal symmetry and the groundstate it unique.  The $+$ sign is conventional $s$-wave. For $g = C_2,C_4,C_6$, the $-$ sign corresponds to $p-,d-,$ and $f-$wave symmetry respectively (i.e. Cooper pairs with angular momentum $\ell = 1,2,3$ mod $2,4,6$) and spin zero. To illustrate that it is possible to achieve these higher angular momentum pairing symmetries in a strictly local model, we first give a minimal 1D example with $p$-wave symmetry.

We consider a 1D SSH-type chain with $s$ and $p$ orbitals at the origin. We pick the single-particle Hamiltonian to be
\bea
h({\bm{k}}) = t (\sigma_0 - \sigma_3 \cos k + \sigma_2 \sin k)
\eea
with the symmetries inversion $D[C_2] = \sigma_3$ and spinless time-reversal $D[T] = K$. The eigenvalues of $h({\bm{k}})$ are $2t,0$, corresponding to two flat bands. The zero energy eigenvector is
\bea
U({\bm{k}}) = e^{i \frac{k}{2}} \bpm - i \cos \frac{k}{2} \\ \sin \frac{k}{2} \epm
\eea
which is periodic on the Brillouin zone $k \in (0,2\pi)$ and is normalized. If we pick $F({\bm{k}}) = \sin k$ (note that there is only one zero energy flat band per spin, so $F({\bm{k}})$ is a $1\times 1$ matrix), we obtain a $p$-wave superconductor described by the interaction
\bea
\label{eq:tauSSHex}
\tau({\bm{k}}) &= \frac{1}{4} \left(
\begin{array}{cc}
  2 \sin k + \sin 2 k &  -i (1- \cos 2 k ) \\
i (1 - \cos 2 k) & 2 \sin k -\sin 2 k \\
\end{array}
\right) = \frac{1}{4} \big(2 \sigma_0 \sin k - \sigma_3 \sin 2k + \sigma_2 (1- \cos 2k) \big)
\eea
which obeys $\sigma_3 \tau(-{\bm{k}}) \sigma_3 = - \tau({\bm{k}})$. A key feature of $\tau({\bm{k}})$ is that is is supported on strictly finite harmonics $e^{i nk}, n = 0, \pm 1, \pm2$, and as such is a compact obstructed atomic limit arising from a model with completely flat bands \cite{2021PhRvB.104t1114S,2023arXiv230906487S,2023arXiv230211608S,2021JPhA...54G5302S}. Although such bands have zero Berry curvature, they exhibit nontrivial many-body phases \cite{PhysRevB.108.195102,PhysRevLett.128.087002} thanks to quantum geometry. This work shows they can also exhibit exotic superconductivity. Since $\tau(\mbf{R}-\mbf{R}')$ is finitely supported, the unprojected Hamiltonian is strictly local. This is usually a desirable property of tight-binding models.

Note that the $s$-wave Cooper pair with $F({\bm{k}}) = 1$ is not a symmetry (see \Eq{eq:Ucomspin}) for the choice of $\tau({\bm{k}})$ in \Eq{eq:tauSSHex}, and thus it is reasonable to expect that the $p$-wave groundstates constructed here are unique. Indeed, Remark (2) (see Ref. \cite{PhysRevLett.62.1201}) of Lieb's first theorem proves this is the case.

Finally, we discuss the excitations above the groundstate. It is important to note that, unlike in the $s$-wave case, $\bar{S}_\mbf{q} \neq \bar{S}^\dag_{-\mbf{q}}$ as is apparent from the non-hermiticity of $S_\mbf{R}$ in \Eq{eq:Srnonherm}. Hence although $[\bar{S}_\mbf{q},\eta^\dag] = 0$, generically $\bar{S}^\dag_\mbf{q}$ and $\eta^\dag$ will \emph{not} commute and thus $\eta^\dag$ is not a symmetry of the Hamiltonian. This means that $\gamma^\dag_{{\bm{k}},m,\sigma} \, \eta^{\dag n} \ket{0}$ will not be exact eigenstates. However, we can use such states to get variational upper bounds for the charge 1 excitations. To see this, note that
\bea
\braket{n|\gamma_{{\bm{k}},m,\sigma} H_{int,flat} \gamma^\dag_{{\bm{k}},m,\sigma}|n} &= U \sum_\mbf{q} \braket{n|\gamma_{{\bm{k}},m,\sigma} \bar{S}^\dag_\mbf{q} \bar{S}_\mbf{q} \gamma^\dag_{{\bm{k}},m,\sigma}|n} \\
&= U \sum_\mbf{q} \braket{n|[\gamma_{{\bm{k}},m,\sigma}, \bar{S}^\dag_\mbf{q}] [\bar{S}_\mbf{q}, \gamma^\dag_{{\bm{k}},m,\sigma}]|n} \\
&= \frac{U}{\mathsf{V}} \sum_{\mbf{q},m'm''} \tau^{\sigma *}_{m'' m}({\bm{k}},\mbf{q}) \braket{n| \gamma_{{\bm{k}}+\mbf{q},m'',\sigma}  \gamma^\dag_{{\bm{k}}+\mbf{q},m',\sigma}|n} \tau^\sigma_{m'm}({\bm{k}},\mbf{q})\\
\eea
where we used \Eq{eq:Sqband} to compute
\bea
\null [\bar{S}_\mbf{q}, \gamma^\dag_{{\bm{k}},m,\sigma}] &= \frac{1}{\sqrt{\mathsf{V}}} \sum_{{\bm{k}}', m' n,\sigma}s^z_\sigma [U^{\sigma\dag}({\bm{k}}')  \tau^\sigma({\bm{k}}')U^\sigma({\bm{k}}'-\mbf{q})]_{nm'} [\gamma^\dag_{{\bm{k}}',n,\sigma}  \gamma_{{\bm{k}}'-\mbf{q},m',\sigma} ,\gamma^\dag_{{\bm{k}},m,\sigma}] \\
&= \frac{1}{\sqrt{\mathsf{V}}} \sum_{n}s^z_\sigma [U^{\sigma\dag}({\bm{k}}+\mbf{q})  \tau^\sigma({\bm{k}}+\mbf{q})U^\sigma({\bm{k}})]_{nm} \gamma^\dag_{{\bm{k}}+\mbf{q},n,\sigma} \\
&= \frac{1}{\sqrt{\mathsf{V}}} \sum_{n}s^z_\sigma \gamma^\dag_{{\bm{k}}+\mbf{q},n,\sigma} \tau^\sigma_{nm}({\bm{k}},\mbf{q}),\qquad [\tau^\sigma({\bm{k}},\mbf{q})]_{mn} = [U^{\sigma\dag}({\bm{k}}+\mbf{q})  \tau^\sigma({\bm{k}}+\mbf{q})U^\sigma({\bm{k}})]_{mn} \ . \\
\eea
We now use the matrix inequality $ \braket{n| \gamma_{{\bm{k}},m'',\sigma}  \gamma^\dag_{{\bm{k}},m',\sigma}|n} \preceq [\braket{n|n} \mathbbm{1}]_{mm'}$ to observe that
\bea
\label{eq:ineqbaby}
\frac{\braket{n|\gamma_{{\bm{k}},m,\sigma} H_{int,flat} \gamma^\dag_{{\bm{k}},m,\sigma}|n}}{\braket{n|n}} &\leq \frac{U}{\mathsf{V}} \sum_{\mbf{q}} [\tau^{\sigma \dag}({\bm{k}},\mbf{q}) \tau^\sigma({\bm{k}},\mbf{q})]_{mm} = [\frac{U}{\mathsf{V}} \sum_{\mbf{q}} \tau^{\sigma \dag}({\bm{k}},\mbf{q}) \tau^\sigma({\bm{k}},\mbf{q})]_{mm} \equiv [R^\sigma({\bm{k}})]_{mm} \ . \\
\eea
Using \Eq{eq:tauSSHex}, we compute
\bea
\tau^\sigma({\bm{k}},\mbf{q}) &= \frac{1+e^{iq}}{2} \sin k, \qquad R^\sigma({\bm{k}}) = \frac{U}{2} \sin^2 k
\eea
which has zeros at ${\bm{k}}=0,\pi$ corresponding to the nodes of the Cooper pair $\eta^\dag$. Since the excitation energies must also be positive semi-definite, the existence of nodes in the upper bound $R^\sigma({\bm{k}})$ proves nodes in the full many-body spectrum. The inequality in \Eq{eq:ineqbaby} is sufficient to prove the desired result in the one-band case considered here, and in general can be improved using the generating function formalism in \cite{2017ScPP....3...43V,herzog2022many} to calculate the normalization $\braket{n|\gamma_{{\bm{k}},m,\sigma} \gamma^\dag_{{\bm{k}},m,\sigma}|n}$ exactly.

\bibliographystyle{apsrev4-1}
\bibliography{ref}